# Experimental Nanoscale Thermodynamics

G. A. D. Briggs* and J. Schmiedmayer†

* University of Oxford, Department of Materials, 16 Parks Road, Oxford OX1 3PH, UK
QuantrolOx, Keilaranta 10 E, 02150 Espoo, Finland
† Vienna Center for Quantum Science and Technology (VCQ), Atominstitut, Technische Universität Wien, Vienna, Austria

Modern fabrication techniques have allowed artificial systems to be constructed on a nanometre length scale. The ideas underlying thermodynamics, a field invented to describe large macroscopic systems, need to be modified to describe much smaller systems. There are implications for our understanding of entropy, cooling, the arrow of time, heat flow, engines, erasure, and mutual information. The theoretical concepts have been implemented and tested in a wide range of experiments, many of which are at low temperatures. Nanoscale thermodynamics is not only a descriptive tool of nanoscale systems, but a design principle that helps us to build new architectures, processors and learning machines.

Advances in microscopy have opened new possibilities for fabricating and operating devices on the nanoscale, alongside advances in the quantum control of systems such as cold gases. The purpose of this article is to present a selection of experiments that illustrate some of the key ways in which nanoscale systems reveal new aspects of thermodynamics which would not be apparent or even accessible in larger systems. This does not overturn established thermodynamics—heaven forbid!—but it does provide new insights and perhaps even new science.

We first introduce the core concepts which we shall be exploring, emphasising the relationship to concepts of information. The body of the paper is then taken up with examples of experiments in which these ideas have been implemented and tested. Rather than attempting a comprehensive review, the aim is to give a selection of case studies which bring the concepts to life. We apologise to anyone whose favourite experiment—whether their own or someone else's—we have negligently omitted, and we apologise too to anyone whose experiment we have not described as they would have wished. We can only echo the defence of Dr Johnson in mistakenly defining *pastern* as the knee of a horse. When a woman asked him why he had made this error, he reportedly replied "Ignorance, Madam, pure ignorance." Thermodynamics was originally stimulated by engineering problems. We therefore conclude the article with a discussion of the applications of nanoscale thermodynamics to classical and quantum information processing.

In the article we shall move between six concepts of entropy:

(i) thermodynamic entropy, defined operationally for a chosen macroscopic or reduced description as a change $\delta S := \int \frac{\delta Q}{T}$, where $Q$ is heat transferred to a system from the surroundings and $T$ is the temperature.
(ii) Boltzmann entropy $S := k_B \ln W$, for a macrostate with $W$ possible microstates of equal probability $p = 1/W$.
(iii) Gibbs entropy $S := -k_B \sum(p \ln p)$, for a microstate with possible microstates of differing probabilities $p$.
(iv) Shannon entropy $S(X) := -\sum_{x\in\mathcal{X}} p(x) \ln p(x)$, where $p(x)$ is the probability of each outcome within the set $\mathcal{X}$.
(v) von Neumann entropy of a quantum state $S(\rho) := -\mathrm{Tr}(\rho \ln \rho)$, where $\rho$ is the density matrix; for an isolated quantum system evolving unitarily, $S(\rho_{\mathrm{total}})$ is constant.
(vi) entanglement entropy of subsystem A in a bipartite quantum state $\rho_{AB}$: $S(\rho_A) := -\mathrm{Tr}_A[\rho_A \ln \rho_A]$, where $\rho_A = \mathrm{Tr}_B[\rho_{AB}]$ is the reduced density matrix obtained by tracing out B. This is the von Neumann entropy of the reduced state; unlike $S(\rho_{\mathrm{total}})$, it can grow under unitary evolution as correlations spread across the partition.

The Boltzmann constant $k_B$ is the empirical bridge between the thermodynamic entropy (in units of J $K^{-1}$) and the information-theoretic entropies (dimensionless, or measured in bits when ln is replaced by $\log_2$). Multiplying the Shannon entropy of the equilibrium distribution of microstates by $k_B$ recovers the Gibbs or thermodynamic entropy; the same conversion applies to the von Neumann and entanglement entropies whenever they enter thermodynamic statements. Most of the bridges between information-theoretic and thermodynamic statements in this paper rest on this identification, applied in either direction depending on context. The reader may find it helpful to keep these distinct but related concepts in mind.

# 1 Thermodynamics as selective ignorance

It sometimes comes as a surprise, even to many trained in mathematical statistics, to learn that probability is subjective. The probability for a *you* may be different from the probability for a *me* because we may have access to different information. Probability depends on what you already know. Suppose a coin is tossed, and before either of us looks I cover it with my hand. What is the probability that it is heads? If everything is fair, then 50%. Now I look at it, but without letting you see it. What is now the probability that it is heads? For *you* it is still 50%. But I now know, and so for *me* it is either 0 or 1 depending on what I saw. This kind of thinking is mathematically embodied in Bayesian probability, where the probability of a prior belief being true can be updated in the light of new information.

Many parameters in thermodynamics are not amenable to direct observation. When we speak of measuring temperature, this is generally shorthand for measuring some observable indicator from which we can deduce the temperature. It might be the length of a mercury column in a capillary tube attached to a bulb which is in equilibrium with whatever it is whose temperature we wish to know. What we actually *measure* is the position of the end of the mercury column, which then enables us to *determine* the temperature of the fluid or body of interest. We are so used to thinking of this as measuring the temperature that we mark a temperature scale on the barrel of our thermometers.

The operational distinction between measurement and determination of parameters becomes all the more significant in thermodynamic experiments concerning entropy and information. In nanoscale thermodynamics this operational restriction is not a pedantic detail: it *defines* the effective state space, and therefore what entropy means. Irreversibility enters when we restrict attention to a reduced description—coarse graining, limited observables, finite resolution, or tracing out inaccessible degrees of freedom—so that reduced entropies (and mutual information) can change.

One way of thinking of conventional thermodynamics is as the powerful science of *selective ignorance*. To adapt William Shakespeare, "Some are born ignorant, some achieve ignorance, and some have ignorance thrust upon them."[1] Pierre Laplace famously postulated that an intelligence that knew the position and momentum of all the atoms in the universe and the forces acting on them could completely predict their future behaviour.[2] Whatever the status of that assertion, it remains utterly impractical to have such knowledge even for a mole of gas as a closed system. It is a triumph of thermodynamics that by abandoning the attempt and restricting consideration to a few key parameters, an enormous amount can be said about the physics and chemistry of order Avogadro number of particles.

Thermodynamics originated in that way. The pioneers of thermodynamics developed a theoretical framework for the constraints on certain kinds of state transformations, where the states were characterised by parameters such as pressure, volume, and temperature. Entropy was not a statistical property, but a macroscopic quantity with units which enabled states to be ordered with respect to specific operational transformations. Subsequently entropy came to be defined in informational terms, in a mathematical way that is independent of any underlying physics. The first formulation is physical and has units of energy divided by temperature. The second is all about probability over discrete or continuous spaces and has no units; some would argue because it is independent of physical things in the world. These two concepts of entropy can be experimentally connected. Thermodynamic entropy appears on one side of the equation, information on the other, connected by the Boltzmann constant $k_B$ with units J $K^{-1}$.

The role of knowledge in thermodynamics is illustrated by what became known as the Gibbs paradox. Suppose I have $N$ molecules of nitrogen confined by a partition to the left half of a container. If I designate the left half as 0 and the right half as 1, then I have one bit of information about the location of each particle; namely 0 for each molecule. If I instantaneously remove the partition, the gas will undergo a Joule expansion. This is an irreversible process which has the effect of reducing the information about the location of each molecule by one bit, because I no longer know whether each is in position 0 or 1. There is therefore an increase in entropy $\Delta S = N\, k_B \ln 2$ (interpretable as: *either* a thermodynamic entropy change inferable from heat in a slow mixing process; *or* $k_B$ times the Shannon entropy of the post-mixing position distribution). Now suppose I start again with $N$ molecules of nitrogen in the left half and this time also with $N$ molecules of oxygen in the right half. If I now remove the partition, each gas undergoes a Joule expansion. The mixing of the two gases is clearly irreversible, this time with increase in entropy $\Delta S = 2N\, k_B \ln 2$. But if instead both sides had contained nitrogen, there would be no entropy change at all. Why does it make a difference to the entropy change what the molecules of the gas consist of? Gibbs himself worried greatly about what to him was a paradox, and it led to a new insight into the profound role played by indistinguishability. It means that states that differ only by a permutation of identical particles should be considered as the same state. This has further implications for the statistics of quantum particles.

We can press the thought experiment further to highlight the role of knowledge. Suppose I do the second experiment above, with $N$ molecules of nitrogen on each side. After I have removed the partition and the previously separated nitrogen molecules have thoroughly mixed, there is no change in entropy. That would be a correct conclusion. But I know something that I did not tell you. In the left compartment I had put $^{14}$N isotopes, and in the right compartment $^{15}$N isotopes. So with a mass spectrometer I can distinguish the molecules. This is no different from the Gibbs paradox, but it emphasises the role of information. On the basis of chemical information, there is no entropy change, but on the basis of isotopic information there is an entropy increase $\Delta S = 2N\, k_B \ln 2$. Insofar as knowledge is subjective,[3] then the entropy change for me is different from the entropy change for you. This can be related to work, if we consider the partition to be a piston which is selectively permeable only to one of the two isotopes.

[1] Shakespeare, William, *Twelfth Night*, Act 2, Scene 5.

[2] Laplace, Pierre Simon, *A Philosophical Essay on Probabilities*, translated into English from the original French 6th ed. by Truscott, F.W. and Emory, F.L., Dover Publications (New York, 1951) p.4.

## 1.1 Maxwell, Szilárd, Brillouin, Landauer, and Bennett

On 11 December 1867 James Clerk Maxwell wrote to his friend Peter Tait about a thought experiment. Maxwell had long since derived his eponymous distribution of the speeds of molecules in a gas, and he wondered what would happen if you could sort the molecules into fast and slow. He imagined "a being whose faculties are so sharpened that he can follow every molecule in its course". The idea was to have gas contained in a chamber with a partition in the middle. The partition would have a hole which the being could open or close depending on the speed of a molecule approaching. The hole would be opened only for swifter molecules arriving from one side and only for slower molecules approaching from the other side. In this way the temperatures of the gases either side of the partition would gradually diverge, and this could provide a hot and a cold bath for a heat engine, thereby seriously violating the Second Law of Thermodynamics. Two years later Maxwell wrote about the same idea in a letter to John William Strutt, who as Lord Rayleigh was to become his successor as the Cavendish Professor of Physics at the University of Cambridge. Lord Kelvin in a letter to *Nature* in 1874 dubbed the being a 'demon', albeit without implying moral attributes (rather as we may speak of wicked problems without implying a moral judgement), and the designation has stuck ever since.

Maxwell's provocative thought experiment stimulated much further debate about how the paradox could be resolved, with several distinguished philosophers jumping in. Leó Szilárd and Léon Brillouin each pointed out that the demon would need to expend energy to make the measurement, which would exceed the energy that could be extracted from the heat engine. This was close, but not quite right. The definitive answer came much later in two steps. First, Rolf Landauer argued that whenever one bit of information is erased, there is an inescapable entropic cost of $k_B \ln 2$, with a corresponding generation of heat $k_B T \ln 2$. That result can be derived by considering a single molecule in a vessel, and moving a partition to halve the volume available to the molecule.

For those who wish to experience Maxwell's demon for themselves, there is a game available on the web in which each reset of the shutter memory costs $k_B \ln 2$ J K$^{-1}$.[4] The game was devised by a former student. Try it and see how close you can get to the thermodynamic limit!

The second insight came from Charles Bennett, who recognised that although a demon might be able to do the operation a few times, and remember each occurrence, the demon would eventually reach the limit of its memory and would have to erase previous information in order to have capacity for further observations. Landauer's principle (the inescapable entropic cost of $k_B \ln 2$ for the erasure of a single bit) provides the quantitative bridge between information loss and thermodynamic irreversibility across all the scales discussed so far.

## 1.2 Many and few: Avogadro scale and nanoscale thermodynamics

Books on thermodynamics and statistical mechanics often open with a statement that this is a subject for handling large numbers of particles. For example, "Statistical physics is devoted to the study of the physical properties of

---

[3] 'There is an implication in such an expression [for the error in an "experimentally determined" probability] that there *is* a "true" or "correct" probability which *could* be computed if we knew enough, and that the observation may be in "error" due to a fluctuation. There is, however, no way to make such thinking logically consistent. It is probably better to realize that the probability concept is in a sense subjective, that it is always based on uncertain knowledge, and that its quantitative evaluation is subject to change as we obtain more information.' *The Feynman Lectures on Physics* I:6-6 (1963)

'Probability does not exist.' [Opening statement of Adrian Smith's translation of *Theory of Probability* by Bruno de Finetti (Wiley 1974)]. 'This may seem extreme and, although it is beyond my pay grade to start discussing what it means for something to "exist", I interpret this as de Finetti simply proclaiming that probability was not an objective probability of the world. I fully absorbed this sentiment in my youth, and in fifty years I have never shifted from the view that probabilities are subjective judgements, even if they are based on considerations of physical symmetries, data analyses or complex models. The only possible exception is at the subatomic quantum level, where it could be argued that true objective and determined probabilities exist.'

[4] https://maxwellsgame.brandonseverin.com/

macroscopic systems, i.e. systems consisting of a very large number of atoms or molecules."[5] Or again, "The subject of thermal physics involves studying assemblies of large numbers of atoms. As we shall see, it is the large numbers involved in macroscopic systems that allow us to treat some of their properties in a statistical fashion."[6]

We use the term *Avogadro scale* for such systems in which the number of accessible microstates is so vast that individual trajectories are permanently beyond reach and thermodynamics must proceed through ensemble averages. A gas in a piston is the prototype: the number of possible states becomes practically uncountable, like the proverbial grains of sand on the seashore. This is the regime for which classical thermodynamics was developed, and in which it works without qualification.

*Nanoscale thermodynamics*, by contrast, concerns systems in which the number of thermodynamically relevant states is small enough to be individually resolved, tracked, or counted, and in which fluctuations around the mean are large enough to be directly observed. Nanoscale here need not refer to a physical size of $\sim 10^{-9}$ m, although often the experimental systems are small, nor a temperature of $\sim 10^{-9}$ K, although often the experiments are performed at $< 1$ K. The defining criteria are combinatorial and statistical, not geometric: we are concerned with the new aspects of thermodynamics which arise when the state count becomes tractable and when deviations from mean behaviour are no longer negligible corrections but may themselves be the signal. The system might be a vibrating nanotube or membrane, where what we care about is the displacement in a fundamental vibrational mode and all other degrees of freedom play no part — and where the thermal fluctuations of that single mode can be followed in real time. A particularly clear example is excess electrons in quantum dots, where you can directly observe both the number and its fluctuations. That is nanoscale for the purpose of thermodynamics.

The change in dominant physics as size decreases is a recurring theme across science. A dust particle suspended in air is visibly deflected by individual molecular collisions that a marble would never notice; an insect walks on water because surface tension, negligible compared with the weight of a human, overwhelms gravity at that scale; a bacterium swimming through water is dominated by viscosity, with inertia playing almost no role. In each case an effect that is a minor correction at large scales becomes the governing phenomenon at smaller ones. In thermodynamics the key transition is from a regime in which states are uncountable and fluctuations are negligible, to one in which individual states can be resolved and deviations from the mean may themselves be the object of study. What is unfeasible for Avogadro-scale systems may become tractable in the nanoscale regime, and questions arise that were never relevant to the designers of steam engines or the statistical physicists who served them.

Long before modern concepts of nanoscience were formulated, Albert Einstein foresaw that nanoscale thermodynamics would be different from the conventional thermodynamics which he had received. On 30 April 1905, his *annus mirabilis*, Einstein completed his doctoral thesis, *Eine neue Bestimmung der Moleküldimensionen* (A New Determination of Molecular Dimensions). This applied a diffusion equation to data from sugar molecules to give an account of Brownian motion and provide a determination of the Avogadro number which helped to convince sceptics that atoms are real. Within two weeks he wrote from Berne,

> "If the movement discussed here can actually be observed (together with the laws relating to it that one would expect to find), then classical thermodynamics can no longer be looked upon as applicable with precision to bodies even of dimensions distinguishable in a microscope: an exact determination of actual atomic dimensions is then possible."[7]

At the nanoscale, thermodynamics is no longer only the theory of average heat and work, but also of which degrees of freedom are accessible, which correlations are resolved, and what irreversibility is created when information is discarded.

This Avogadro/nanoscale boundary sharpens into four specific contrasts that permeate this review:

---

[5] *Statistical Physics* F. Mandl (Wiley, 2nd Edn 1988) p. 1.

[6] Stephen J. Blundell and Katherine M. Blundell, *Concepts in Thermal Physics* (Oxford University Press, Second Edition 2009) p. 2.

[7] A. Einstein, On the movement of small particles suspended in a stationary liquid demanded by the molecular-kinetic theory of heat (Berne, 1905). English translation in *Investigations on the Theory of the Brownian Movement* pp. 1-2 (Ed. R. Fürth, Trans A. D. Cowper, Methuen & Co. Ltd. London, 2026)

1. *How is small different from big?* As nanoscale size is approached, so the effect of individual disturbances becomes significant and measurable. That is what the Scottish botanist Robert Brown observed in 1827, when with a microscope he saw pollen particles jiggling about in water. In his 1905 *Annus Mirabilis* Einstein published his derivation from Brownian motion of order Avogadro number, which helped to convince the world that molecules are real. How do such observations affect interactions?
2. *How is few different from many?* It is unfeasible to track of order Avogadro number of molecules. But as the number of particles is reduced, it may become possible to forgo ignorance of individual positions and momenta. What does that do to the concept of entropy? Does entropy become an observable quantity? What difference does charge quantisation make, when a partial derivative of energy with respect to particle number cannot be defined?
3. *How is quantum different from classical?* What difference does it make to the statistics when the quantisation of energy levels becomes significant? How do quantum fluctuations augment or overtake thermal fluctuations? How do superposition and entanglement contribute to mutual information, and how does that affect entropy? What is work in a closed quantum system, and how do the First and Second Laws apply?
4. *How is cold different from hot*? Many, or perhaps most, nanoscale thermodynamics experiments are carried out at low temperatures. What thermodynamic phenomena begin to dominate as thermal excitations diminish? What difference does it make when quantum noise becomes greater than thermal noise?

These four contrasts are not mutually exclusive — a system can simultaneously be small, few-particle, quantum, and cold. Often the most revealing experiments are those in which all four conditions conspire. But they are conceptually distinct, and distinguishing them helps to clarify which feature of a given experiment is doing the thermodynamic work. Avogadro-scale results — the Clausius inequality, the equipartition theorem, the Gibbs entropy — remain valid at the nanoscale as ensemble averages. What the nanoscale adds is resolution: the ability to observe the distribution around those averages, to follow individual trajectories, and to test the equalities (Jarzynski, Crooks) that underlie the averaged inequalities. The Second Law is not overturned; it is sharpened into a probabilistic statement whose full content is only visible when fluctuations are accessible.

Rather than addressing these questions in isolation, we have selected specific experiments that illuminate them, starting with the fundamental link between thermodynamics and information.

## 1.3 From Selective Ignorance to Accessible Information

To understand how thermodynamics shifts as we descend from the Avogadro scale to the nanoscale, we must re-examine the role of information. Conventional thermodynamics can be viewed as the *powerful science of adopted ignorance*. In macroscopic systems, it remains *utterly impractical* to track the position and momentum of $6 \times 10^{23}$ particles. The triumph of statistical mechanics lies in abandoning this attempt and restricting consideration to a few key parameters, effectively washing out microscopic details.

However, *what is unfeasible for large numbers of particles may become more tractable in smaller systems*. As we approach the nanoscale, the separation of scales that justifies adopted ignorance breaks down. The effect of individual disturbances becomes significant and measurable, and fluctuations can no longer be ignored. In these regimes, the probability distributions are not sharply peaked, and individual fluctuations can—for a fleeting moment—appear to defy the Second Law, even if the law remains robust in the long term.

Consequently, entropy in small systems transforms from a fixed state function into an operational quantity that depends on the observer's resolution. It is no longer sufficient to define entropy through global variables; we must account for specific constraints, such as charge quantization. In single-electron devices, for instance, the number of charge carriers is quantized, meaning the partial derivative of energy with respect to particle number—the standard definition of chemical potential—becomes undefined. In Coulomb-blockaded few-electron devices one therefore works with a discrete electrochemical potential (addition energy) $\mu_N \equiv E(N) - E(N-1)$, which is tunable by gate voltages. We shall use $\mu$ in this standard 'addition-energy' sense for quantum dots, and $\mu_S, \mu_D$ for the reservoir electrochemical potentials.

This shift necessitates an information-theoretic framework. If entropy measures our lack of information, then acquiring information (measurement) and discarding it (erasure) are thermodynamic acts. Losing one bit of Shannon information carries a minimum thermodynamic entropic cost $k_B \ln 2$—we shall return to this as Landauer erasure. In a quantum system, ignoring something has a cost.

Historically, thermodynamic entropy was introduced operationally (by Clausius) as a macroscopic state function inferred from heat and temperature, without any appeal to probability or 'missing information'. The later statistical entropies of Boltzmann, Gibbs, and Shannon are mathematical functionals of probability distributions. Information provides a unifying *operational* bridge: what experiments directly access are distributions over outcomes and correlations between subsystems, from which entropic quantities can be inferred. The Boltzmann constant provides the physical conversion factor that turns such entropic measures into thermodynamic costs (e.g. Landauer erasure).

A question often pressed in the foundations of thermodynamics is whether these two entropies are *the same thing*, or whether they are distinct quantities that happen to coincide numerically when $k_B$ is supplied as a conversion factor. The Clausius entropy is defined operationally as a state function inferred from heat and temperature, with no commitment to any underlying probability distribution. The Shannon and Gibbs entropies are mathematical functionals of probability distributions, indifferent to whether the distribution describes a physical system at all. Our position throughout this article is operational rather than metaphysical: we treat the identification $S_{\text{thermo}} = k_B S_{\text{Shannon}}$ (and its quantum analogue with the von Neumann entropy) as an *empirical bridge* whose validity is established by precisely the experiments we review. Quantum-dot entropy measurements (§4) recover a thermodynamic state function from charge statistics; Landauer-bound erasure (§7.1) converts a bit of Shannon information into $k_B T \ln 2$ of heat; the generalised decomposition of §7.2 equates $k_B$ times the destroyed mutual information with the irreversible entropy production. The bridge is not asserted; it is the experimental content of nanoscale thermodynamics, and it is what makes the language of "information as a thermodynamic resource" a quantitative rather than a metaphorical claim.

In the quantum regime, this connection broadens to include correlations between subsystems. This can be characterized by the mutual information between subsystems A and B. It can be thought of as the overlap in a Venn diagram between the isolated entropies $S_A$ and $S_B$ of two systems, where $S_i = -\mathrm{Tr}_i\, \rho_i \ln(\rho_i)$ is the von Neumann entropy of state $\rho_i$.[8] If the total entropy of the interacting systems is $S_{AB}$, then the mutual information is $I(A{:}B) = S_A + S_B - S_{AB}$, where $S_{AB}$ is the von Neumann entropy of the joint state. It measures the extent to which entropies fail to add: it is zero for uncorrelated systems and positive whenever A and B share correlations, classical or quantum. Mutual information is more than a measure of correlation — it is a thermodynamic resource. Established correlations can drive heat spontaneously from a cooler to a hotter body (§6.4), and in the quantum many-body regime the dominant contribution to irreversible entropy production is not heat dissipation but the destruction of mutual information, $\Delta I(A{:}B)$ — the quantum generalisation of the classical Landauer cost.

## 1.4 Entropy as a consequence of partitioning

In conventional thermodynamics, we partition the universe into a "system" and an "environment" based on spatial boundaries. However, in small quantum systems, *partitioning is a fundamental choice between observed and unobserved degrees of freedom*. It is therefore an operational restriction, a division that is often forced upon us by the complexity of the system, between degrees of freedom that are accessible to measurement/control and those that are not.

This distinction is not merely technical; it defines the thermodynamic variables. We effectively partition the full microscopic reality into:

1. **Emergent Degrees of Freedom:** The specific variables we can control, measure, and manipulate (e.g. the occupation of a specific cavity mode, the charge state of a quantum dot, or the centre-of-mass motion of a levitated nanoparticle).

[8] Michael A. Nielsen and Isaac L. Chuang, *Quantum Computation and Quantum Information*, Cambridge University Press (10th Anniversary Edition 2010) p. 508

2. **Ignored Degrees of Freedom:** The vast background of unobservable modes (e.g., the internal phononic vibrations of the nanoparticle, or the entangled partner in a quasiparticle pair) that act as a bath.

The emergent degrees of freedom arise through coarse graining, whereby short-lived or microscopic variables are integrated out, yielding collective modes that capture the physics at accessible energy and length scales. This procedure defines an effective theory but does not yet imply openness or entropy production.

In favourable cases, emergence simplifies the thermodynamics dramatically. A Bose–Einstein condensate provides a paradigmatic example. Although the microscopic Hamiltonian is strongly interaction-dominated, the low-energy sector is described by weakly interacting collective excitations, the phonons. Here, the emergent phononic modes provide a natural set of experimentally accessible variables. If we restrict attention to these modes and ignore residual microscopic fluctuations, we effectively define them as the relevant system. The remaining microscopic modes then act as a bath after we trace over them. In this situation, scale separation and system–bath partition approximately coincides and the hierarchy: microscopic → emergent → system–bath approximately aligns.

However, this alignment is not universal. In more complex quantum systems, the emergent degrees of freedom remain strongly interacting and highly correlated. We shall present an example in §7 of tunnel coupled one-dimensional (1D) quantum fluids where the coarse-grained emergent physics is the Sine-Gordon model, a strongly correlated quantum field theory (QFT) supporting topological excitations (solitons/anti-solitons).

In such systems, emergence does not lead to isolated or only weakly coupled modes, and there is no physically obvious separation between a complete description of an emergent (coarse-grained) system (simple modes) and a background bath. The partition between emergent and ignored degrees of freedom is therefore not dictated by the Hamiltonian, but by operational constraints: which observables we can measure, reconstruct, and analyse with available resources.

Thus, a unified view emerges: entropy arises from the information we choose—or are able—to track. For small, coherent systems, the relevant entropy — Shannon for classical distributions, von Neumann for quantum states — is the metric of the information explicitly accessible to the experimenter. Sometimes information can play the role of a cold reservoir, or the geometry of correlations can determine the flow of heat. Whether we are counting ticks in a nanomechanical clock or separating quasiparticles in a Bose gas, thermodynamics describes the exchange between information (which may be in the form of correlations) and energy. The experiments in the following sections—ranging from autonomous clocks to quantum field simulators—explore this frontier.

# 2 Fluctuations

In systems with large numbers of particles, the distribution function over microstates is very sharply peaked; *individual fluctuations* are then negligible and make little difference to the whole *mean behaviour*. However, in small systems, fluctuations can be dominant. It becomes possible for individual fluctuations to result in a reduction of entropy — trajectories that, for a *fleeting moment*, appear to defy the Second Law of Thermodynamics. The Second Law must be regarded as statistical ($\langle \Delta S_{\text{tot}} \rangle \geq 0$). Individual single-shot negative values do not fundamentally violate it, because they are simply part of the distribution that creates the average.

## 2.1 The Jarzynski equality

A system that can be driven away from equilibrium by one or more control parameters can be described by the **Jarzynski equality**.[9] This relates non-equilibrium work to equilibrium free energy differences.

$$\langle \exp\left(-\frac{W}{k_B T}\right) \rangle = \exp\left(-\frac{\Delta F}{k_B T}\right), \tag{1}$$

where $\Delta F$ is the free-energy difference between the initial and final states, $W$ is the work performed on the system, and the average is taken over all possible ways of transitioning from the initial state to the final state. The Jarzynski equality relates non-equilibrium work statistics directly to equilibrium state variables.

[9] C. Jarzynski. Hamiltonian Derivation of a Detailed Fluctuation Theorem. *Journal of Statistical Physics* 98, 77-102 (2000)

Crucially, this relation implies more than just the stability of the Second Law; it mandates the existence of apparent violations. Since the average work $\langle W \rangle$ must exceed $\Delta F$ (the Second Law), in the Jarzynski equality most trajectories contribute a value $e^{-\frac{W}{k_B T}} < e^{-\frac{\Delta F}{k_B T}}$ to the average. To balance the equation to unity, there must exist rare trajectories where $W < \Delta F$. In these instances, heat fluctuations from the bath contribute to trajectories with negative entropy production. While the Second Law is robust in the long-term average, the Jarzynski equality reveals that apparent disobedience is not an error—it is a statistical necessity required to satisfy thermodynamic equilibrium. It therefore implies that while the average work performed must satisfy the Second Law, $W \geq \Delta F$, individual trajectories with negative entropy production occur, though positive entropy production is exponentially more probable. This confirms that the Second Law remains robust in the long term, even if disobedience seems possible on the scale of single experimental shots.

Crooks' Fluctuation Theorem provides a generalization of the Jarzynski equality to give the precise probability of these fleeting moments.[10] If σ is the entropy produced, the probability of seeing an apparent violation (−σ) compared to a normal event (+σ) is

$$\frac{P(-\sigma)}{P(+\sigma)} = e^{-\sigma/k_B} \tag{2}$$

This explains why apparent violations are fleeting and rare. As the system gets larger or is driven harder, small values of σ become less likely and hence the probability of an apparent violation decays exponentially.

The Jarzynski equality has been tested in the quantum regime in experiment with a single $^{171}Yb^{+}$ ion trapped in a harmonic potential.[11] The initial thermal distributions of phonons were determined from projective measurements. A laser-induced force was then applied to the equilibrium eigenstates, and the transition probabilities to the far-from-equilibrium eigenstates were measured. When the work was done relatively slowly (taking either 45 µs or 25 µs), the response was fairly linear and could be accounted for by the fluctuation-dissipation theorem described in §2.3 below. But when the work was performed more quickly (in 5 µs), the free energy differences could be accounted for only by the Jarzynski estimate. The method could be used to test further aspects of quantum thermodynamics which have hitherto been restricted to theory.

## 2.2 The Clausius inequality

The Clausius inequality states that in any thermodynamic system undergoing a closed cycle involving external reservoirs, the entropy of the reservoirs increases or does not change, and cannot decrease. This can be seen to arise from the Jarzynski equality as a corollary via convexity. The Jarzynski equality is exact even for strongly nonequilibrium processes. The Clausius inequality applies to ensemble-averaged entropy production, which cannot be violated in sufficiently long runs.

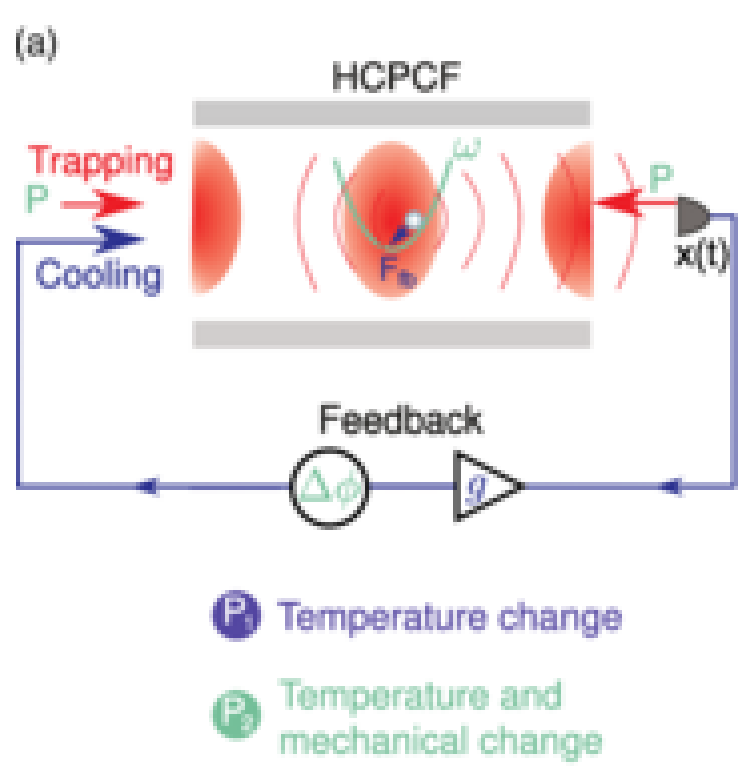

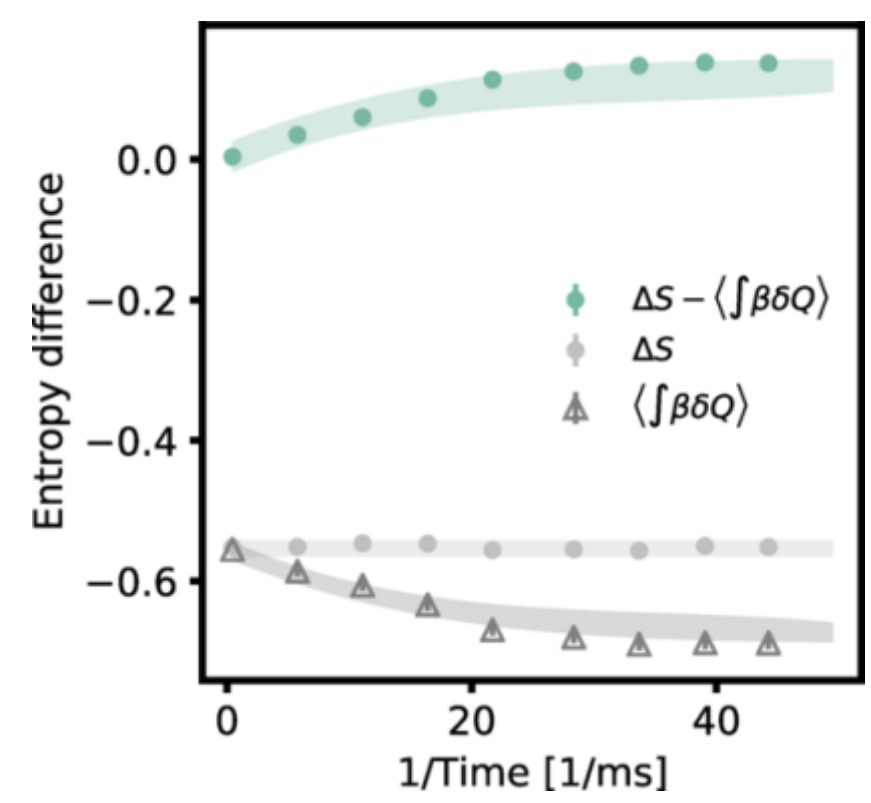

[10] G. Crooks, Entropy production fluctuation theorem and the nonequilibrium work relation for free energy differences, *Phys. Rev. E* **60**, 2721 (1999) https://doi.org/10.1103/PhysRevE.60.2721

[11] S. An, *et al.*, Experimental test of the quantum Jarzynski equality with a trapped-ion system. *Nature Physics* **11**, 193-199 (2015) https://www.nature.com/articles/nphys3197

**Fig. 1(a)** An experiment to test the Clausius inequality by trapping a 969 nm silica particle at the intensity maximum of the standing wave formed inside a hollow-core photonic crystal fibre (HCPCF) by two counter-propagating laser beams (red arrows) of wavelength 1064 nm. The scattered light (red lines) gives the centre-of-mass motion $x(t)$ of the particle. An additional feedback laser beam (blue arrow) cools the centre-of-mass motion of the particle. The velocity contribution of the feedback force $F_{fb} = g\gamma_p m \sin(\Delta\phi)x'$ depends on the feedback gain $g$ and the trapping laser power $P$ via the feedback phase $\Delta\phi$. Thermal change (purple) is implemented by a linear increase of the feedback gain $g$ at constant optical trap power. Thermal and mechanical change (turquoise) is realized by changing the optical trap power $P$.[12]

**Fig. 1(b)** Comparison between the measured heat exchanged with the environment $\langle\int\beta\,\delta Q\rangle$ (triangles) and the entropy variation of the system $\Delta S$ (grey full circles), as a function of the inverse protocol time for thermal and mechanical changes followed by an isothermal thermalization (entropy is in units of $k_B$; $\beta$ is the inverse temperature $1/k_BT$). The inequality, $\Delta S - \langle\int\beta\,\delta Q\rangle \geq 0$, is verified for any protocol speed. Shaded areas are theoretical predictions that include the uncertainty on the underlying experimental parameters. Error bars are determined via the standard deviation of the respective value over 15,000 runs and are smaller than the symbol size.[12]

The Clausius inequality has been tested for both thermal and mechanical changes using small particles which are optically confined. A silica microsphere of diameter 969 nm was trapped in a hollow-core photonic crystal fibre (HCPCF) in an approximately harmonic potential in the intensity maximum of a standing wave formed by two counter-propagating laser beams, as shown in Fig. 1.[12] The effective spring constant could be controlled by varying the optical trap power, and thermal control in the form of cold damping could be implemented by an additional laser that counteracted the motion of the particle. In this way, it was possible to investigate first the effect of mechanical control, and then the effect of combined mechanical and thermal control. These were investigated as a function of speed, expressed as the inverse protocol time. In both cases, when the experiment was done very slowly the entropy difference $\Delta S - \langle\int\beta\,\delta Q\rangle$ approached zero. At higher speeds $\Delta S - \langle\int\beta\,\delta Q\rangle \geq 0$, and the integrated heat absorbed by the system divided by temperature never exceeded the entropy change. But because this is a small system, individual fluctuations were not bound in the same way.

## 2.3 From Linear Response to Thermodynamic Uncertainty

To understand why relations like the Jarzynski equality are necessary at the nanoscale, we must appreciate how the relationship between fluctuation and dissipation fundamentally changes with scale.

In macroscopic systems close to equilibrium, this relationship is codified by the celebrated **Fluctuation-Dissipation Theorem (FDT)**.[13] Originally formulated by Nyquist and later generalized by Callen and Welton, the FDT states that the spontaneous fluctuations of a system in equilibrium are proportional to its dissipative response to an external force. In frequency space, one common form is $S_{xx}(\omega) = \frac{2k_BT}{\omega}\,\mathrm{Im}\,\chi_{xx}(\omega)$, relating equilibrium fluctuation spectra $S_{xx}(\omega)$ to linear response, where $\chi_{xx}(\omega)$ is the susceptibility or response function. The **power spectral density** $S_{xx}(\omega)$ of the equilibrium fluctuations of $x$ tells you how strongly the observable fluctuates at frequency ω. The **linear susceptibility** $\chi_{xx}(\omega)$ tells you how much the expectation value of $x$ changes when you apply a weak force conjugate to $x$; its **real part** is the reactive, non-dissipative response, while its **imaginary part** is the dissipative part.

A classic example is the Einstein relation for Brownian motion, where the diffusion coefficient (fluctuation) is effectively set by the fluid's viscosity (dissipation). At the Avogadro scale, this is an equality: if you know the friction, you know the noise. The system's response is linear, and fluctuations are merely small perturbations around a stable mean.

Fluctuation–dissipation theorems and the relations derived from them are statements about **equilibrium** systems in the **linear-response** regime, relying on **detailed balance** (time-reversal symmetry of the effective dynamics). They therefore cease to apply whenever any of these assumptions fails—when the system is driven away from equilibrium, when perturbations probe nonlinear response, or when forward and reverse transitions are systematically biased.

[12] M. Rademacher *et al*. Nonequilibrium control of thermal and mechanical changes in a levitated system. *Phys. Rev. Lett.* **128**, 070601 (2022). Figures sourced from the publication and licensed under CC BY 4.0

[13] R. Kubo, The fluctuation-dissipation theorem. *Rep. Prog. Phys*. **29** 255 (1966)

As one moves to mesoscopic and nanoscale devices, these conditions are violated in two practically unavoidable ways:

- **Fluctuations become non-perturbative.** The noise is no longer a small correction to the mean; it can be comparable to the signal itself. In this regime, linearisation around a single equilibrium state can be quantitatively misleading because the dynamics sample nonlinear features of the landscape and rare events contribute measurably to statistics.
- **Nonequilibrium steady operation becomes the norm.** Many nanoscale devices—single-electron clocks, molecular motors, autonomous engines—are maintained by continuous driving in states that break detailed balance. In such conditions the equilibrium fluctuation–response proportionality (and simple equilibrium relations such as the Einstein relation $D = k_B T/\gamma$ in its usual setting) no longer provides the natural organising framework; one must use nonequilibrium response relations.

This breakdown does not imply *lawlessness*. Rather, the *rigid equalities* characteristic of equilibrium linear response are replaced by *universal inequalities and trade-offs* for thermodynamically consistent stochastic dynamics. The most prominent of these are the **Thermodynamic Uncertainty Relations (TURs)**, which bound the precision of sustained currents by the entropy produced to maintain them: to keep a current (for example, the ticks of a clock) stable against fluctuations, the dynamics must generate a minimum amount of dissipation. The key point is not smallness *per se*, but the presence of appreciable fluctuations and nonequilibrium currents—conditions that become far more visible, and often unavoidable, in nanoscale systems.

## 2.4 Thermodynamic uncertainty relations: Fluctuations, Information, and Precision:

The results of the preceding sections — Jarzynski, Crooks, the Clausius inequality — can all be traced to a single path-level symmetry of stochastic thermodynamics. Consider a system coupled to a heat bath at inverse temperature $\beta$, driven by some time-dependent protocol. A *trajectory* $\Gamma$ is the complete record of the system's microscopic evolution — every position and jump — over the duration of the protocol, and its *time reverse* $\tilde{\Gamma}$ is the same sequence of states traversed in the opposite order under the time-reversed protocol. The total entropy produced along $\Gamma$ is then defined by the log-ratio of the probability of seeing $\Gamma$ under the forward protocol to the probability of seeing $\tilde{\Gamma}$ under the reverse:

$$\Delta S_{tot}[\Gamma] = k_B \ln \frac{P[\Gamma]}{P[\tilde{\Gamma}]}. \tag{3}$$

The right-hand side is $k_B$ times a Shannon-type log-probability ratio; its identification with thermodynamic entropy production is justified by local detailed balance, which sets the proportionality between forward/reverse trajectory probabilities and the heat exchanged with the bath.

Equation (3) is not an approximation; it is the *definition* of trajectory-level entropy production in stochastic thermodynamics, and it encodes the intuition that an entropy-producing trajectory is one that is more probable in the forward direction than in reverse. The integral fluctuation theorem follows by direct averaging. Inverting the definition gives $e^{-\Delta S_{tot}[\Gamma]/k_B} = P[\tilde{\Gamma}]/P[\Gamma]$, and averaging over all forward trajectories $\Gamma$ weighted by $P[\Gamma]$, one obtains

$$\langle e^{-\Delta S_{tot}/k_B} \rangle = 1 \tag{4}$$

because the sum over forward trajectories, each reweighted by the ratio $P[\tilde{\Gamma}]/P[\Gamma]$, is equivalent to summing $P[\tilde{\Gamma}]$ over all reverse trajectories — which is normalised to unity. Equation (4) is exact and holds arbitrarily far from equilibrium.

For a system initially in thermal equilibrium at temperature $T$, the total entropy production can be decomposed as $\Delta S_{tot} = \beta(W - \Delta F)$, where W is the work performed on the system and $\Delta F = F_{\text{final}} - F_{\text{initial}}$ is the free-energy difference between the equilibrium states corresponding to the initial and final values of the control parameter. Substituting into equation (4) and factoring out the state-function part gives

$$\langle e^{-\beta W} \rangle = e^{-\beta \Delta F}, \tag{5}$$

which is the Jarzynski equality already introduced in equation (1). The decomposition $\Delta S_{\mathrm{tot}} = \beta(W - \Delta F)$ also makes clear why $\langle W\rangle \geq \Delta F$ (the Second Law): Jensen's inequality applied to the convex exponential in equation (4) immediately yields $\langle \Delta S_{\mathrm{tot}} \rangle \geq 0$, and hence $\langle W\rangle \geq \Delta F$.

Equations (3) – (5) constrain the *full distribution* of entropy production and work. A separate family of results constrains the *precision* of time-integrated currents in nonequilibrium steady states. For any such time-integrated current $J_{\mathrm{T}}$ — for example, the number of electrons transferred through a device, or the number of ticks emitted by a clock — the Thermodynamic Uncertainty Relations (TURs) [14,15] state that

$$\frac{Var(J_{\mathrm{T}})}{\langle J_{\mathrm{T}}\rangle^2} \geq \frac{2k_B}{\langle \Delta S_{\mathrm{tot}}\rangle} \quad . \tag{6}$$

The physical content is that maintaining a precise current (small relative fluctuations) against thermal noise requires a minimum rate of entropy production. Unlike the fluctuation–dissipation theorem, which is an equality valid only in equilibrium, a TUR is an *inequality* valid continuous-time classical Markov jump processes, and for related Langevin systems under the usual local-detailed-balance assumptions — precisely the non-equilibrium regime where the FDT fails. TURs thus replace the rigid equalities of equilibrium with universal trade-offs between precision and dissipation, which bound the relative fluctuations by entropy production.

A conceptually distinct but related bound governs information processing. When one bit is changed from an unknown state containing information to a known state containing no unknown information, the *Shannon* entropy changes by ln 2. Landauer's principle states that irreversibly erasing one bit of classical information produces a minimum *thermodynamic* entropy change in the environment

$$\langle \Delta S_{\mathrm{env}}\rangle \geq k_B \ln 2. \tag{7}$$

Equation (7) follows from (4) by the same Jensen step that gave the Second Law: $\langle \Delta S_{\mathrm{tot}}\rangle \geq 0$. Writing $\Delta S_{\mathrm{tot}} = \Delta S_{\mathrm{sys}} + \Delta S_{\mathrm{env}}$ and noting that erasure is defined by a system transition from a maximally uncertain bit (Shannon entropy ln 2, thermodynamic entropy $k_{\mathrm{B}} \ln 2$) to a known state (entropy 0), the deterministic system contribution is $\langle \Delta S_{\mathrm{sys}}\rangle = -k_{\mathrm{B}} \ln 2$, and the inequality on the total transfers to the environment as $\langle \Delta S_{\mathrm{env}}\rangle \geq k_{\mathrm{B}} \ln 2$. The integral fluctuation theorem equation (4) guarantees that the average entropy production cannot be negative; the magnitude $k_{\mathrm{B}} \ln 2$ is the initial Shannon entropy of the unknown bit, converted to thermodynamic units by $k_{\mathrm{B}}$.

Together, equations (3) – (7) form a coherent hierarchy. The path-level symmetry (3) is the foundation. Averaging yields the integral fluctuation theorem (4) and hence the Jarzynski equality (5), which governs driven processes. The TUR (6) governs the precision of sustained currents. Landauer's bound (7) governs information erasure. Dissipation biases the arrow of time and stabilises information, while residual stochasticity remains unavoidable unless entropy production diverges.

In the classical domain the thermodynamic uncertainty relation (Eq. 6) imposes a hard cost–precision trade-off, based on classical Markovian dynamics with local detailed balance. But quantum systems are not bound in the same way, because quantum measurement is never passive. Continuous monitoring (e.g. homodyne detection) introduces backaction, and the achievable precision is then set by the quantum Fisher information $F_Q$, which provides a measure of how much information the monitored dynamics carry about the quantity being estimated through the *quantum* Cramér–Rao bound

$$\frac{\mathrm{Var}(J)}{\langle J\rangle^2} \geq \frac{1}{F_Q}. \tag{8}$$

Because quantum coherence enlarges $F_Q$ beyond what an equivalent classical (population-only) description can carry, coherent systems can attain a precision below the classical *kinetic* (dynamical-activity) bound.[16]

[14] A. C. Barato and U. Seifert, Thermodynamic uncertainty relation for biomolecular processes. *Phys. Rev. Lett.* **114**, 158101 (2015)

[15] T. R. Gingrich, J. M. Horowitz, N. Perunov, and J. L. England, *Physical Review Letters* **116**, 120601 (2016)

[16] Y. Hasegawa, Quantum Thermodynamic Uncertainty Relation for Continuous Measurement, *Phys. Rev. Lett.* **125**, 050601 (2020). https://doi.org/10.1103/PhysRevLett.125.050601; Y. Hasegawa, Thermodynamic uncertainty relation for general open quantum systems. *Phys. Rev. Lett.* **126**, 010602 (2021). https://doi.org/10.1103/PhysRevLett.126.010602

Separately, and through a different route, the entropy-production bound of $2k_B$ itself can be violated once the dynamics leave the populations-only Markovian regime. Coherent systems can attain a precision the classical bound would forbid, and the $2k_B$ limit can be violated. This occurs through more than one mechanism rather than a single quantum effect: coherent transport that leaves the populations-only regime, via higher-order tunnelling or interference between near-degenerate levels, can lift the precision–dissipation product above the classical limit, [17] while replacing a stochastic transport step with coherent unitary evolution can suppress current fluctuations and relax even the power–efficiency–constancy trade-off of heat engines.[18] The common thread is that coherence supplies a resource with no classical counterpart such as extra distinguishability or reduced trajectory stochasticity. The classical inequality is not eliminated but rather superseded by a modified quantum bound. The trade-off is thus relaxed rather than removed, since the information must still be acquired and its acquisition carries a thermodynamic cost.

Thus, while quantum mechanics does not offer a free lunch—because a trade-off between information and energy persists—it offers a more flexible currency. Coherence allows us to purchase precision at a discount compared to purely classical systems, provided we can protect the system from decoherence.

# 3 Connecting entropy to experience: The arrow of time

The arrow of time is a universal human experience. We do not need to know the mechanisms in order to observe aging, in ourselves or anything else. This human experience of the arrow of time is all the more remarkable when we consider that almost all the laws of physics are symmetric in time; the phenomena work just as well if you substitute –*t* for *t*. Newton's laws, Maxwell's equations, general relativity, and the Schrödinger equation all work with either sign for *t*. Almost everything, that is, except for work and heat. Any amount of work can be completely converted to heat, but no amount of heat can be completely converted to work. There are ways of showing why this should be so without knowing the mechanisms. For example, the efficiency of an ideal engine can be derived without any knowledge of the constituent materials or components.[19]

As numbers get smaller, it becomes possible to follow each individual particle. Consider a film of a collision between two billiard balls. If the coefficient of restitution is close to 1 and the rolling friction is negligible, then you cannot tell whether the film is running forwards or backwards. But now consider a film of the first shot in a game of snooker, in which the white ball strikes the prearranged triangle of 15 red balls. Because only 16 balls are involved, it is not difficult to track them all. And in this case, it is readily intuitive whether the film is playing forward or reverse. In simple terms, this is because there the red balls move from an ordered state (if they are indistinguishable then there is only one way to arrange them in the initial triangle) to a disordered state (there are countless ways in which they can be distributed following the initial impact).

The concept of entropy is inextricably linked to the arrow of time. In nanoscale systems, where we can track individual particles, the arrow becomes statistical rather than absolute. The asymmetry of conversion between work and heat requires careful re-interpretation when the heat corresponds to energy exchange with only a small number of degrees of freedom and becomes strongly fluctuating.

We use economic and fluid and spatial metaphors to describe the experience of time's arrow (itself a metaphor). We speak of spending, saving, or investing time. And we speak of time flowing or passing. The passage of time is measured by clocks. Conventional clocks can be made more accurate by reducing the energy loss. This can be thought of simply as increasing the quality factor, $Q$ (the ratio of the energy in the system to the energy dissipated per radian of the oscillation), or alternatively in terms of the fluctuation dissipation theorem, whereby the level of dissipation is matched by the level of disturbances of the oscillation.[13] At the nanoscale, our ability to track time relies on suppressing fluctuations.

This brings us to the thermodynamics of clocks, where the arrow of time becomes operational and accuracy demands an entropic cost. All clocks, whether classical or quantum, are irreversible systems which produce entropy.[20] Clocks therefore not only tell the instantaneous time, but they also give an arrow of time between successive readings.

[17] B. K. Agarwalla and D. Segal, Assessing the validity of the thermodynamic uncertainty relation in quantum systems. *Phys. Rev. B* **98**, 155438 (2018).

[18] K. Ptaszyński, Coherence-enhanced constancy of a quantum thermoelectric generator. **Phys. Rev. B 98**, 085425 (2018).

[19] *The Feynman Lectures Physics* I, 44-6 (Addison-Wesley 1965)

## 3.1 Classical nano-clock

Periodic clocks work by measuring an oscillating system each time it passes a point in its phase space. We can call such a measurement a tick. The escapement mechanism of a traditional pendulum clock makes two such measurements for each cycle: hence "tick tock". If the average interval between ticks is $t_{\text{tick}}$ with an uncertainty $\Delta t_{\text{tick}}$, then the uncertainty in the time of the $n^{\text{th}}$ tick is $\sqrt{n}\,\Delta t_{\text{tick}}$ (the relative uncertainty is $\frac{1}{\sqrt{n}}\frac{\Delta t_{\text{tick}}}{t_{\text{tick}}}$). We can define the dimensionless *clock coherence* as the number of ticks at which accumulated random-walk jitter has grown to one period, $N_{\text{cc}} \equiv (t_{\text{tick}}/\Delta t_{\text{tick}})^2$. In close analogy with coherence-cycle counts in laser physics, this is the number of ticks over which the clockwork preserves coherent phase. Clock coherence relates to the output quality of the clock and should be distinguished from the dynamical quantum coherence to be discussed in §3.4, which describes whether the clockwork's internal evolution is unitary or dissipative. $N_{\text{cc}}$ is an intrinsic property of the clock and carries no observation time. Operational figures of merit such as the Allan deviation $\sigma_y(\tau)$, discussed in §8.1, depend separately on the observation time $\tau$.

The minimum entropy generated per tick, $\Delta S_{\text{tick}}$, is sometimes called the *thermodynamic precision bound*. For a clock which uses a maximum-likelihood estimation of a sinusoidal signal,[21]

$$\Delta S_{\text{tick}} \geq \frac{k_B}{2\pi^2} N_{cc} \tag{9}$$

The prefactor in (9) is not universal. For a clock decoded as a counting renewal process the bound can be up to 40 times larger, $\Delta S_{\text{tick}} \geq 2\, k_{\text{B}}\, N_{\text{cc}}$, as can be seen from the TUR (6); this applies to an Erker autonomous quantum clock[22] and a Barato–Seifert biomolecular clock.[14] Different clock architectures are subject to different bounds, and improving the decoding scheme can reduce the entropic cost per tick. A classical optomechanical clock beats the naive counting-current limit by $4\pi^2$ by using a smarter estimator.

Why must a clock produce entropy at all? A perfect, frictionless oscillator — a pendulum in vacuum, an electron orbiting a nucleus — can maintain a stable frequency indefinitely without dissipation. But such a system has no output: it keeps time internally without producing any record that an external observer can read. To be useful, a clock must produce *ticks* — discrete, irreversible events that distinguish *before* from *after*. Each tick is a measurement that collapses the continuous oscillation into a countable classical record. And each such measurement is thermodynamically irreversible. By the fluctuation theorem Eq. (3), if a process produces zero entropy, the forward and time-reversed trajectories are equally probable, and there is no way to distinguish the direction of time. A tick that could equally run backwards is no tick at all. The more precisely the tick must be located in time — the higher the accuracy — the more strongly the forward trajectory must dominate over the reverse, and therefore the more entropy must be produced. This is the content of the thermodynamic uncertainty relation Eq. (6) applied to clock ticks, and it is the reason why Eq. (9) is a fundamental *lower bound*, not a practical limitation that better engineering could circumvent.

There is thus an irreducible entropy cost of measuring intervals of time, and this cost is proportional to the precision to be obtained.[22] This may seem counterintuitive with respect to the history of mechanical clocks, where steady improvements in accuracy were achieved by reducing frictional and other losses. This is because those clocks, and still most practical clocks today, operate far from the thermodynamic limit.

A pioneering experiment is illustrated in Fig. 2.[21] A silicon nitride membrane was driven by an electrical signal that consisted of white noise. This constituted an effective thermal bath which raised the mechanical mode temperature of the membrane in its fundamental flexural mode of vibration. The apparatus was at room

[20] G. J. Milburn. The thermodynamics of clocks. *Contemporary Physics* **61**, 69-95 (2020)

[21] A. N. Pearson *et al*. Measuring the thermodynamic cost of timekeeping. *Phys. Rev. X* **11**, 021029 (2021) https://journals.aps.org/prx/abstract/10.1103/PhysRevX.11.021029. Figures sourced from the publication and licensed under CC BY 4.0

[22] P. Erker, M. T. Mitchison, R. Silva, M. P. Woods, N. Brunner, and M. Huber, Autonomous quantum clocks: Does thermodynamics limit our ability to measure time? *Phys. Rev. X* **7**, 031022 (2017)

temperature with a pressure 5 x $10^{-5}$ Pa. The membrane was capacitively coupled to a radiofrequency cavity circuit whose output served as a measure of instantaneous displacement. The entropy cost arises from the power dissipated in the detection circuit in the optomechanical sidebands that contain the displacement information, which can be measured from that part of the power spectrum. The noise temperature is calculated from the average spectral density at other frequencies. A tick of the clock was identified by an upward zero crossing in the cavity output signal fed to the oscilloscope in Fig. 2.

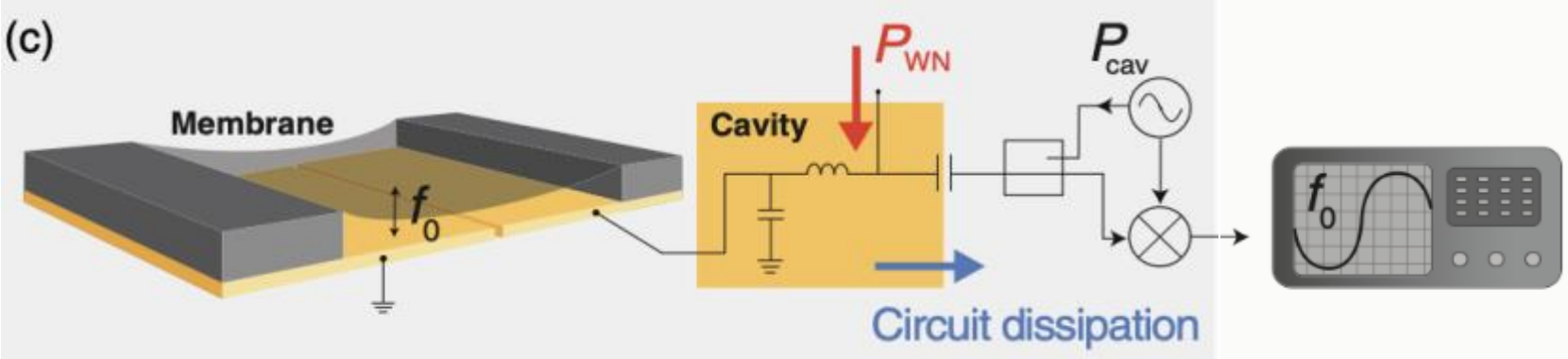

**Fig. 2** Electromechanical system acting as a clock. A membrane is driven by a white-noise signal (power $P_{WN}$). Its vibrations are probed by a rf-cavity driven with a signal of power $P_{cav}$. The cavity output signal, and thus the clock ticks, are registered by an oscilloscope.[21]

The experimental results in Fig. 3(a) show how the accuracy increases with increasing power. At lower powers, the stronger drive leads to larger vibration amplitudes, so that the timing of the ticks is less obfuscated by noise, thereby reducing the uncertainty in when the tick occurs. At higher powers the rate of improvement of accuracy is not maintained, perhaps due to leakage of the heating tone and also departure of the membrane oscillations from simple harmonic motion. The variation of the observed accuracy with the accuracy predicted by theory differed by a factor of 10, as shown in Fig. 3(b). This is not inconsistent with theory, since the predicted entropy generation is a lower limit. If the whole measured signal were to be used, rather than a single crossing point as a trigger, then the accuracy for a given power input might more closely approach the theoretical limit.

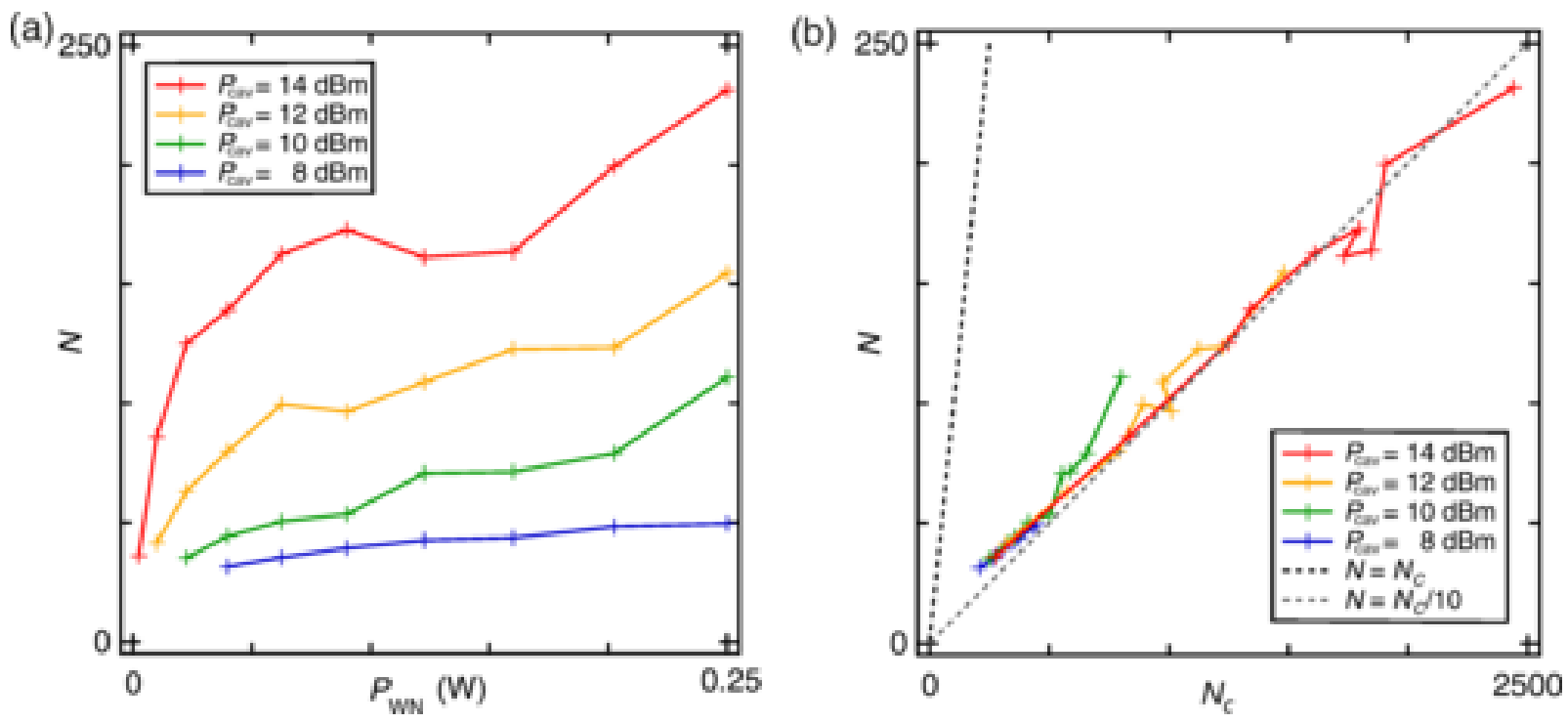

**Fig. 3** Experimental measurements of the clock in Fig. 2. **(a)** Accuracy $N$ versus white-noise power $P_{WN}$ for $P_{cav}$ in the range 8–14 dBm. **(b)** Accuracy $N$ versus the accuracy predicted from the classical model $N_C$ for $P_{cav}$ in the range 8–14 dBm. The black dotted line is a guide to the eye showing the slope corresponding to $N = N_C$. The grey dotted line is chosen to approximate the slope of the experimental results.[21]

How is the rigid penalty of the entropic demands of accuracy to be reconciled with the probabilistic looseness of fluctuations? It seems that the dissipation in the clock is necessary to overcome the fleeting apparent violations of the Second Law in order to produce, in the long term, reliable ticks. In this way the *necessity* of the clock's entropic cost would be tied to the *suppression* of the Jarzynski-style fluctuations.

Dissipation enforces *time-directional bias*. In stochastic thermodynamics this appears as the exponential suppression of time-reversed trajectories by inverting Equation (3),

$$\frac{P[\Gamma]}{P[\tilde{\Gamma}]} = e^{\Delta S_{\text{tot}}[\Gamma]/k_B} \tag{10}$$

Large entropy production does *not* eliminate fluctuations, but it makes backward ticks exponentially small. This is formalized in the *thermodynamic uncertainty relations* discussed above, leading to:

$$\frac{Var(N)}{\langle N\rangle^2} \gtrsim \frac{2k_B}{\langle \Delta S\rangle}, \tag{11}$$

where $N$ is now simply the number of ticks in a given time window. This is the precise sense in which dissipation overcomes fleeting apparent lawlessness (i.e. apparent violations of the Second Law). Dissipation is not a penalty imposed despite fluctuations—it is the mechanism that ensures fluctuations do not dominate long-term timekeeping.

## 3.2 Quantum clock

The entropy generation as a function of accuracy has been investigated in a quantum clock which consisted of a superconducting coplanar waveguide resonator, which served as a cavity, dispersively coupled to transmon qubit, operated at temperature 22 mK.[23] The qubit was driven and measured by microwave signals. At smaller drive amplitudes the system manifested Rabi oscillations, which provided the ticks. At larger amplitudes quantum jumps from one energy level to another occurred, which provided a different kind of ticks.

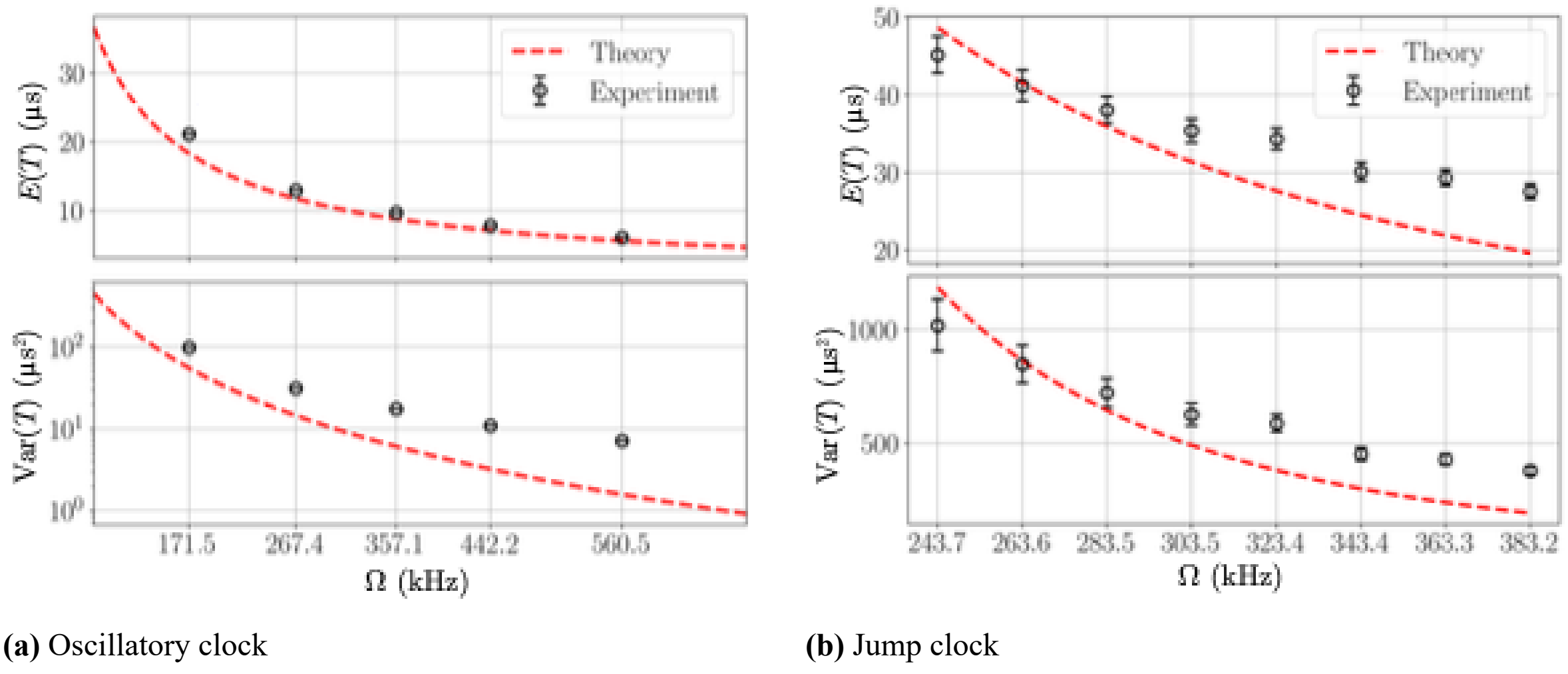

**(a)** Oscillatory clock

**(b)** Jump clock

**Fig. 4** Superconducting quantum clock. Statistics of the period $T$ as a function of the qubit drive amplitude $\Omega$, compared with theory based on kinetic uncertainty relations, for: **(a)** oscillatory clock; **(b)** jump clock.[23]

For each mode, the mean $E(T)$ and the variance $\mathrm{Var}(T)$ of the period were determined and compared with theory based on kinetic uncertainty relations, which describe the trade-off between power and efficiency in heat engines.[23] In the oscillatory regime these were determined directly from the measured time-domain signal. In the jumps regime, the time-domain measurement yielded a random telegraph signal, in which each upward jump was taken as a tick. The results are presented in Fig. 4. Within experimental uncertainty, in each case $E(T)$ and $\mathrm{Var}(T)$ are not less than the theoretical bound. The excess variance in the oscillatory case is attributed to unidentified systematic noise. Further experimental confirmation of the theoretically predicted linear growth of the precision in the quantum coherent regime would demonstrate that precision can be maintained or enhanced despite the degradation caused by quantum backaction.

A complementary line of theoretical work has shown that the linear bound (9) is not the ultimate quantum limit. Meier and co-workers have proposed a fully autonomous quantum many-body *ring clock* in which a single excitation propagates coherently around a tailored spin chain with dissipation confined to a single link.[24] The clockwork is *not* fully dissipative; coherent transport carries the dynamics, and only the tick-registration step injects entropy. Where Eq. (9) requires the entropy per tick to grow *linearly* with the accuracy $N$, the ring clock achieves the same accuracy with entropy per tick growing only *logarithmically*:

$$\Delta S_{\text{tick}} \sim \frac{k_{\text{B}}}{c} \ln N_{\text{cc}} \tag{12}$$

for a system-dependent constant $c > 0$; equivalently, the accuracy grows exponentially in the entropy budget per tick. Equation (12) demonstrates that Eq. (9) is the saturating relation for fully dissipative clockworks, rather than an absolute thermodynamic limit. The gap between extant clockworks and the true fundamental limit is therefore not yet known, and may be substantially larger than the linear framework would suggest.

[23] X. He *et al*. Effect of measurement backaction on quantum clock precision studied with a superconducting circuit. *Phys. Rev. Applied* **20**, 034038 (2023). Figures sourced from the publication and reused by permission.

[24] Meier, F., Minoguchi, Y., Sundelin, S., Apollaro, T. J. G., Erker, P., Gasparinetti, S., & Huber, M. (2025). Precision is not limited by the second law of thermodynamics. *Nature Physics*, **21**, 1147–1152. https://doi.org/10.1038/s41567-025-02929-2

## 3.3 Classical ticks from a quantum clock

There is no thermodynamic barrier to a lossless oscillatory system that in principle would dissipate no energy. Such a system could constitute a perfect clock—except that it would have no hands and could not tell the time. To be useful a clock must have an output. A quantum clock must be observed in order for a classical user to catch their train or meet their date. A theme that we shall return to in the concluding §8 is that the entropy cost of the ancillary mechanisms can vastly exceed the minimum thermodynamic requirements. This has been illustrated with a single-electron quantum clock.[25]

The clock consisted of a double quantum dot device, in which the dots were separated from each other by a quantum tunnelling barrier, and the left and right hand dots were respectively separated from the source and the drain also by quantum tunnelling barriers. The device and the corresponding triple point in the charge stability diagram are illustrated in Fig 5(a). The chemical potentials of the dots were adjusted so that three possible states had the same energy: no excess electron in either dot; an excess electron in the left dot; or an excess electron in the right dot. The device also contained a charge sensor in the form of a single electron transistor, which was adjusted to discriminate between the three states. The charge sensor could be read either by the current through it or via radiofrequency (rf) reflectometry. The tunnel barriers were set so that transitions between these states occurred slowly with respect to the measurement time. Time traces by the two measurement methods are shown in Fig. 5(b).

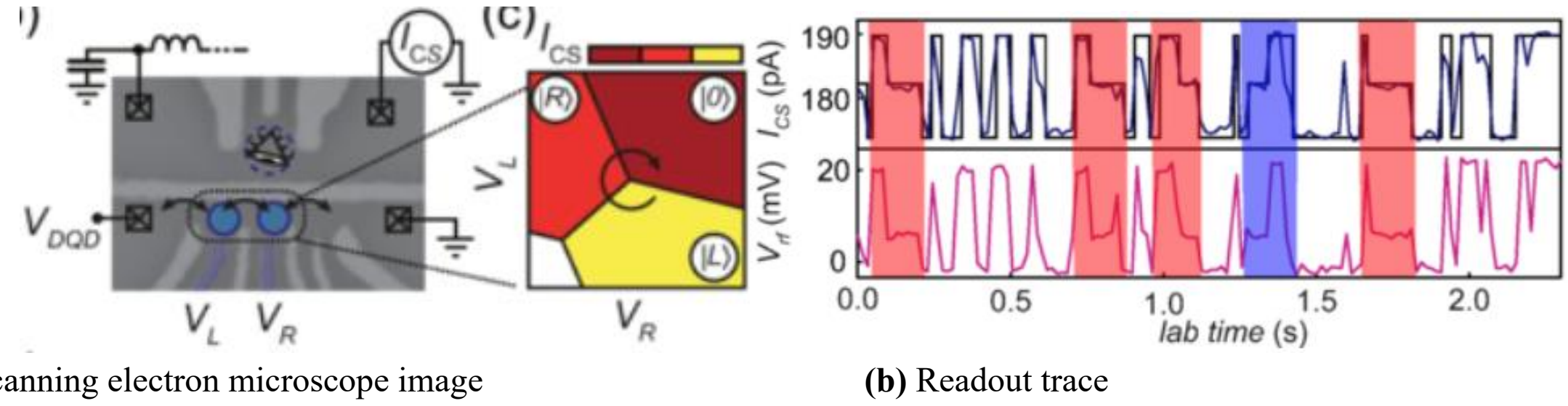

**(a)** Scanning electron microscope image **(b)** Readout trace

**Fig. 5** A quantum clock formed by a double quantum dot device. **(a)** Scanning electron microscope image. Blue circles represent the DQD and the eye represents the charge sensor dot used for classical readout. Gate electrode voltages $V_L$ and $V_R$ control the number of charges in the left and right dot, respectively. Charge transport can occur from source to drain, or drain to source, or any intermediary process (black arced arrows), influenced by an applied source-drain voltage $V_{\mathrm{DQD}}$ across the DQD. The charge stability diagram illustrates the charge state as a function of $V_L$ and $V_R$. Intersection of charge occupation states $|0\rangle$, $|L\rangle$ and $|R\rangle$ is a triple point. Fluctuations about this point cause a cycling of states; the black arrow corresponds to a forward tick. **(b)** Readout trace from both dc (top) and rf reflectometry (bottom), with respect to lab time; $T = 180$ mK. Black trace is the result of the level-identification algorithm. Coloured regions indicate ticks: red for forward and blue for backward.[25]

The three distinct levels in the time traces correspond to the three states. Low–high–intermediate–low is counted as a forward tick, while the reverse is counted as a backward tick. In the absence of bias, with $V_{\mathrm{DQD}} = 0$, forward and backward ticks occur with equal probability, although they can be distinguished by post processing of the signal. Two methods of counting ticks were used for measuring time over an averaging period of 30 minutes. In the first the net number of transfers $N_{\mathrm{net}}$ is determined from the number of forward ticks minus the number of backward ticks. Dividing this by average transfer rate $\nu$ gives a time estimator $\Theta_{\mathrm{net}} = N_{\mathrm{net}}/\nu$. In the second, an improved optimal estimate is obtained by summing over states: $\Theta_{\mathrm{opt}} = \sum_s n_s\, t_s$, where $n_s$ is the number of visits to state $s$ and $t_s$ is the mean dwell time in that state. In each case we define the precision as $\mathcal{S} = \frac{\langle\Theta\rangle^2}{\mathrm{Var}(\Theta)}$, which captures the stability of the time estimator.

The entropic cost per tick within the double dot device is the dissipation of energy associated with one electron flowing between source and drain divided by the temperature.

$$\Delta S_{\mathrm{tick}} = e\left|V_{DQD}\right|/T. \tag{13}$$

[25] V. Wadhia *et al*. Entropic costs of extracting classical ticks from a quantum clock. *Phys. Rev. Lett.* **135**, 200407 (2025) https://journals.aps.org/prl/abstract/10.1103/5rtj-djfk. Figures sourced from the publication and licensed under CC BY 4.0

Results from a series of experiments are presented in Fig. 6. The precision determined by simply counting net ticks is zero when the entropy per tick is too small. The optimal estimator can yield finite precision even when the entropy produced by the DQD clockwork itself tends to zero, because the measurement record carries its own hidden entropy cost in the detection circuit, The entropy per tick, up to a saturation limit, yields an increase in precision, just like the earlier cases discussed above.

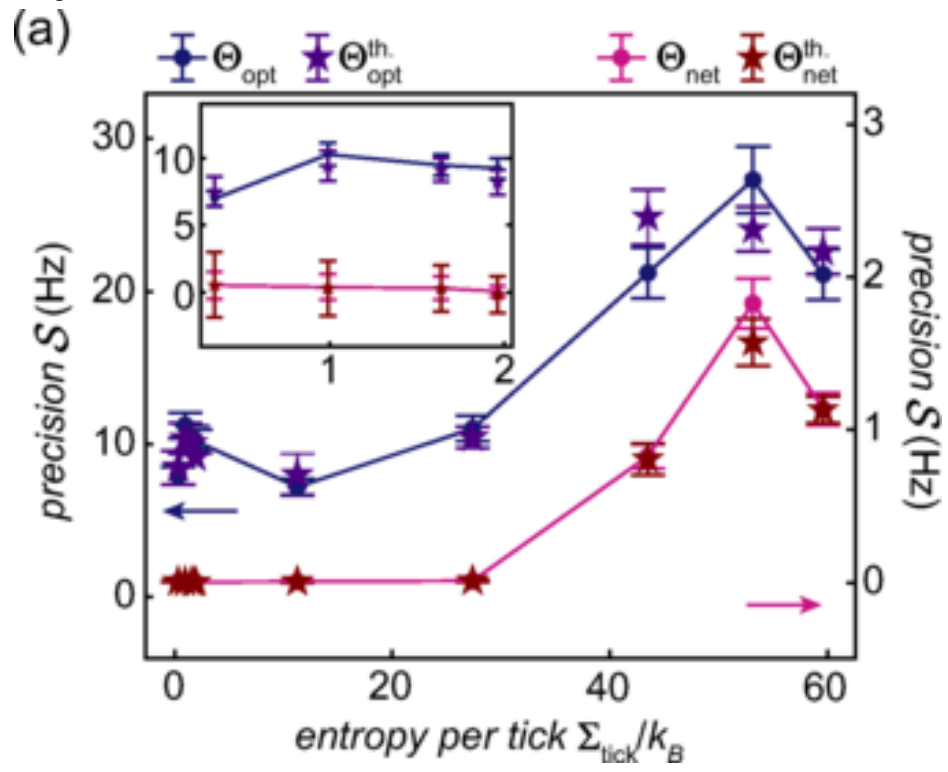

**Fig. 6** The precision $\mathcal{S}$ of the double quantum dot clock as a function of entropy per tick. Left scale: precision determined from $\Theta_{\text{net}} = N/\nu$. Right scale: precision determined from $\Theta_{\text{opt}} = \sum_s n_s t_s$. Dots indicate empirical estimates for precision; stars indicate theory prediction $\Theta^{\text{th}}$ from a Markovian model.[25]

Separate estimates were made of the entropy generation of the detection circuit. This was vastly greater than the entropy of the double-dot device itself. For the first method the power dissipation was estimated by integrating the voltage drop multiplied by changes in the current above a baseline. For the second method the power dissipation was simply taken as the difference between the rf power in and the rf power out. Dividing by the temperature, which for these experiments was $T$ = 180 mK, yielded the entropic cost of the read-out. The first method gave an entropic cost nine orders of magnitude greater than the cost of the double-dot device mechanism. The second method was better but still yielded a cost that eclipsed the intrinsic cost of the basic device.

However hard we work to make a classical or quantum device approach its theoretical lower bound of entropy generation, other factors can generate entropy that greatly exceeds the thermodynamic limit. In the case of quantum clocks, this generally occurs in the conversion from quantum to classical information that constitutes read-out.

### 3.4 Relation to other thermodynamic precision bounds

It is worth placing the bounds of Equations (9) and (12) in their wider theoretical context. They belong to a family of thermodynamic-precision relations governing parameter estimation under nonequilibrium dynamics. The classical thermodynamic uncertainty relation of Barato and Seifert[14] and its quantum extension by Hasegawa[16] can both be expressed as Cramér–Rao inequalities in which the relevant Fisher information is bounded by the entropy production,[26] and the clock case above is recovered as one instance of this scheme.

The bounds (9) and (12) are uniquely sharp within this wider family because timekeeping has no equilibrium realization. A clock that produces no ticks is not a clock; the limit $\Delta S_{\text{tick}} \to 0$ cannot be approached without destroying the device. For other observables — length, force, magnetic field, temperature — the precision–dissipation trade-off applies whenever the measurement is continuous, but the bound enters indirectly. For any compliant sensor, the fluctuation–dissipation theorem of §2 imposes a thermal-noise floor proportional to the damping rate (what the optomechanics literature refers to as *the thermodynamic limit* for position resolution[27]). At the precision frontier, the SI definitions route length, force, and so on through the speed of light $c$ to the time unit of the second. Modern length metrology is therefore not exempt from bounds such as (9) and (12); it inherits it through the optical-frequency reference that defines the metre. What is special about the clock case is not that other measurements are unconstrained, but that the clock bound is *constitutive* of the device, while bounds on *derived* observables are *inherited* from the apparatus that realises them.

[26] Y. Hasegawa and T. Van Vu, "Uncertainty relations in stochastic processes: An information inequality approach," *Phys. Rev. E* **99**, 062126 (2019). DOI: 10.1103/PhysRevE.99.062126.

[27] F. Zhou, Y. Bao, R. Madugani, D. A. Long, J. J. Gorman, and T. W. LeBrun, "Broadband thermomechanically limited sensing with an optomechanical accelerometer," *Optica* **8**, 350–356 (2021). DOI: 10.1364/OPTICA.413117.

The thermodynamic precision bound (9) and its coherent many-body counterpart (12) apply to a specific class of clocks: those whose ticks emerge from a stochastic process driven by entropy-producing dynamics. We shall call these *dissipative clockworks*. The noise-driven membrane clock (§3.1), the transmon jump clock (§3.2), the double-dot clock (§3.3), and biochemical oscillators all belong to this class, with the tighter logarithmic bound (12) applying to coherent many body architectures like the ring clock (§3.2).

A distinct class of clocks references a natural quantum oscillator — the electronic transition of an atom or ion, in atomic and optical lattice clocks. For these *reference clocks* the oscillator itself is subject only to spontaneous emission and is not constrained by (9) or (12). For clocks using hyperfine transitions the spontaneous emission time may be so long that spontaneous emission can be neglected. The precision is then set by quantum estimation of a fixed natural frequency, bounded by the standard quantum limit or its Heisenberg-scaled refinements for systems harnessing entanglement, and the accuracy is limited only by static effects such as Stark shifts.

The distinction between the two clock classes reflects a more general decomposition. Any physical clock incurs two thermodynamically distinct costs: a **maintenance cost** paid to sustain coherent oscillation against losses, with power scaling as $\hbar\omega_0/\tau_{\mathrm{obs}}$ (per oscillator for systems with, for example, many atoms), where $\tau_{obs}$ is the usable observation time; and a **readout cost** with energy of order $k_B T \ln 2$ per bit, paid each time the phase is consulted and converted into a classical record. In a dissipative clockwork the two are inseparable — each tick is simultaneously a loss event and a measurement, and the TUR (9) governs their joint cost. A reference clock decouples them: the natural-oscillator with quality factor $Q$ acts as a coherence resource that amortizes a one-time preparation cost across many sparse readouts, recovering Landauer scaling per query in the regime where the number of cycles per observation $n \ll Q$. The thermodynamic costs of a reference clock are located not in the oscillator but in the apparatus that interrogates and stabilizes it. We shall return to this distinction in §8.1. The bounds (9) and (12) constrain only the maintenance cost of a dissipative clockwork; the readout cost is bounded separately by Landauer's principle (§7) and is what dominates the entropy budget of a reference clock.

A more ambitious vision of clocks not requiring continuous dissipation was proposed by Wilczek in the form of *time crystals* — many-body systems whose ground state would spontaneously break time-translation symmetry and oscillate persistently in equilibrium.[28] No-go theorems established that such equilibrium time crystals cannot exist;[29,30] what survives is a family of *non-equilibrium* time crystals:[31] discrete, Floquet-driven, boundary, and continuous dissipative. The dissipative variants (boundary and continuous), together with bath-stabilised discrete time crystals, dissipate energy and fall within the scope of the bound (9); the closed Floquet (many-body-localised) case is driven rather than self-sustained, and its cost lies instead in the drive. Their performance as clocks has been analysed within the autonomous-clockwork framework;[32] time crystals are genuine quantum clocks, and the spontaneous breaking of time-translation symmetry quantitatively enhances clock performance compared with non-symmetry-broken alternatives. Like the ring clock of §3.2, time crystals exploit many-body coherence to navigate within the slack of the bounds (9) and (12) rather than against it.

### 3.5 The arrow of time as a brute fact

The great physicist Sir Arthur Eddington wrestled with the tension between the time-reversibility of fundamental physics and irreversibility of entropy in thermodynamics. He sought to relate these to universal human experience of past and future. "… clocks measure time-like intervals, but they conceal time's arrow. Time enters into our consciousness in a different way: consciousness declares that this private time possesses an arrow … entropy-gradient is then the direct equivalent of the time of consciousness in both its aspects … Temporal relation in physics has no arrow, but consciousness does insist on an arrow."[33] The experimental results here confirm that, far from concealing time's arrow, clocks reveal it by the entropic cost associated with their accuracy. Time's arrow does not depend on consciousness; it is apparent in any physical process that produces an irreversible record. The observer is not a *cause* of irreversibility, but a description-level *restriction*: finite resolution, finite bandwidth, limited access to degrees of freedom, and the necessity of a classical record. Once a reduced description is fixed, the corresponding entropy production and loss of accessible correlations become objective within that description.

---

[28] F. Wilczek, "Quantum time crystals," *Phys. Rev. Lett.* **109**, 160401 (2012). DOI: 10.1103/PhysRevLett.109.160401.

[29] P. Bruno, "Impossibility of spontaneously rotating time crystals: A no-go theorem," *Phys. Rev. Lett.* **111**, 070402 (2013). DOI: 10.1103/PhysRevLett.111.070402.

[30] H. Watanabe and M. Oshikawa, "Absence of quantum time crystals," *Phys. Rev. Lett.* **114**, 251603 (2015). DOI: 10.1103/PhysRevLett.114.251603.

[31] M. P. Zaletel, et al, "Colloquium: Quantum and classical discrete time crystals," *Rev. Mod. Phys.* **95**, 031001 (2023). DOI: 10.1103/RevModPhys.95.031001.

[32] L. Viotti et al., "Quantum Time Crystal Clock and its Performance," *Phys. Rev. Lett.* (2026), DOI: 10.1103/dj21-gmdj. arXiv:2505.08276.

Every action with an entropic cost leads inexorably to time's arrow. How time's arrow arises from time-symmetric physical laws remains to be adequately resolved. The arrow of time seems to be a brute fact of the Universe which is irrevocably tied to increases in entropy and flows of information. While the *measure* of subsystem entropy (Shannon, von Neumann, or entanglement) depends on the observer's decision about the partition, the *loss of information* (correlations spreading into inaccessible degrees of freedom) becomes an objective physical process once the partition is chosen. The arrow of time corresponds to the awareness that because of complex many-body evolution and the spread of entanglement, we cannot track everything. The brute fact is not the absolute value of the entropy, but rather the inevitability of correlations spreading beyond our knowledge in one particular direction in time.

# 4 Direct measurement of entropy

Having considered how entropy in small systems is to be attributed according to which states are observed, we now turn to experiments that measure or deduce entropy (or entropy changes) directly in nanosystems. In electronic nanosystems, the granular nature of matter is not a correction term but the dominant feature, forcing us to abandon macroscopic calorimetry in favour of monitoring discrete charge states.[34] We illustrate this from measurements of entropy in single electron nanodevices, in which the occupation of a quantum dot can be varied by changing its chemical potential.

## 4.1 Charge state measurements

In a device containing electrons in a quantum dot, the entropy of the system can be determined from the behaviour of the charge state. In solid state devices such as single electron transistors, the quantum dot is connected by tunnel barriers to a source and a drain, and the electrostatic energy of the dot can dominate. When the electrostatic energy is much larger than the thermal energy, Coulomb blockade can occur. The number of excess electrons or holes then becomes quantised and is no longer a continuous variable. Therefore, the partial derivative of any quantity with respect to particle number becomes undefined, and it is necessary to use the mean particle number

$$\underline{\overline{N}} = p_N N + p_{N+1}(N+1) = N + n, \tag{14}$$

where $n = p_{N+1}$ is defined as the mean excess number of electrons.

In a typical device, the chemical potential of the quantum dot is controlled by a plunger gate. There is generally a capacitive voltage drop between the plunger gate and the dot, and the ratio of the changes in the electrostatic potential of the dot and the gate is known as the lever arm. In a seminal experiment, the quantum dot was tunnel coupled to a reservoir whose temperature could be modulated by an oscillating current through a nearby quantum point contact.[35] The chemical potential of the quantum dot was varied by applying a voltage $V_\mathrm{p}$ to a plunger gate with a calibrated lever arm, and recorded in terms of the difference between $V_\mathrm{p}$ and an intermediate gate voltage $V_\mathrm{mid}$ at which the probabilities are equal of finding $N$ and $N + 1$ electrons on the quantum dot. The charge on the quantum dot was sensed by measuring the conductance $G_\mathrm{sens}$ of a further quantum point contact. The chemical potential at which a charge transition occurred on the quantum dot was then determined as a function of temperature. Changes in entropy can be deduced by an entropy-to-charge conversion using the Maxwell relation

[33] A. S. Eddington. *The Nature of the Physical World*. Cambridge University Press 1948, p. 35

[34] E. Pyurbeeva, J. Mol, and P. Gehring. Electronic measurements of entropy in meso- and nanoscale systems. *Chem. Phys. Rev.* **3**, 041308 (2022)

[35] N. Hartman *et al*. Direct entropy measurement in a mesoscopic quantum system. *Nature Physics* **14**, 1083–1086 (2018) https://www.nature.com/articles/s41567-018-0250-5. Figures sourced from the publication and reused by permission.

$$\left(\frac{\partial S}{\partial \bar{N}}\right)_{p,T} = -\left(\frac{\partial \mu}{\partial T}\right)_{p,N}, \tag{15}$$

where $\overline{N}$ is the average particle number. The particle number can fluctuate between $N$ and $N + 1$ with occupation probabilities $p_N$ and $p_{N+1}$ that depend on external parameters (temperature, chemical potential, magnetic field, etc.) similarly to $N$ in continuous-charge devices.

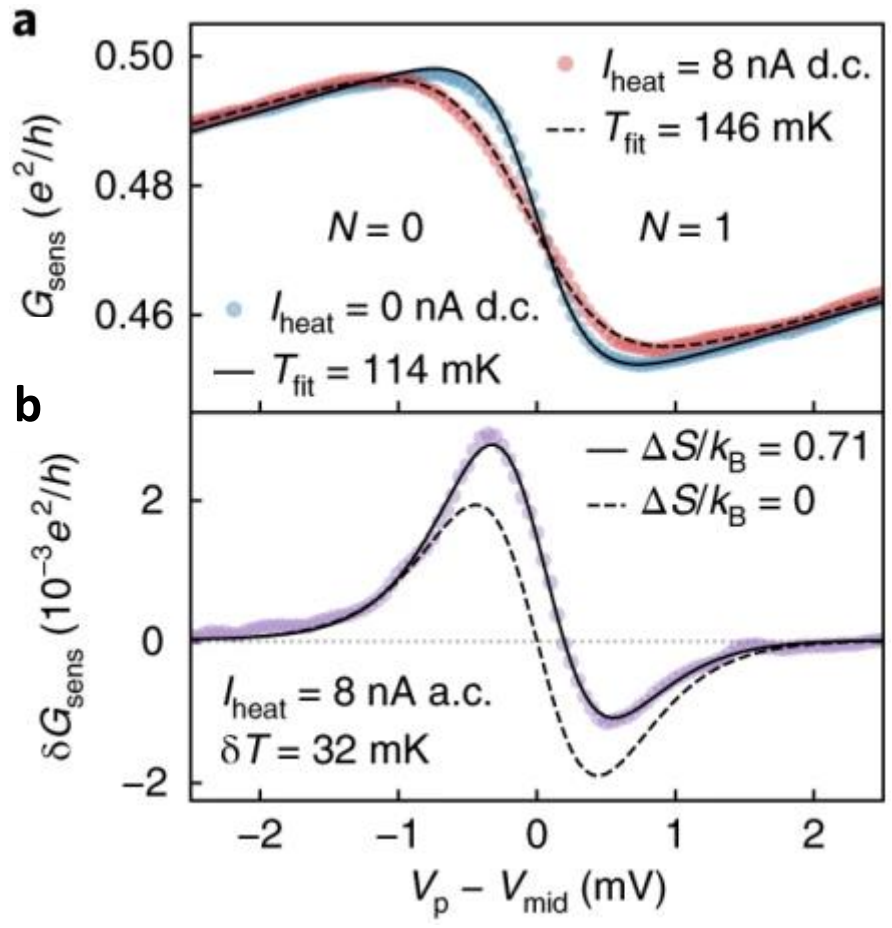

**Fig. 7** Measurements of a quantum dot device as a function of the difference between the voltage of the plunger gate and the voltage to give equal probabilities of $N = 0$ and $N = 1$. **a** Conductance from charge sensor data for $N = 0 \rightarrow 1$ at two temperatures set by the current $I_{heat}$ through the QPC heater. **b** Conductance difference from lock-in measurement of $\delta G_{sens}$ with $\delta T = 32$ mK.[35]

Fig. 7**a** shows the results of experiments with two different d.c. currents through the heater quantum point contact. The data in 7**b** were measured with an oscillating temperature amplitude 32 mK. The fitted conductance difference curve indicated that the entropy change for $N = 0 \rightarrow 1$ was $\Delta S = 0.71\ k_B$. This is comparable with the entropic cost of erasing one bit of information. Operationally, the Maxwell relation Eq. (15) lets us convert the experimentally measured charge-sensor response into a state-space entropy of the dot — Shannon for a classical occupation distribution, or its von Neumann generalisation when quantum coherences matter.

## 4.2 Charge flow measurements

The entropy of a quantum dot device subject to a temperature gradient can be determined from the thermoelectric effect. If a temperature difference $\Delta T$ is established between the source and drain of a quantum dot device, for example by the use of a localised heater, then a thermocurrent can be generated. It is possible to deduce the entropy difference between two charge states of the quantum dot by scanning the plunger gate voltage, $V_p$, and measuring both the conductance, $G = \partial I/\partial V$, and the thermal response, $L = \partial I/\partial \Delta T$. Since $L$ depends on both the entropy and the conductance, $\Delta S$ can then be determined by fitting a suitable model to the data.[36]

The results shown in Fig. 8 illustrate both the strength and the challenge of the method, since there are several fitting parameters including the temperature (which is difficult to determine directly). The deduced entropy change is different for the two Coulomb peaks, and this is attributed to the first peak corresponding to a transition into a four-fold degenerate state, and the second peak to a transition either from a four-fold to a two-fold degenerate state or from a two-fold to a non-degenerate state.

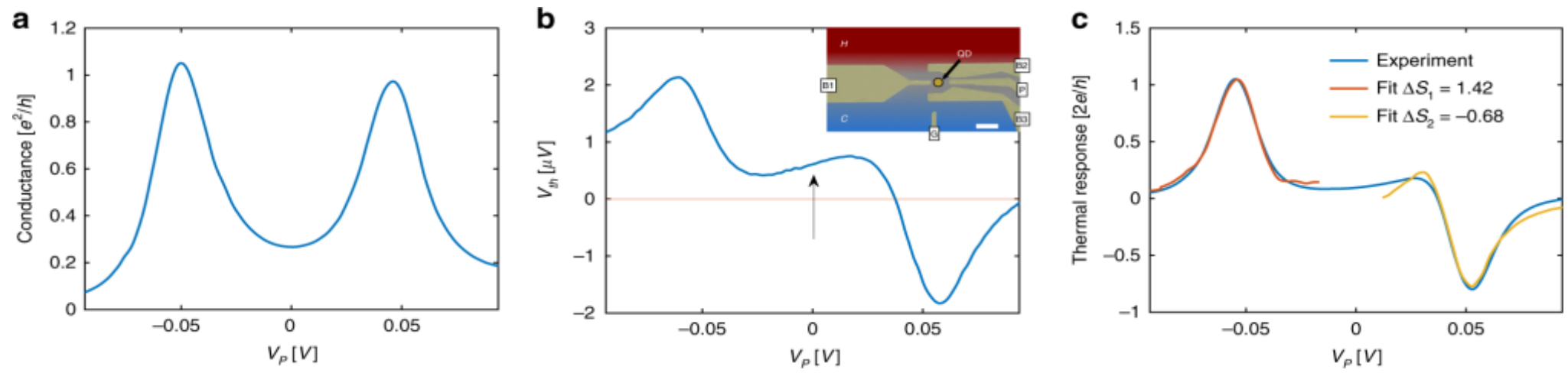

[36] Y. Kleeorin, How to measure the entropy of a mesoscopic system via thermoelectric transport. *Nat Commun* **10**, 5801 (2019) https://doi.org/10.1038/s41467-019-13630-3. Figures sourced from the publication and licensed under CC BY 4.0

Fig. 8 Experimental measurements of the QD device depicted in false colour in the inset to (**b**); electron temperature $T = 230$ mK. The horizontal axis corresponds to the QD energy, obtained from multiplying the plunger gate voltage with gate lever arm, and shifting the point of zero energy to the centre of the Coulomb-blockade valley. **a** Conductance. **b** Thermovoltage, with a non-zero value in the middle of the valleys around the apparent particle-hole symmetry point (arrow). **c** Fitting procedure, performed directly on the experimental data where each peak was fitted separately[36]

# 5 Heat pumps

In parts of Africa where electricity is not readily available, it is common to see a terrace irrigated from a stream below it by an autonomous water pump. Visitors sometimes gaze in amazement. How does that work? The pump exploits a small difference in the water level of the stream, either naturally occurring or induced by a dam, to drive a large piston which powers a much smaller piston that pumps the water up to the terrace. It is not difficult to derive a mechanical equation for the maximum achievable ratio of water pumped to water used to drive the pump, which in the limit of 100% efficiency is simply ratio of relevant heights. There is an obvious analogy with the thermodynamic limit for a perfect Carnot engine. Classical heat pumps have long been made in the form of refrigerators powered by a gas flame for situations where there is not a ready supply of electricity. Albert Einstein and Leó Szilárd collaborated on designs between 1926 and 1933, and even filed no less than 45 patents in six countries for three different models.[37]

## 5.1 Cooling a quantum dot

A simple quantum dot refrigerator can be constructed by using quantum dots as selective energy filters.[38] A device is illustrated in Fig. 9. "Centre" is the region to be cooled, and electrons enter one at a time through quantum dot A and exit one at a time through quantum dot B. The charging energy of Center was characterised by a third quantum dot.

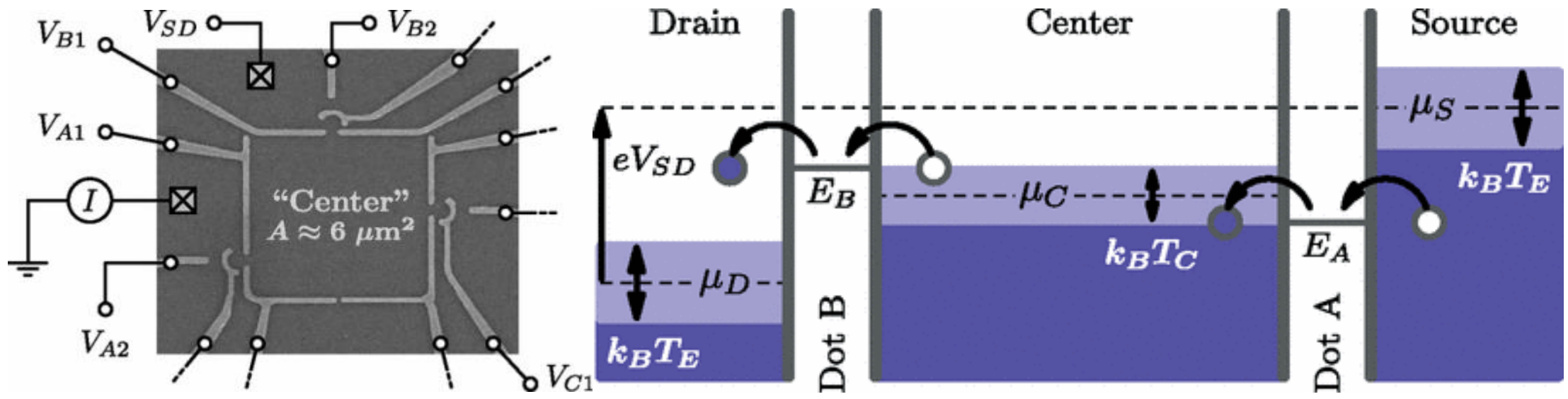

**Fig. 9** Quantum dot refrigerator (left) and the corresponding energies (right). Thermal broadening in the three two-dimensional electron gases (2DEGs; Source, Center, and Drain) is shown by the light shading around their electrochemical potentials ($\mu_S$, $\mu_C$, $\mu_D$). The net flow of an electron from Source to Drain removes an energy $E_B - E_A$ from Center. $E_A$ ($E_B$) is the ground state addition energy of dot $A$ ($B$).[38]

Dots A and B are adjusted so that they each have only one available state that falls within the energy window between the full states of the source and the empty states of the drain, as shown in Fig. 9. Dots A and B can be thought of as energy filters for the electrons flowing into and out of Center. A and B are set so that mostly cooler electrons enter Center, and mostly warmer electrons leave Center. This provides a cooling effect. Instead of providing an independent thermometer, the temperature could be inferred from the current, in this case as a function of $V_{B2}$. It was possible to cool Center to 177 mK from an ambient temperature of 280 mK, with an estimated cooling power of 0.5 fW.

[37] The Einstein Szilárd Refrigerators—Two visionary theoretical physicists joined forces in the 1920s to reinvent the household refrigerator by Gene Dannen. *Scientific American* **276**, 90-95 (1997) https://static.scientificamerican.com/sciam/cache/file/294B0F99-1A70-4B50-B07F94CA6EB4350C.pdf

[38] J. R. Prance *et al.* Electronic refrigeration of a two-dimensional electron gas. *Phys. Rev. Lett.* **102**, 146602 (2009) https://doi.org/10.1103/PhysRevLett.102.146602. Figures sourced from the publication and reused by permission.

## 5.2 Cooling a superconducting device

A quantum autonomous refrigerator has been developed for the practical aim of re-initialising superconducting qubits.[39] An early design for a quantum adsorption refrigerator had been proposed in 2012.[40] The experimental implementation is illustrated in Fig.10.

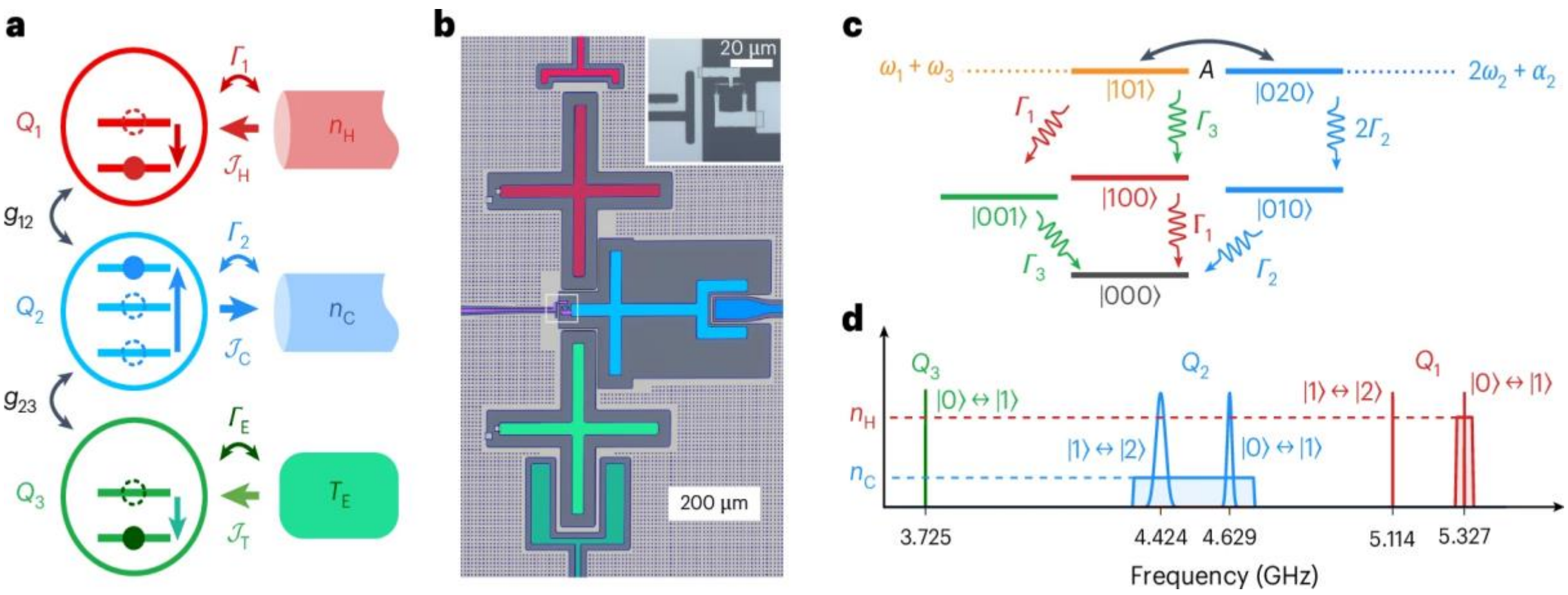

**Fig. 10** Quantum refrigerator comprised of three superconducting devices. **a** Conceptual scheme with two qubits and a three-level qutrit. Qubit $Q_1$ couples directly to a waveguide at a rate $\Gamma_1$; and qutrit $Q_2$, to another waveguide at a rate $\Gamma_2$. Qubit $Q_3$ couples undesirably to an uncontrolled bath in its environment. This bath keeps $Q_3$ at an effective temperature $T_E$. The coupling rate $\Gamma_E$ determines the natural energy-relaxation time of $Q_3$. The waveguides can operate as heat baths containing photons of average numbers $n_H$ and $n_C$. The inter qudit couplings ($g_{12}$, $g_{23}$) engender a process in which one excitation in $Q_1$ and one excitation in $Q_3$ are simultaneously exchanged with a double excitation in $Q_2$. This exchange helps reset $Q_3$. When heat baths drive this process, the system operates as an autonomous quantum refrigerator. The average heat currents are depicted by wide arrows from the baths of the hot ($J_H$), cold ($J_C$) and target ($J_T$) systems. By energy conservation, $J_H + J_T = J_C$. **b** False-colour micrograph of the device implemented with superconducting circuits. $Q_2$ is frequency-tuneable due to a flux current line and two parallel Josephson junctions, as magnified in the inset. **c** Level diagram showing tensor products $|q_1q_2q_3\rangle$ of the qudits' energy eigenstates. $|101\rangle$ and $|020\rangle$ are resonant if $\omega_1 + \omega_3 = 2\omega_2 + \alpha_2$. At resonance, a three-body interaction couples the states at a rate A. **d** Distributions over the qudits' experimentally observed transition frequencies. The Lorentzian distributions' widths represent spectral widths. The red-shaded box depicts the spectral density $n_H$ of photons injected into the waveguide coupled to $Q_1$. This synthesized noise realizes the refrigerator's hot thermal bath. Analogous statements concern the blue box, $n_C$, $Q_2$ and the cold bath.[39]

The device to be cooled is the superconducting qubit $Q_3$, which receives heat through undesirable but inevitable leakage from the environment in the cryostat. A similar qubit, $Q_1$ is connected to the hot heat bath, which provides the thermal energy to drive the process. A further device, $Q_2$, is connected to the cold heat bath. $Q_2$ is operated as a qutrit, which can be thought of as an oscillator with three levels. The potential of the qutrit is not perfectly harmonic, and it can be tuned so that the energy of its second excited state is the sum of the energies of the first excited states of $Q_1$ and $Q_3$, thereby allowing energy from $Q_1$ and $Q_3$ to be resonantly transferred to $Q_2$, whence it can be dissipated to the cold bath. By this process, the superconducting device could be cooled to close to 22 mK, less than half the temperature of the cold bath at 45 mK. As a result the effective energy-relaxation time of $Q_3$ decreased by a factor of >70, from 16.8 $\mu$s to 230 ns. This could provide a practical application for nanoscale thermodynamics in quantum technologies.

# 6 Heat engines

Lord Kelvin's statement of the second law of thermodynamics is that no process is possible whose sole result is the complete conversion of heat into work. A consequence is that it is not possible to make a heat engine using only a single thermal reservoir, or equivalently more than one reservoir at the same temperature. We shall see below that work can be extracted from a single reservoir if there is an additional resource of information, albeit with an entropic cost associated with processing the information. With more than one reservoir, a heat engine can in principle operate reversibly with no entropic cost, although practical engines never achieve this limit.

---

[39] M. A. Aamir, *et al.* Thermally driven quantum refrigerator autonomously resets a superconducting qubit; *Nature Physics* **21**, 318–323 (2025) https://doi.org/10.1038/s41567-024-02708-5. Figures sourced from the publication and licensed under CC BY 4.0

[40] A. Levy and R. Kosloff. Quantum absorption refrigerator, *Phys. Rev. Lett.* **108**, 070604 (2012). https://doi.org/10.1103/PhysRevLett.108.070604

As an alternative to reservoirs at different temperatures, an engine can be operated with reservoirs at different chemical potentials. On the macroscopic scale this is how ordinary electric motors are powered. On the nanoscale it is possible to power mechanical motion by the movement of electric charge in novel ways which provide additional insight into the role played by information. A striking example of this is the generation of self-sustained mechanical oscillations in a carbon nanotube, driven purely by DC bias and information flow.[41]

## 6.1 Self-oscillations driven by information

Figure 11 illustrates what may be the world's smallest and coldest guitar string.[42] A similar single-walled carbon nanotube with a narrow semiconducting band gap and a diameter of about 2 nm was stamped across metal electrodes to give a vibrating length of about 800 nm with a frequency of about 230 MHz. Between the source and drain electrodes were several gate electrodes. The sample was cooled to about 25 mK, which is less than 1/100 of the temperature of the background radiation in outer space. The electrode voltages were set to create tunnel barriers near each contact, so as to form a single electron transistor (SET) whose motion could be detected either by direct current or by rf reflectometry, as illustrated in Fig. 12.

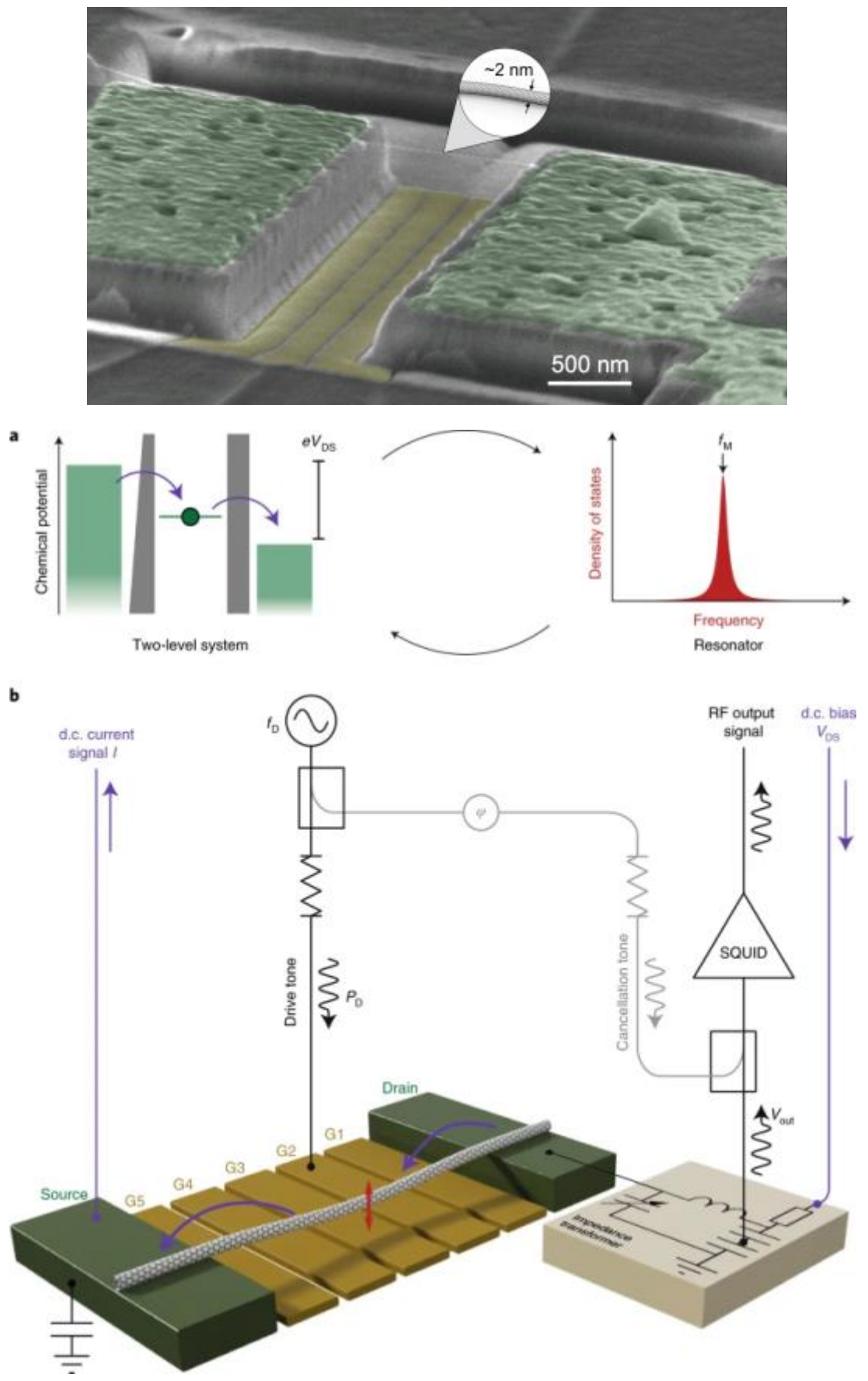

**Fig. 11** A single walled carbon nanotube suspended between source and drain electrodes, similar to the one used in the experiment described here. The electrodes between the source and the drain can be used to control tunnel barriers and apply a rf signal.[42]

**Fig. 12** A vibrating nanotube acting as a mechanical resonator configured to be a single electron transistor (SET). **a** The SET acts as a two-level system, while the resonator is a phonon cavity at mode frequency $f_M$. Tunnelling of electrons with charge $e$ through the SET leads to a non-equilibrium population distribution that pumps the oscillator. **b** The nanotube is suspended between contact electrodes (green) and above gate electrodes (yellow). The SET is biased by a drain–source voltage $V_{DS}$, and the motion is measured via the electrical current, which is monitored at d.c. ($I$, current path indicated by blue arrows), and simultaneously via an RF circuit for time-resolved measurements ($V_{out}$, signal path marked by undulating arrows). The resonator can be driven directly by a tone with power $P_D$ at frequency $f_D$, part of which is routed via a cancellation path (with tunable phase φ and attenuation) to avoid saturating the amplifiers.[41]

The results of three successive experiments are presented in Fig. 13. In the stability plot (a), the current is displayed as a function of the bias voltage (vertical scale) and the gate potential (horizontal scale). The irregular Coulomb diamonds confirm that the nanotube is behaving as a single electron transistor, in which the dark regions indicate Coulomb blockade where there is insufficient energy to change the integer electron charge. In the second plot (b), the vertical scale is the frequency of a tone which was applied to one of the electrodes, G2. The nanotube in its role as a single electron transistor now acts as a charge sensor which responds to the deflection of the nanotube from the gate electrode. The radio-frequency component of the current passes through an impedance transformer and is then amplified by an ultra-low-noise superconducting quantum interference device (SQUID) to produce a signal corresponding to the presence of mechanical vibrations. The resonant frequency shows a dependence on the gate voltage corresponding to the spacing of the Coulomb diamonds, which is therefore associated with changes in the charge between the tunnel barriers because the effective spring constant softens in the vicinity of a charge transition. There is also a steady average change with gate voltage, which we attribute to an increase in tension rather like tuning a guitar string. For the third experiment (c), no oscillating excitation was applied. The only source of energy arose from the potential difference of the source and drain electrodes on which the nanotube was mounted. The output signal was fed to a spectrum analyser to detect the response of the nanotube. At sufficient bias, above an onset voltage of 0.5 mV, oscillations spontaneously appear at frequencies that depend on the gate voltage in almost the same way as the induced oscillations in Fig. 13b.

[41] Y. Wen *et al*. A coherent nanomechanical oscillator driven by single-electron tunnelling. *Nature Physics* **16**, 75–82 (2020) https://www.nature.com/articles/s41567-019-0683-5, Figures sourced from the publication and used with permission.

[42] E. A. Laird *et al*. A high quality factor carbon nanotube mechanical resonator at 39 GHz. *Nano Lett.* **12**, 193–197 (2012) https://pubs.acs.org/doi/full/10.1021/nl203279v. Figure sourced from the publication and used with permission.

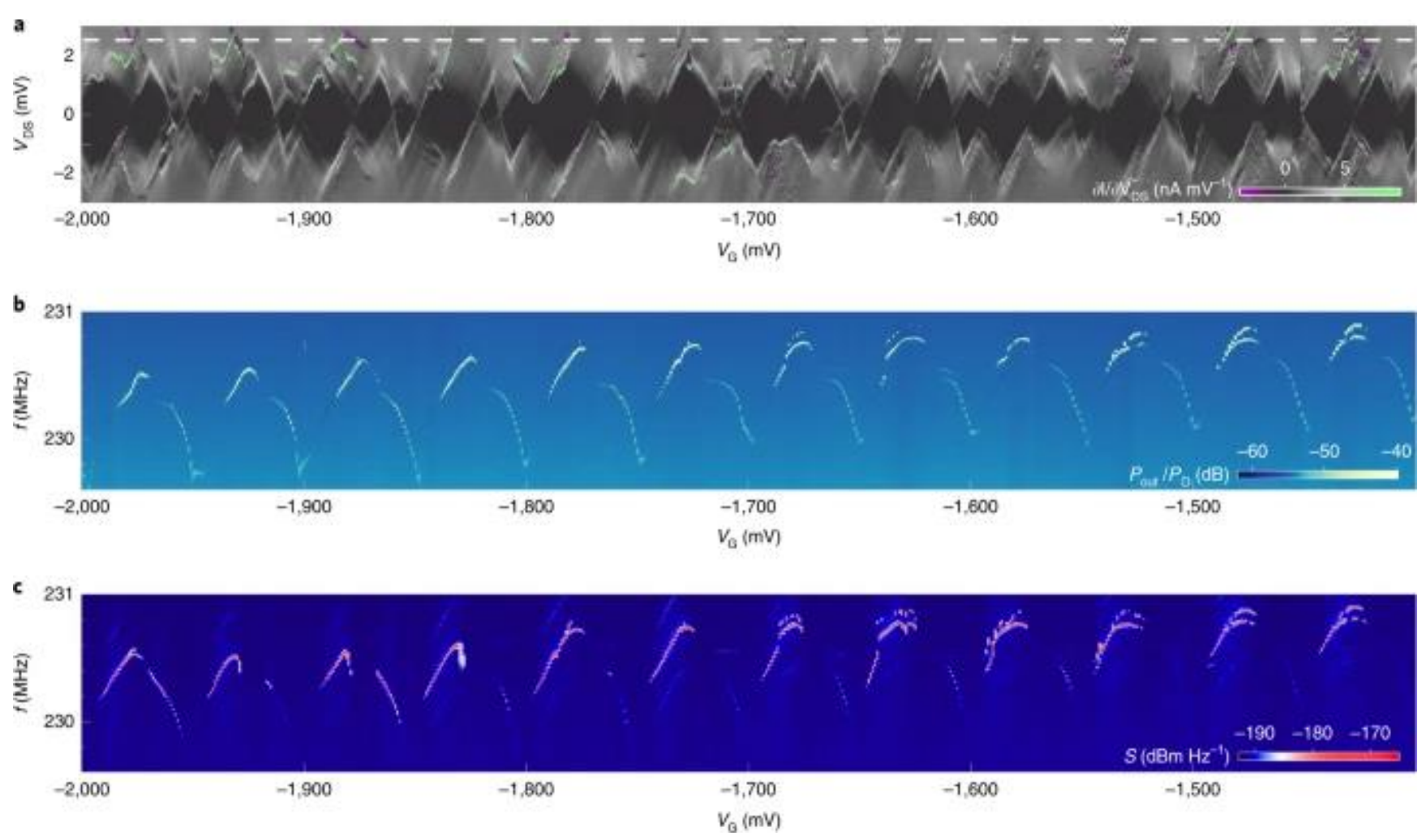

**Fig. 13. a** Differential conductance of the nanotube as a function of gate voltage $V_G$ and bias $V_{DS}$, with no driving applied. The diamond pattern is characteristic of Coulomb blockade in an SET, with some irregularity due to electrostatic disorder. Superimposed on the expected diamond pattern are sharp lines of strongly positive or negative differential conductance (green or purple in this colour scale), indicating thresholds for self-oscillation. **b** Mechanical resonance detected in a transmission measurement. The nanotube is biased with $V_{DS}$ = 2.5 mV (dashed line in a) and driven with power $P_D$ = −99 dBm to gate G2. The transmission is plotted as the emitted power $P_{out}$ into the amplifier chain. Each mechanical resonance appears as a sharp spectral peak, or occasionally as a faint dip when RF leakage interferes destructively with the mechanical signal. **c** Emission spectrum density *S* as a function of frequency *f* under the same conditions but with no RF drive. Spectral peaks are still present, at almost exactly the same frequencies as in **b**. This indicates self-driven mechanical oscillations. [41]

How do these self-oscillations arise? The appearance of the oscillations over a wide range of gate voltages precludes positive feedback through conventional stimulated phonon emission. The presence of oscillations on both sides of the Coulomb peaks rules out an electrothermal mechanism, which might be caused by a time lag in thermal expansion due to ohmic heating. It seems rather that the damping term normally associated with Coulomb blockade in a vibrating device becomes negative in some regimes, so that instead of damping the oscillations it amplifies them. This comes about through an asymmetry in energy dependence of the tunnelling rate through the tunnel barriers which form the confined region of the nanotube. Suppose that the gate has a positive voltage which in its rest position gives the central section of the nanotube a probability 0.5 of an additional electron, with the rates of tunnelling on and off equal. If now a fluctuation of the nanotube away from the electrode changes the relationship between the central section and the tunnel barriers in such a way that the rate of electrons tunnelling *on* becomes greater than the rate of electrons tunnelling *off*. The negative charge will increase, and the nanotube will be attracted to the positively charged gate. As the motion continues, the opposite will happen, and the nanotube will experience a disproportionate net force away from the gate. In this way a limit cycle will arise, manifesting itself in the oscillations observed in Fig. 13c.

This invites comparison with Feynman's celebrated ratchet-and-pawl,[43] the paradigmatic example showing that no mechanical device can rectify the thermal fluctuations of a single bath: the pawl is subject to the same fluctuations as the ratchet, and over long times the rectification and the back-slip cancel exactly. The suspended nanotube appears to evade this prohibition — directed oscillations emerge spontaneously from a configuration that looks at first sight like equilibrium — but the appearance is misleading. The DC bias $V_{DS}$ places the device in a non-equilibrium steady state with source and drain reservoirs at different electrochemical potentials, and the directed motion is powered by the asymmetry of tunnelling rates across that bias, not by the fluctuations of a single bath. The equivalent version of Feynman's ratchet would need to have the pawl at a lower temperature than the ratchet, so that the thermal bounces of the pawl were less than the thermal jiggling of the wheel. This could then constitute a heat engine that could do work. There are suggestions that biological engines may operate in such a way.[44] The self-oscillating nanotube provides an electromechanical demonstration that information about the charge state, combined with a chemical-potential difference, can produce the directed motion that Feynman's mechanical original could not.

We can begin to imagine this effect in terms of information. A similar result could be obtained by using a charge sensor to detect whether there is an excess electron on the nanotube. This could be used to control the tunnel barriers, in a way that in principle might be fully reversible and therefore generate no entropy. Operationally, the behaviour of the nanotube might then be similar to Fig. 10c. The binary information about the presence or absence of an excess electron would then generate entropy when the sensor is reset for the next measurement, along the lines of a Szilárd engine and the Landauer principle. The suspended nanotube thus provides a platform for future experiments on nanoscale thermodynamics of information.

## 6.2 Experimental realization of a Szilárd engine with a single electron

In 1929, the Hungarian born physicist Leó Szilárd conceived of an engine in which information can be used to pump heat from an isothermal environment and transform it into work. There is an intimate connection between the amount of energy which a Szilárd engine can convert from heat into work using one bit of information and the minimum thermal cost of erasing one bit of information. Both results can be derived from thought experiments involving a single molecule in a box with a movable partition. In a Landauer erasure work $k_B T \ln 2$ *is done* moving the partition from one end to the middle, and in a Szilárd engine the partition *does* work $k_B T \ln 2$ as the molecule pushes it from the middle to one end.

Szilárd's thought experiment is a refinement of Maxwell's earlier demon concept. There is again a chamber with a partition containing a hole which can be opened or closed, but now there is only one molecule. After an observation has been made to determine which side the molecule is on, a piston is placed on the other side of the partition. The hole is opened, and the molecule pushes the piston, which is connected to a mechanism that can do work such as lifting a weight. The kinetic energy which the molecule loses from each collision can be refreshed by thermal energy from the environment, and the process can be repeated cyclically to do further work. An experiment with a Brownian particle and an optical feedback mechanism demonstrated how information could be converted to work in this kind of way, and at the same time demonstrated how the generalised Jarzynski equality of §2.1 converged when the number of cycles grew to 100,000 or more.[45]

An electronic Szilárd engine has been implemented in a nanoscale system of two quantum dots, see Fig. 14.[46] Two metallic islands connected by a tunnel junction form a single-electron box (SEB). Neglecting the underlying sea of electrons, the probability of one additional electron being in either the left or the right dot can be controlled by the gate voltage $V_g$ applied to one of the dots (the left dot in the picture). A single-electron transistor (SET) in the vicinity of one of the dots (the right dot in the picture) serves as an electrometer to detect which dot the excess electron is on. Initially $V_g$ is set so that the excess electron is equally likely to be on either dot. The average charge associated with the gate capacitance, $n_g$, is then half-integer. The SET detects which dot the excess electron is on, and then $V_g$ is rapidly switched to stabilise the electron on that dot. In a conventional Szilárd engine this would correspond to inserting a partition and connecting a weight on which the molecule does work as it pushes against the partition. In the SEB experiment, the voltage $V_g$ is allowed to ramp slowly back to the equilibrium value, thereby enabling the excess electron to extract energy from the heat bath and feed it as electrical work into the circuit. By conservation of energy, the heat extracted from the reservoir must be equal to the electrical work done, which can be calculated from the changes in charge and the relevant capacitances, allowing also for the work done by the gate voltage source.

[43] R. P. Feynman, R. B. Leighton, and M. Sands, *The Feynman Lectures on Physics*, Vol. I, Ch. 46 (Addison-Wesley, 1963).

[44] Peter M. Hoffmann, *Life's Ratchet: How Molecular Machines Extract Order from Chaos*. Basic Books (2012)

[45] Toyabe, S., Sagawa, T., Ueda, M. *et al.* Experimental demonstration of information-to-energy conversion and validation of the generalized Jarzynski equality. *Nature Phys* **6**, 988–992 (2010). https://doi.org/10.1038/nphys1821

[46] J. V. Koski, V. F. Maisi, J. P. Pekola, and D. V. Averin, Experimental realization of a Szilárd engine with a single electron, *Proceedings of the National Academy of Sciences* **111**, 13786 (2014). Figures sourced from PNAS; permission not required.

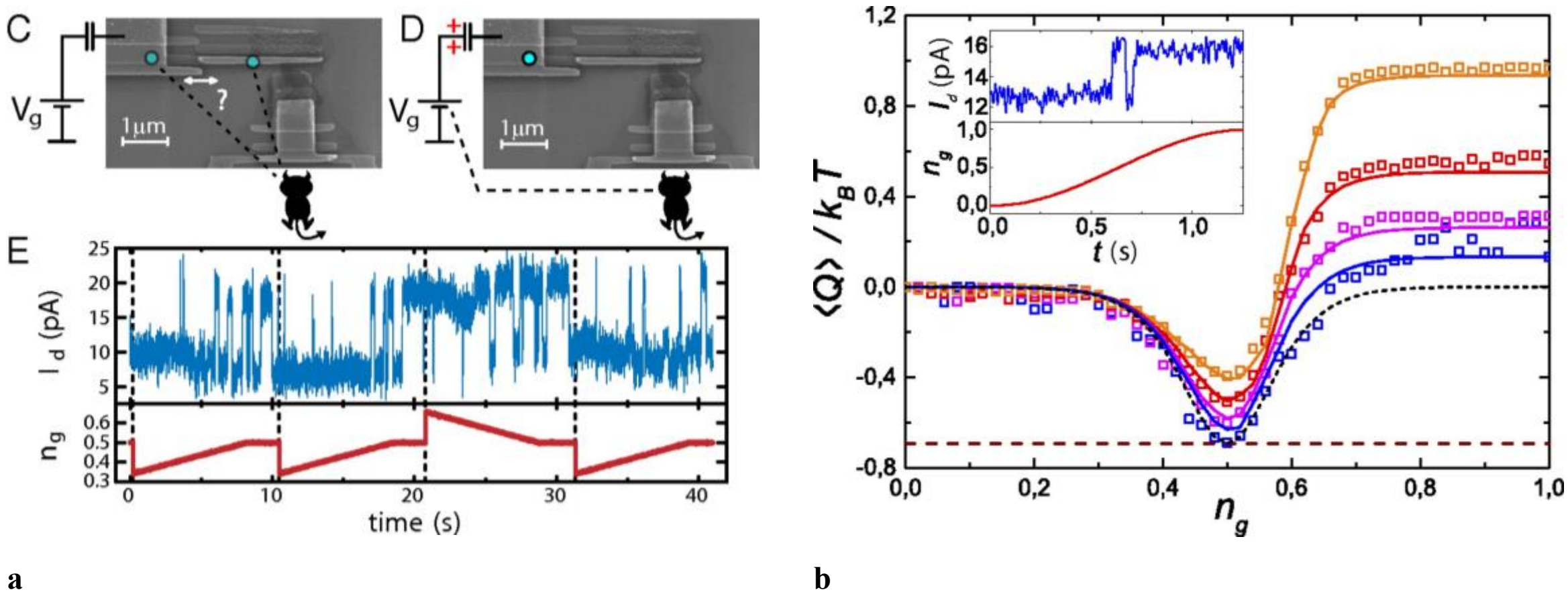

**Fig. 14**. Experimental realization of a Szilárd engine in a single-electron box. **a** An excess electron is located on one of the two metallic islands. An SET electrometer detects the electron, while the gate voltage $V_g$ is applied to control the tunnelling of the extra electron, thus trapping it capacitively. As $V_g$ is slowly driven back to the original value, the net extracted work $k_B T$ ln 2 is produced by thermal activation. The time trace shows the excess electron location, signalled by the SET current $I_d$; the bottom trace shows the charge associated with the gate capacitance $n_g = C_g V_g / e$, where $C_g$ is the coupling capacitance between the gate electrode and the gated box island. **b** The average total heat transferred to the reservoir in a ramp from $n_g = 0$ up to the value on the $n_g$ axis (negative values denote transfer from the reservoir). Symbols show the measured and solid lines the theoretical results. Dashed curve gives the theoretical quasi-static limit; dashed horizontal line gives the fundamental $-k_B T$ ln 2 limit. The maximum drive rates are $dn_g/dt = 0.22\ \Gamma_0$ for orange, $0.11\ \Gamma_0$ for red, $0.055\ \Gamma_0$ for magenta, and $0.027\ \Gamma_0$ for blue, where $\Gamma_0 = 22$ Hz is the tunnelling rate at degeneracy. The averages are taken over $n$ = 2,105, 1,764, 333, and 160 repetitions, respectively. (Inset) Example of an actual measurement.[46]

The entropic cost of operating the SEB Szilárd engine can be compared directly with the minimum entropy per observation. Each tunnelling event produces heat corresponding to the change in charging energy across the total capacitance of the two dots. The change in charge is determined from the SET current $I_d$, as shown in Fig. 14a. Hence the average heat ⟨Q⟩ transferred from the reservoir can be deduced, as shown in Fig. 14b. The corresponding decrease in entropy is then $\langle Q\rangle / k_B T$. Figure 14b presents data showing that if the engine starts from $n_g = 0.5$, corresponding to equal probability of the electron being on either dot, then if the experiment is performed sufficiently slowly the average heat transferred from the reservoir is $k_B T$ ln 2, corresponding to the maximum extraction of heat from the reservoir into the electrical circuit. It is a common theme in the application of the second law that equality is approached only in the limit of a very slow operation. As the engine is operated faster, the excess entropy increases, as can be seen in Fig. 14b, where the asymmetry may be attributed to the asymmetry in the application of $V$g.

The cartoon demons in Fig 14a represent the observation and activation activities of Maxwell's demon in his thought experiment, in which neither the observation nor the operation of the shutter necessarily consume energy or generate entropy. But eventually the entropic price has to be paid when the demon resets its memory ready for the next observation (infinite memory is not allowed!). This corresponds to the limit of $-k_B T$ ln 2 observed in Fig. 14b.

The entropy of the reservoir decreases as heat is taken from it, with a minimum *decrease* $k_B$ ln 2. This is offset when reset occurs, with a minimum *increase* $k_B$ ln 2. We thus see the connection between the work per cycle of an information engine which uses one bit of information to extract heat from a reservoir to do work, and the cost of erasure of that information ready for the next cycle.

## 6.3 Information as an ingredient in the Second Law

The second law of thermodynamics was memorably expressed by the British musical comedy duo Michael Flanders and Donald Swann. “Heat won’t pass from a cooler to a hotter. You can try it if you like but you far better notter.”[47] Einstein was explicit about the enduring validity of thermodynamics.

> [A law] is more impressive the greater the simplicity of its premises, the more different are the kinds of things it relates, and the more extended its range of applicability. Therefore, the deep impression which classical thermodynamics made on me. It is the only physical theory of universal content, which I am convinced, that within the framework of applicability of its basic concepts will never be overthrown.[48]

Sir Arthur Eddington, one of the great 20th Century advocates of thermodynamics, expressed the robustness of the Second Law even more memorably.

> The law that entropy always increases — the second law of thermodynamics — holds I think, the supreme position among the laws of Nature. If someone points out to you that your pet theory of the universe is in disagreement with Maxwell's equations — then so much worse for Maxwell’s equations. If it is found to be contradicted by observation — well these experimentalists do bungle things sometimes. But if your theory is found to be against the second law of thermodynamics, I can give you no hope; there is nothing for it but to collapse in deepest humiliation.[49]

More recently the MIT physicist Seth Lloyd expressed the same sentiment succinctly, ‘Nothing in life is certain except death, taxes, and the second law of thermodynamics.’[50]

Well, yes, provided you include information. In a heat engine, which normally requires a hot and a cold reservoir, information can play the role of the cold reservoir, as it does in a Szilárd engine. In a heat pump, which normally requires energy input, information can provide the resource to drive heat from the cooler to the hotter.

Recall from §1.3 that the mutual information $I(\mathrm{A{:}B}) = S_\mathrm{A} + S_\mathrm{B} - S_\mathrm{AB}$ measures the correlations between two systems, is zero for uncorrelated systems, and increases as they thermalise, $\Delta I(\mathrm{A{:}B}) \geq 0$. If instead initial correlations are established between systems at different temperatures, that mutual information becomes a resource that can drive heat from the cooler to the hotter.

## 6.4 Heat driven uphill by mutual information

Heat flow from the cooler to the hotter driven by mutual information has been demonstrated using magnetic resonance applied to two spin ½ systems, as illustrated in Fig. 15.[51] These were the nuclear spins of $^{13}$C and $^{1}$H in a $^{13}$C-labeled $CHCl_3$ liquid sample, with the $^{13}$C isotope to provide nuclear spin, diluted in Acetone-d6 (Fig. 15b). The sample was placed in a strong magnetic field and r.f. pulses were applied to establish effective temperature differences between the carbon and hydrogen spins. The spins were then either left uncorrelated, or through further pulses brought into a state which gave a correlation term in the density matrix $\chi_\mathrm{AB} = \alpha|01\rangle\langle10| + \alpha^*|10\rangle\langle01|$.

[47] M. Flanders and D. Swann, “First and Second Law.” https://en.wikipedia.org/wiki/Flanders_and_Swann

[48] Albert Einstein, quoted in M.J. Klein, *Thermodynamics in Einstein's Universe*, in *Science*, **157** (1967), p. 509.

[49] Sir Arthur Stanley Eddington, in *The Nature of the Physical World*. Macmillan, New York (1948) p. 74.

[50] S. Lloyd, Going into reverse. *Nature* **430**, 197-200 (2004)

[51] K. Micadei *et al*. Reversing the direction of heat flow using quantum correlations. *Nature Commun.* **10**, 2456 (2019) https://www.nature.com/articles/s41467-019-10333-7. Figures sourced from the publication and licensed under CC BY 4.0

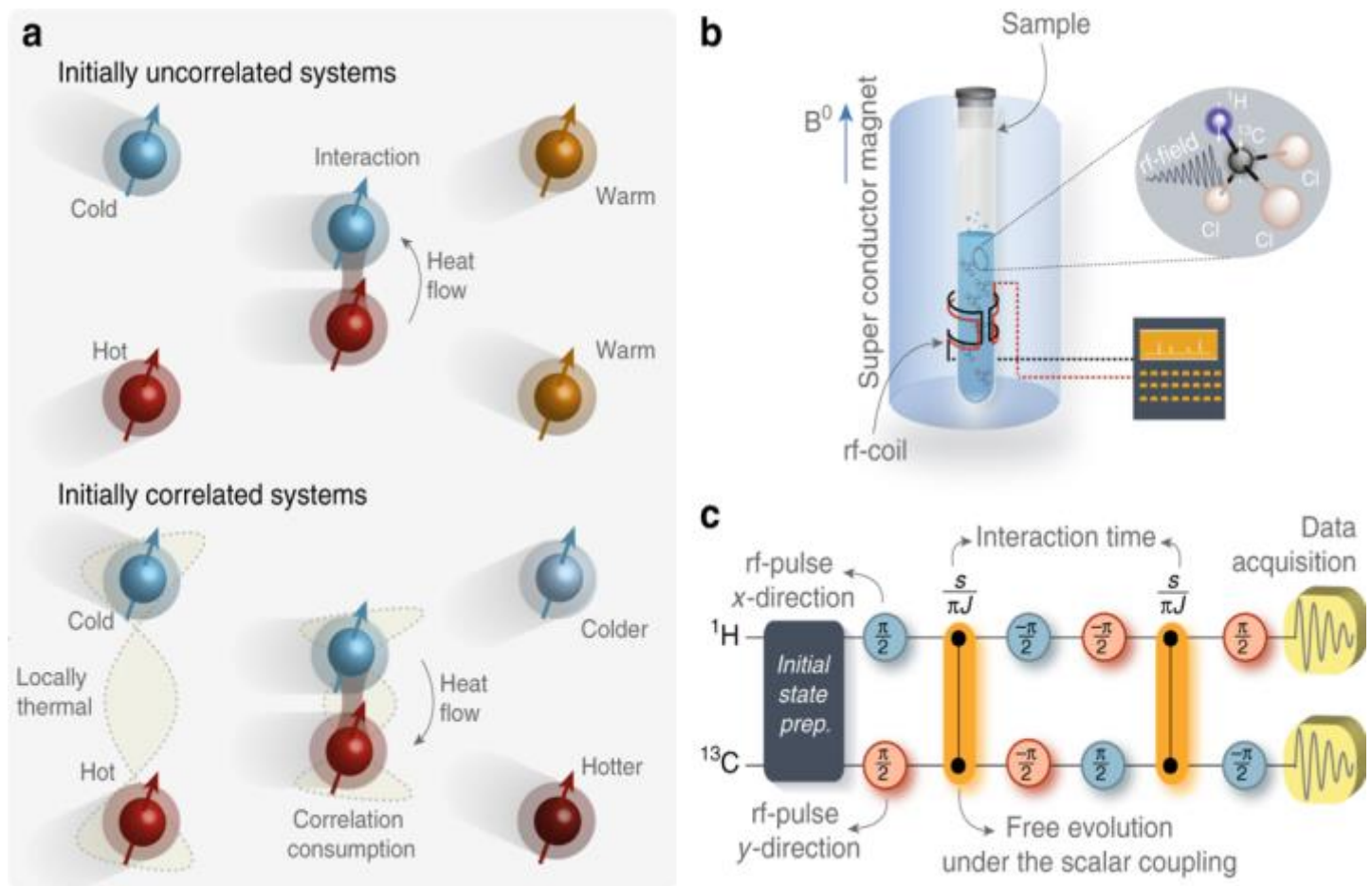

**Fig. 15** Two initially quantum-correlated spin-1/2 systems prepared in local thermal states at different effective temperatures. **a** Heat flows from the hot to the cold spin (at thermal contact) when both are initially uncorrelated. This corresponds to standard thermodynamics. For initially quantum-correlated spins, heat is spontaneously transferred from the cold to the hot spin. The direction of heat flow is here reversed. **b** The magnetometer: a super-conducting magnet, producing a high-intensity magnetic field $B_0$ in the longitudinal direction, is immersed in a thermally shielded vessel in liquid helium. The sample is placed at the centre of the magnet within the rf-coil of the probe head. **c** Experimental pulse sequence for the partial thermalization process. The blue (black) circle represents $x$ ($y$) rotations by the indicated angle. The orange connections represent a free evolution under the scalar coupling between the $^{1}H$ and $^{13}C$ nuclear spins.[51]

The measured parameters of the subsequent evolution are shown in Fig. 16. In the uncorrelated case α = 0. The hotter spin, A, gradually cools and the cooler spin, B, gradually heats, in accordance with the second law as Einstein and Eddington knew it. There is little change in the mutual information or the quantum discord, which is a measure of the non-classical correlations between the two spins. In the correlated case, in which α = 0.19 ± 0.01, the energy flow is from B to A, against the temperature differential. This is driven by the consumption of the correlation for a little more than 1 ms, after which the correlations begin to increase and the energy flow reverts to its conventional direction.

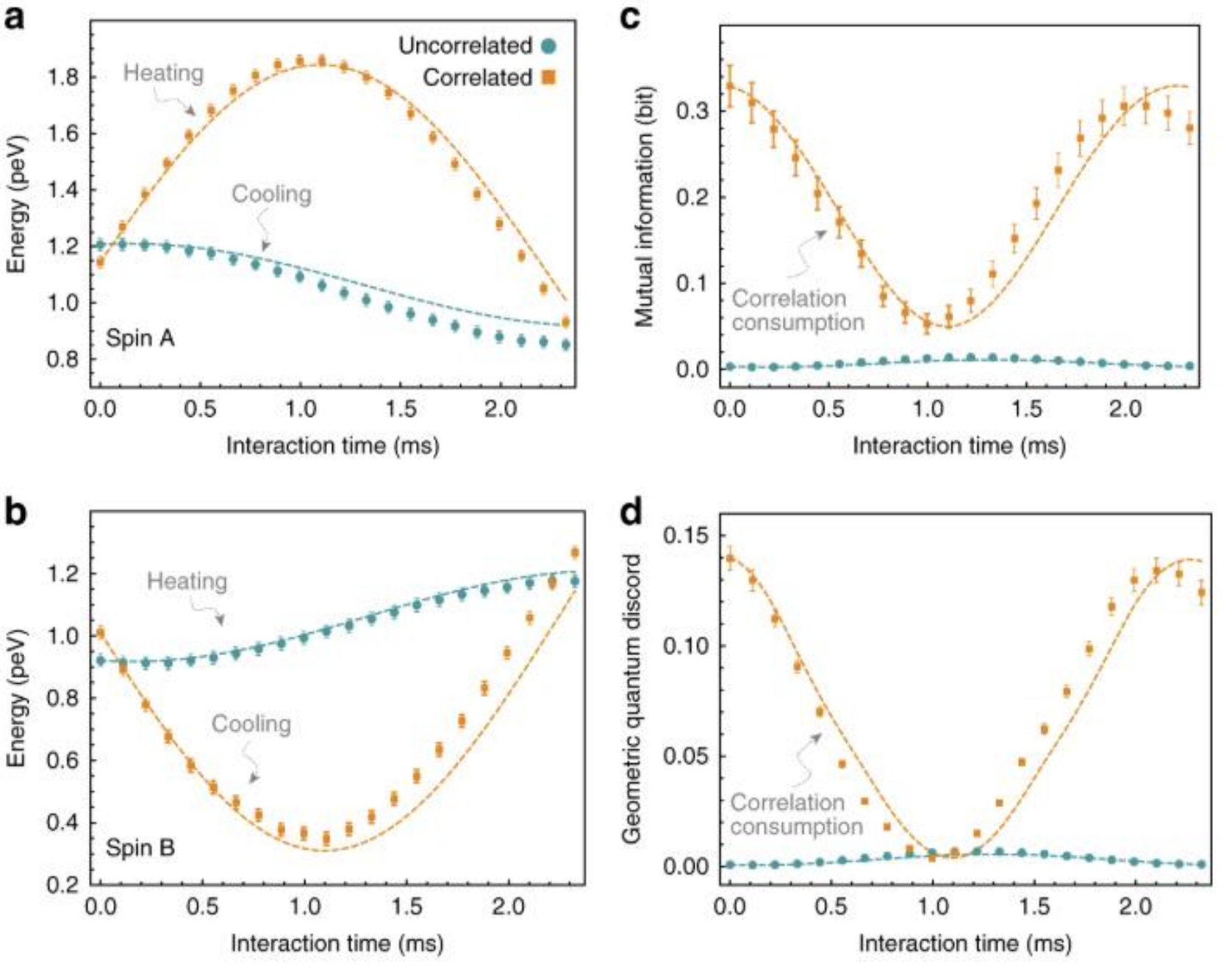

**Fig. 16** Dynamics of heat, correlations, and entropic quantities. **a** Internal energy of qubit A along the partial thermalization process. **b** Internal energy of qubit B. In the absence of initial correlations, the hot qubit A cools down and the cold qubit B heats up (cyan circles in panel **a** and **b**). By contrast, in the presence of initial quantum correlations, the heat current is reversed as the hot qubit *A* gains and the cold qubit B loses energy (orange squares in panel **a** and **b**). This reversal is made possible by a decrease of the mutual information **c** and of the geometric quantum discord **d**.[51]

This experiment provides a further demonstration of how information can play a significant role in nanoscale thermodynamics, just as it does in the resolution of the apparent paradox presented by Maxwell's demonic thought experiment and in the Landauer erasure of a correlated quantum system.

## 7 Landauer's principle

In 1961 Rolf Landauer discovered the principle now named after him. When an irreversible change occurs in the information stored in a physical system such as a computer, there is an irreducible dissipation of heat to the surroundings. The erasure of 1 bit of classical information incurs a minimum entropic cost $k_B \ln 2$. This result can be derived by considering a cylinder containing one molecule of gas. The cylinder can be divided in two by a partition in the middle. The molecule is either in the left half or the right half, which we can consider as corresponding to a 0 or a 1, and therefore carries one bit of information, albeit not known to the observer (this is crucial). The information can be forever erased by removing the partition and then with a moveable piston compressing the molecule isothermally into, say, the left half. Straightforward analysis shows that this requires work $k_B T \ln 2$. Since the compression is isothermal, the work is dissipated as heat into the environment. Landauer's principle generalises this, and states that any erasure of one bit of information generates a minimum amount of heat

$$Q = k_B T \ln 2 \quad (16)$$

Like much in thermodynamics, the Landauer limit can be approached only if the operation is done very slowly. An engine that produces useful power will inevitably have heat gradients and other irreversible processes, and so will not reach Carnot efficiency. An early experimental test of Landauer erasure trapped a microscopic silica bead in a laser beam trap with two valleys in which the particle could sit, representing a 0 and a 1.[52] By manipulating the laser the energy barrier between the valleys could be varied and the relative energy of the valleys could be tilted, and by measuring the position and speed of the particle before and after resetting the memory it was possible to deduce how much energy was dissipated. The result was speed dependent, and by using ever longer switching cycles the Landauer bound could be approached.

## 7.1 Erasure in a single-electron device

The cost of erasure in a nanoscale system has been measured in a device in which the information is embodied in the presence ($n = 1$) or absence ($n = 0$) of a single electron in a quantum dot.[53] The device is illustrated in Figure 17, in which QD1 is the dot which may contain one additional electron, and QD2 forms a single electron transistor which senses the presence or absence of the additional electron. The source and drain regions of the device which are tunnel coupled to QD1 form two external reservoirs which in the presence of bias have different chemical potentials.

The single-particle energy of the quantum dot can be controlled by the plunger gate voltage $V_{G2}$. This is equivalent to varying $\mu$ if it is given the conventional definition of the energy required to add one additional electron. Initially $V_{G2}$ is adjusted to give average occupation $\langle n \rangle = ½$ in QD1, thereby giving a maximum difference between the uncertainty before erasure and the certainty following erasure. The instantaneous occupation is deduced from a random telegraph signal sensed by QD2, as illustrated in Fig. 17(c). These measurements are much faster than changes in $\mu$, so that the mean occupation of the dot can be monitored at each stage in Fig. 17(e). Electrons on the dot are tunnel coupled to source and drain electrodes which serve as baths. Erasure to 1 is achieved by lowering $\mu$ and erasure to 0 by raising $\mu$. The work cost of erasure is proportional to the area of the red or blue regions, for resetting to 1 or 0 respectively. The asymmetry in Fig. 17(d) arises from inevitable differences between the tunnel barriers either side of QD1 and also the degeneracy of the quantum dot state,[36] so that for non-zero bias the chemical potential for average ½ occupation is not midway between the chemical potentials of the source and the drain.

[52] Bérut, A., Arakelyan, A., Petrosyan, A. *et al*. Experimental verification of Landauer's principle linking information and thermodynamics. *Nature* **483**, 187–189 (2012).https://www.nature.com/articles/nature10872

[53] J. Dexter. DPhil thesis, University of Oxford (2025) https://ora.ox.ac.uk/objects/uuid:fcf4f19b-29e9-4778-a675-7830e7deebf4/files/d5712m746b. Figures sourced from the thesis by kind permission of Jon Dexter.

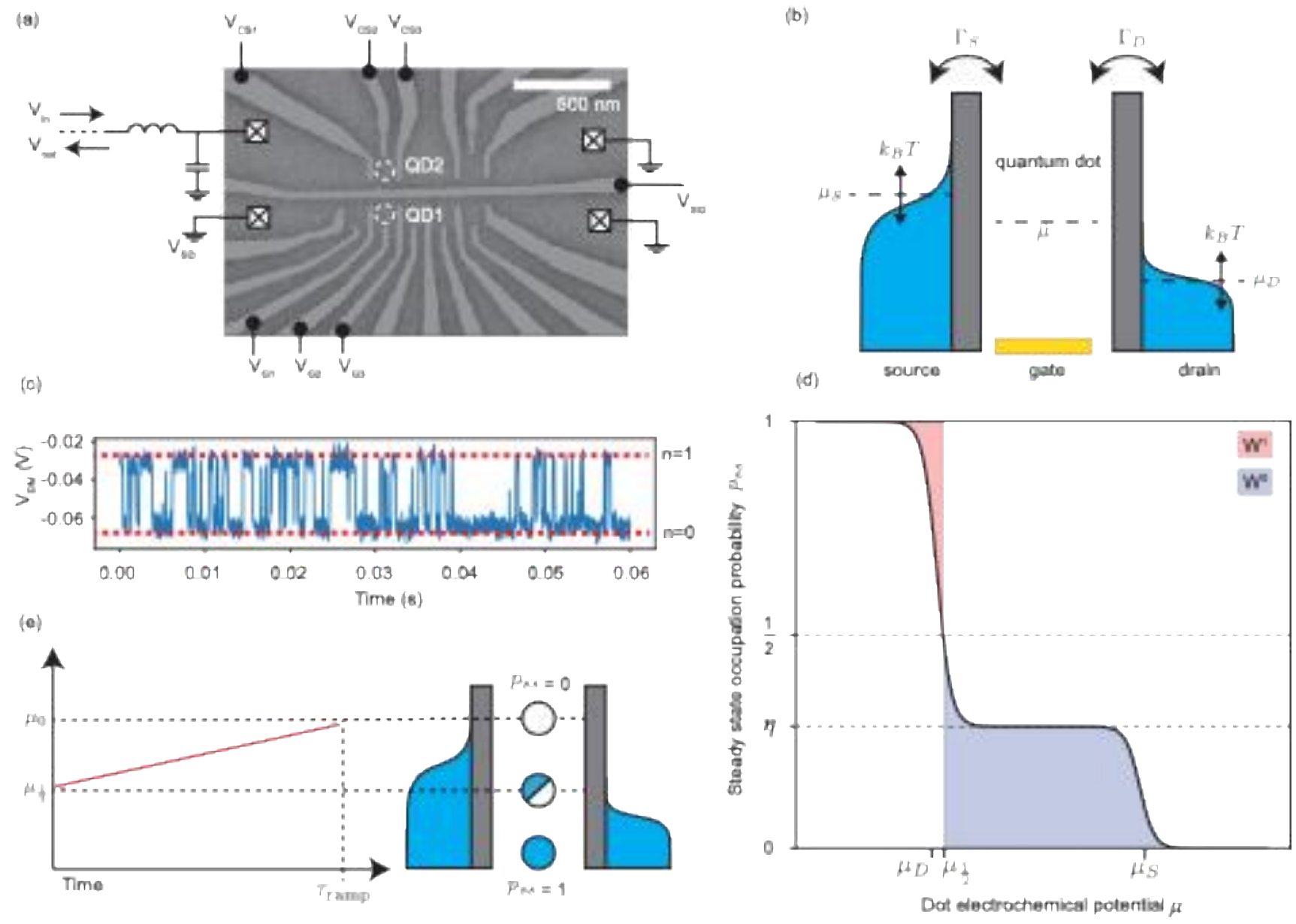

**Fig. 17** Erasure of information in a single-electron memory. **(a)** Scanning electron microscope image of the device. QD1 encodes the information erased in the experiment and is defined using voltages $V_{G1-G3}$. A potential difference can be applied across QD1 via the voltage $V_{SD}$ applied to an ohmic contact. QD2 forms a single electron transistor defined using gates $V_{CS1-CS3}$. Capacitive coupling between QD1 and QD2 is controlled using the splitter gate voltage $V_{SG}$. **(b)** QD1, with electrochemical potential $\mu$, exchanges charge with source and drain reservoirs via tunnelling events at characteristic rates $\Gamma_S$ and $\Gamma_D$. The charge occupations of the source and drain reservoirs are described by Fermi functions with electrochemical potentials $\mu_S$ and $\mu_D$ and temperature $T$. The gate electrode controls the electrochemical potential of the dot. **(c)** A representative time trace showing a measurement of the reflected charge sensing voltage $V_{out}$ following mixing with a reference signal. The demodulated voltage, $V_{DM}$, shows two distinct levels corresponding to the occupied (n = 1) and unoccupied (n= 0) charge occupation of the QD1. **(d)** Steady state occupation probability of QD1 as a function of electrochemical potential. The black curve is computed with $\Delta\mu_{S,D} = 30\ k_BT$ and $\Gamma_S/\Gamma_D = 6$ giving $\eta = 0.25$. The red and blue regions represent the work required to reset to one, $W_1$, and zero, $W_0$, respectively. **(e)** Protocol for erasure to zero. The dot's electrochemical potential $\mu$ is ramped linearly over a time $\tau_{ramp}$, starting from $\mu_{½}$ (where the steady-state occupation probability $p_{ss}(\mu) = 0.5$) and ending at $\mu_0$ (where $p_{ss}(\mu) = 0$).[53]

To determine the cost of erasure, the average charge on QD1 multiplied by the change in $\mu$ is integrated between $\mu_{1/2}$ and a value at which erasure is complete within the limit of experimental uncertainty. A histogram of results for zero bias is shown in Fig. 18(a). A crucial aspect of nanoscale thermodynamics is that fluctuations become significant, and the scatter in the results reflects a Gaussian Jarzynski-Crooks distribution. The mean of the bell-shaped curve coincides nicely with the Landauer bound, thereby providing further experimental evidence that this is indeed the minimum cost of erasure. However, all realistic electronic devices operate with a source-drain bias $\Delta\mu_{SD}$. As the bias is increased, so the cost of erasure increases, in ways that differ for erasure to 1 and to 0 because of the difference between the source and drain tunnelling rates $\Gamma_S$ and $\Gamma_D$. Figure 17(d) illustrates this in terms of the different areas shaded red and blue. In Fig. 18(b) erasure to 1 and to 0 are plotted separately. At low bias (shaded green) the work cost is temperature limited, as illustrated by the brown dashed line. At higher bias (shaded red) the work cost becomes bias limited, as shown by the green dashed line for the average of erasure to 1 and erasure to 0.

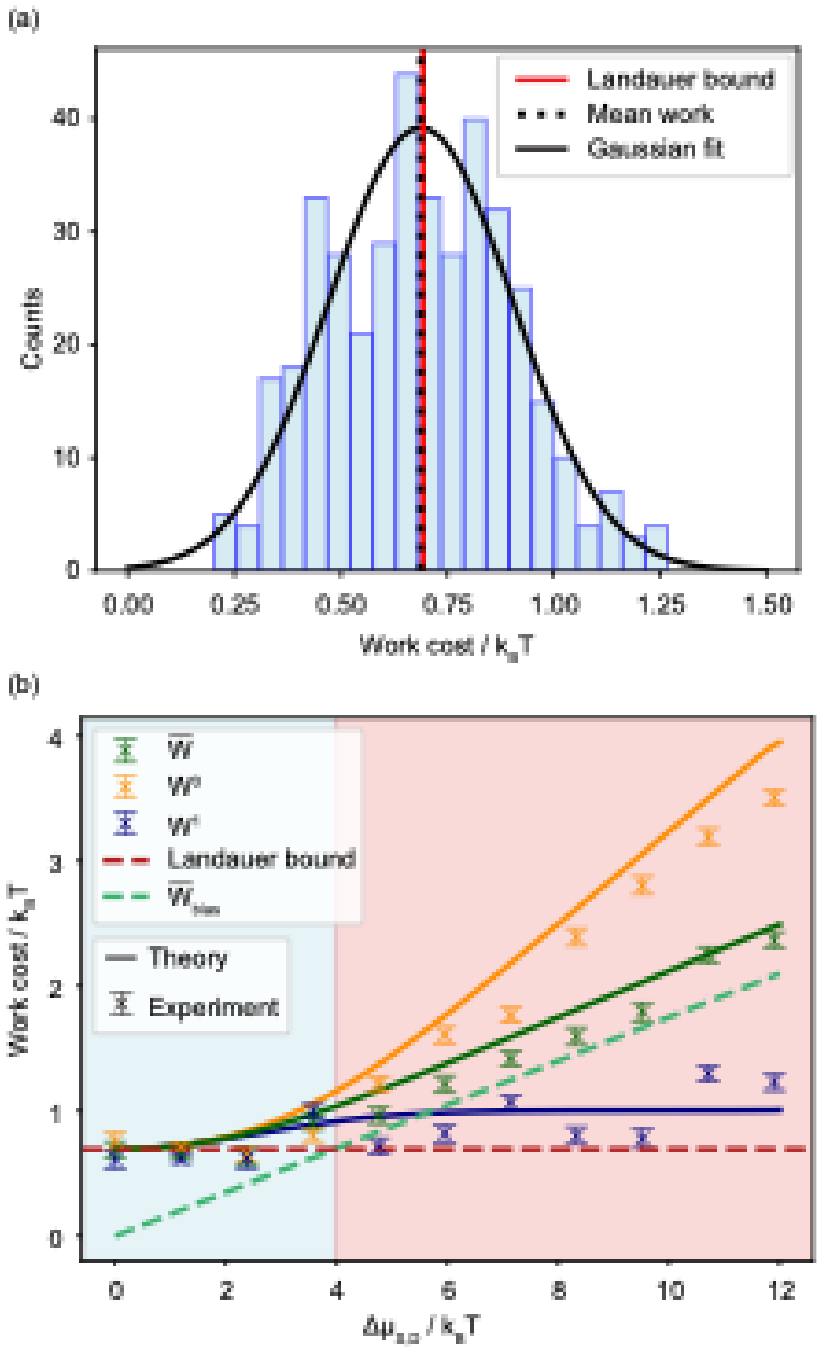

**Fig. 18** Measurements of Landauer erasure. **(a)** Histogram showing the work for 200 quasi-static ($\tau_{ramp}$ = 1 s with $\tau_{eq}$ = 0.1 s) realisations of the erasure to 1 and erasure to 0 protocols at zero bias ($\Delta\mu_{SD}$ = 0). The solid vertical line indicates the LB, while the dotted line represents the mean work. The black solid curve is a Gaussian scaled to match the mean and standard deviation of the measured work distribution. **(b)** Comparison of predicted and measured work costs for erasure to zero $W_0$, erasure to one $W_1$, and their average W for a quasistatic ramp with $\tau_{ramp}$ = 1 s. Crosses with error bars represent the measured work cost and solid lines theoretical predictions. The green and red dashed lines represent $W_{bias}$ and the LB, respectively. In the blue region $W_{bias}$ < LB and temperature dominates the erasure process. In the red region $W_{bias}$ > LB and erasure is bias dominated.[53]

For quasi-static erasure there can also be a *lifetime-broadening* term which is set by the total tunnel rate $\Gamma_{tot} = \Gamma_S + \Gamma_D$, giving a quantum uncertainty in the dot's energy level proportional to $\hbar\Gamma_{tot}$.[54] In devices with strong tunnel coupling, the lifetime-broadening can greatly exceed the other quasi-static contributions to the cost of erasure.

A further factor which can increase the cost of erasure is speed, as for many other thermodynamic processes such as engine cycles. If the ramp time $\tau_{ramp}$ is too short to allow the tunnelling process to maintain an equilibrium average charge as the chemical potential of the dot changes, then the plunger gate voltage $V_{G2}$ has to do more work on the charge in the dot than would be necessary if it could follow the change in potential. Figure 19 shows how the average effects of bias and speed appear to be simply additive:

$$\langle W \rangle \approx W_\infty + a/\tau_{ramp}, \quad (17)$$

where $W_\infty$ is the cost of quasi-static erasure for a given bias, and $a$ is a coefficient which depends only weakly on $\Delta\mu_{SD}/k_BT$, as confirmed by the Fig. 19 inset. Together these results show how the Landauer bound can be experimentally demonstrated, and also how in practice the cost of erasure will generally be greater than $k_BT \ln 2$.

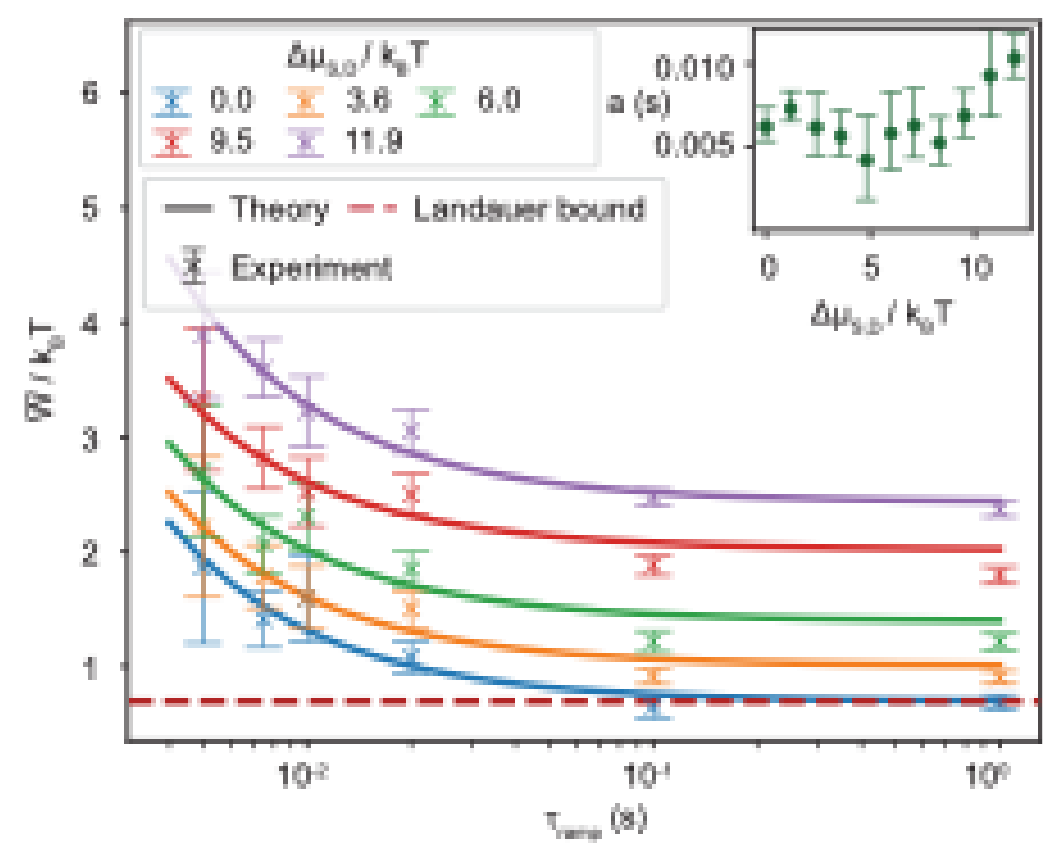

**Fig. 19** The average cost of erasure W plotted as a function of $\tau_{ramp}$ for a range of biases. Crosses with error bars represent the measured work cost obtained by averaging over 200 realisations of the erasure protocol. Solid curves represent a fit to Eq. (17) with $a$ as a fitting parameter. Inset shows that $a$ is only weakly dependent on $\Delta\mu_{SD}/k_BT$.[53]

[54] J. Dunlop, et al., Extra cost of erasure due to quantum lifetime broadening. *Phys. Rev. A* **112**, L010601 (2025). https://doi.org/10.1103/pc2t-ybtz

The experiments above thus embody the classical Landauer limit, and together they show how the cost of erasure scales with the bias and speed of the operation. But they leave open fundamental questions for quantum systems. What is a *bit* in a many-body quantum system, possibly described by an emerging quantum field? In such a system information is not necessarily stored in a single discrete state but may be distributed as mutual information across quantum correlations between subsystems — correlations that can be long-ranged and that cannot be localised to a single electron or reservoir. And if the system is isolated, there is no external thermal bath: the 'environment' is simply the unobserved remainder. The answer is a generalised Landauer principle in which the cost of erasure is $k_B$ *times the mutual information destroyed*, $\Delta I$(A:B) — the classical $k_B T \ln 2$ per bit being the special case where $I$(A:B) = ln 2.

## 7.2 Erasure in a many-body system with quantum correlations

Landauer's principle has evolved far beyond the original $k_B T \ln 2$ bound for erasing a single classical bit.[55] In the quantum many-body regime, the concept of erasure undergoes a fundamental shift. Unlike single-electron devices where heat is dumped into macroscopic leads, an isolated quantum field has no external heat bath. Instead, the "environment" is simply the part of the system we choose not to observe. Landauer's principle becomes a statement about the thermodynamic cost of severing the quantum correlations that bind subsystems together.

**Partitioning.** In a quantum system, *ignoring something has a cost*. When we choose to measure only a subset of variables—the relevant degrees of freedom—we effectively perform a partial trace over the irrelevant degrees of freedom, treating them as a bath.

$$\rho_{rel} = \mathrm{Tr}_{\mathrm{irr}}[\rho_{\mathrm{tot}}]. \tag{18}$$

We thus convert entanglement with the ignored sector into entropy. In simple macroscopic systems this procedure leads to entropy that scales with the volume of a system. In quantum systems, however, entanglement structure can scale as the boundary area or follow more intricate scaling behaviours, making the consequences of partitioning directly observable.[56] Such an area scaling was first encountered in black-hole thermodynamics, where the Bekenstein–Hawking entropy is proportional to the horizon area and arises from tracing over field modes inside the horizon.[57,58,59]

An experimental continuous quantum field simulator using ultracold atoms consisted of two parallel, one-dimensional Bose gases of $^{87}$Rb (Fig. 20).[60] Initially these two gases were tunnel-coupled, locking their relative phase — a state described in field-theory terms as having massive excitations — and thereby simulating the strongly correlated Sine-Gordon quantum field theory[61], or for strong coupling the simpler Gaussian Klein–Gordon quantum field theory.

[55] D. J. Bedingham and O. J. E. Maroney, The thermodynamic cost of quantum operations. *New J. Phys.* **18** 113050 (2016) https://iopscience.iop.org/article/10.1088/1367-2630/18/11/113050/pdf

[56] J. Eisert, M. Cramer and M. B. Plenio, Colloquium: Area laws for the entanglement entropy. *Rev. Mod. Phys.* **82**, 277–306 (2010). https://doi.org/10.1103/RevModPhys.82.277

[57] J. D. Bekenstein, Black holes and entropy. *Phys. Rev. D* **7**, 2333–2346 (1973). https://doi.org/10.1103/PhysRevD.7.2333

[58] S. W. Hawking, Particle creation by black holes. *Commun. Math. Phys.* **43**, 199–220 (1975). https://doi.org/10.1007/BF02345020

[59] R. M. Wald, The thermodynamics of black holes. *Living Rev. Relativ.* **4**, 6 (2001). https://doi.org/10.12942/lrr-2001-6

[60] S. Aimet *et al*., Experimentally probing Landauer's principle in the quantum many-body regime. *Nature Physics* **21**, 1326-1331 (2025) https://www.nature.com/articles/s41567-025-02930-9. Figures sourced from the publication and licensed under CC BY 4.0

[61] T. Schweigler, *et al*., Experimental characterization of a quantum many-body system via higher-order correlations, *Nature* **545**, 323 (2017)

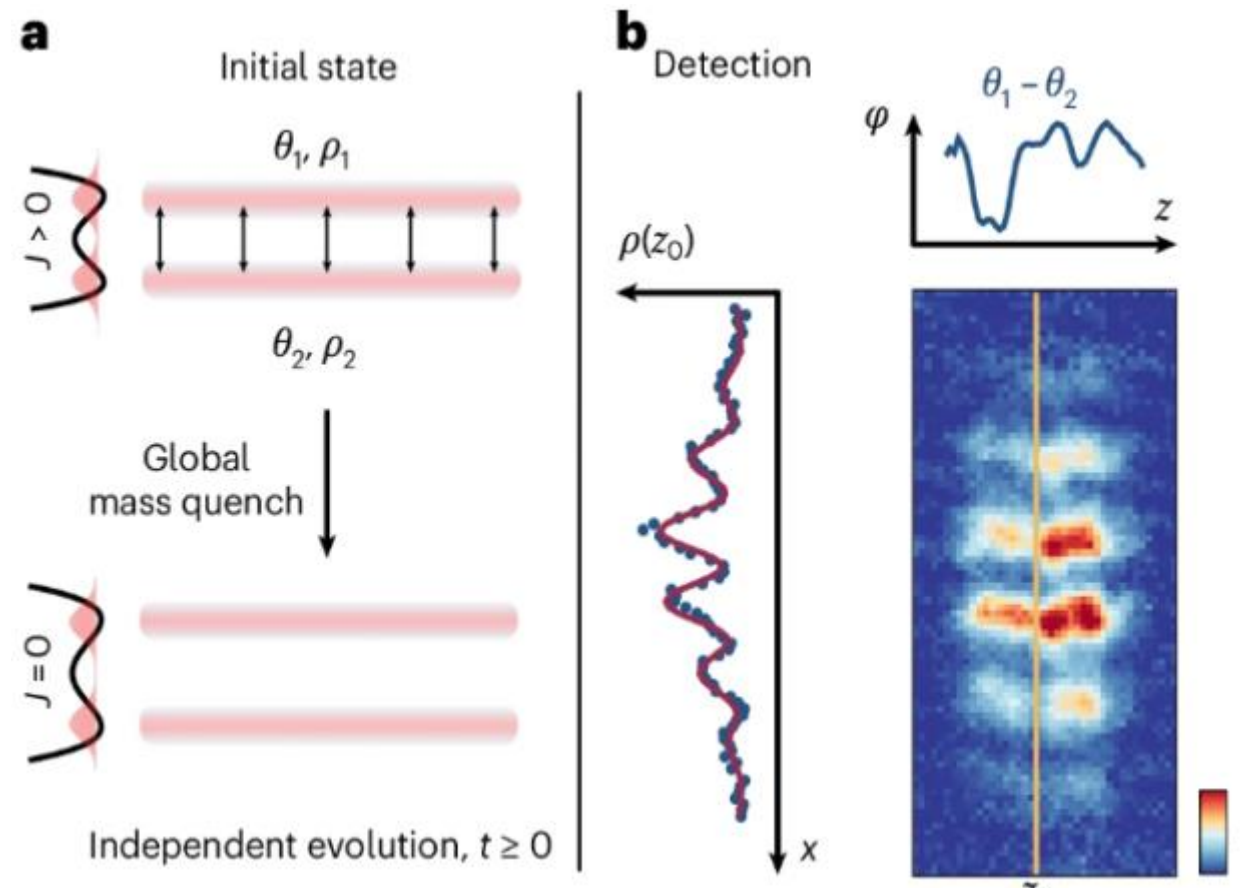

**Fig. 20** An ultracold Bose gas as a quantum field simulator. **a** The experimental system consists of two tunnelling-coupled ultracold $^{87}$Rb gases prepared in an initial global thermal state. Depending on the tunnel coupling $J$ one prepares a strongly correlated Sine-Gordon field or, for large $J$, a Klein–Gordon field. Ramping up a barrier, performs a global mass quench. For $J = 0$ the condensates evolve independently under the post-quench massless Klein–Gordon Hamiltonian. **b** to measure the quantum fields the atomic clouds are released, and they interfere as they expand. The observed interference pattern yields the relative phase profiles which provide a measurement of the phase quadrature of the quantum field.[60]

**The information to be erased: area-law correlations.** Before the quench, the system is in thermal equilibrium. What is the structure of the correlations that the erasure process will disrupt? An initial experiment on the same platform measured the mutual information $I(\mathrm{A{:}B}) = S_\mathrm{A} + S_\mathrm{B} - S_\mathrm{AB}$ between two spatial partitions of the 1D Klein-Gordon quantum field.[62] The experiment verified the area law of mutual information: unlike the thermal von Neumann entropy ($S_\mathrm{A}$, $S_\mathrm{B}$, $S_\mathrm{AB}$) which scales with volume, the mutual information between two subsystems is determined solely by their shared boundary. In equilibrium, quantum correlations are localised at the cut — the interface where we partition the field into observed and unobserved sectors.

The experiment extracted mutual information from the von Neumann entropy by full reconstruction of the correlation matrix — a procedure that relies on the underlying model being a non-interacting field theories such as the Klein–Gordon model. For strongly correlated systems, such as the Sine-Gordon model realised by the same platform at intermediate tunnel coupling, full many-body tomography becomes intractable. A recent experiment and analysis overcomes this limitation by extracting mutual information from strictly information-theoretic measures that do not require reconstruction of the density matrix or correlation matrix.[63] This extends the experimental verification of the area law to interacting quantum field theories with tuneable coupling strength, confirming that the area scaling of shared information — and thus the localisation of correlations at the partition boundary — persists beyond the Gaussian regime.

Such experiments illustrate that in finite quantum systems the area law is not strictly sharp: correlations penetrate a shell of thickness set by the correlation length (coherence length) on either side of the boundary.[61,62] This is a further aspect of the nanoscale regime — when the system size approaches the correlation length, the shell can no longer be neglected, and the information cost of partitioning becomes sensitive to the full geometry of the system rather than just the dimensionality of the cut.

**The erasure protocol: a mass quench.** The erasure was performed by rapidly raising the potential barrier between the two gases, thereby decoupling them. In field-theory language, this is a global mass quench: the mass of the relative-phase excitations is changed suddenly from finite (massive) to zero (massless). This jolts the system, launching pairs of entangled quasiparticles (phonons) that propagate in opposite directions at the speed of sound. The resulting *acoustic cones* of spreading correlations have been directly observed in earlier work on the same platform through the decay of long-range order and the emergence of exponential (thermal) correlations.[64]

**The generalised Landauer decomposition.** To analyse the thermodynamic cost, the 1D system of length $L$ is partitioned into a *system* of length $L_\mathrm{S}$ and an *environment* of length $L_\mathrm{E} = L - L_\mathrm{S}$.

[62] Tajik *et al*., Experimental verification of the area law of mutual information in a quantum field simulator. *Nature Physics* **19**, 1022–1026 (2023)

[63] M.T. Jarema, *et al*., Observation of area laws in an interacting quantum field simulator, arXiv:2510.13783

[64] T. Langen *et al*., Local emergence of thermal correlations in an isolated quantum many-body system. *Nature Physics* **9**, 640-643 (2013)

Crucially, this environment is not a thermal reservoir in the traditional sense; it is simply the rest of the quantum wire. As discussed in §1.4, in an isolated quantum system entropy arises from the choice of partition; tracing over the unobserved sector converts entanglement into entropy. The generalised Landauer principle for this scenario decomposes the total entropy production ΔΣ into distinct information-theoretic components:[65]

$$\Delta\Sigma = \beta_{\mathrm{E}}\Delta E_{\mathrm{E}} + \Delta S = \Delta I + \Delta D \qquad (19)$$

Where $\beta_{\mathrm{E}}\Delta E_{\mathrm{E}}$ is the energy dissipated into the environment, $\Delta S$ is the change in the system's von Neumann entropy, $\Delta I$ the change in mutual information between system and environment and $\Delta D$ the relative entropy of environment from a Gibbs state. The second decomposition reveals the two information-theoretic drivers of irreversibility:

1. ΔI (Loss of mutual information). This represents the breaking of quantum correlations. As the entangled quasiparticle pairs generated by the quench travel ballistically apart, they define a light cone of correlations. For an observer monitoring only the subsystem $L_{\mathrm{S}}$, information is lost at the precise moment one partner of an entangled pair exits the observation window and enters the environment $L_{\mathrm{E}}$. This geometric loss of correlation is the physical mechanism of forgetting in a quantum field.
2. $\Delta D$ (Relative entropy, or *athermal* dissipation). This term measures how far the environment is driven from a thermal Gibbs state. The quench does not merely heat the environment; it fills it with non-equilibrium excitations — specifically, squeezed states. $\Delta D$ quantifies the ordered energy dumped into the environment that has not yet relaxed into random thermal energy.

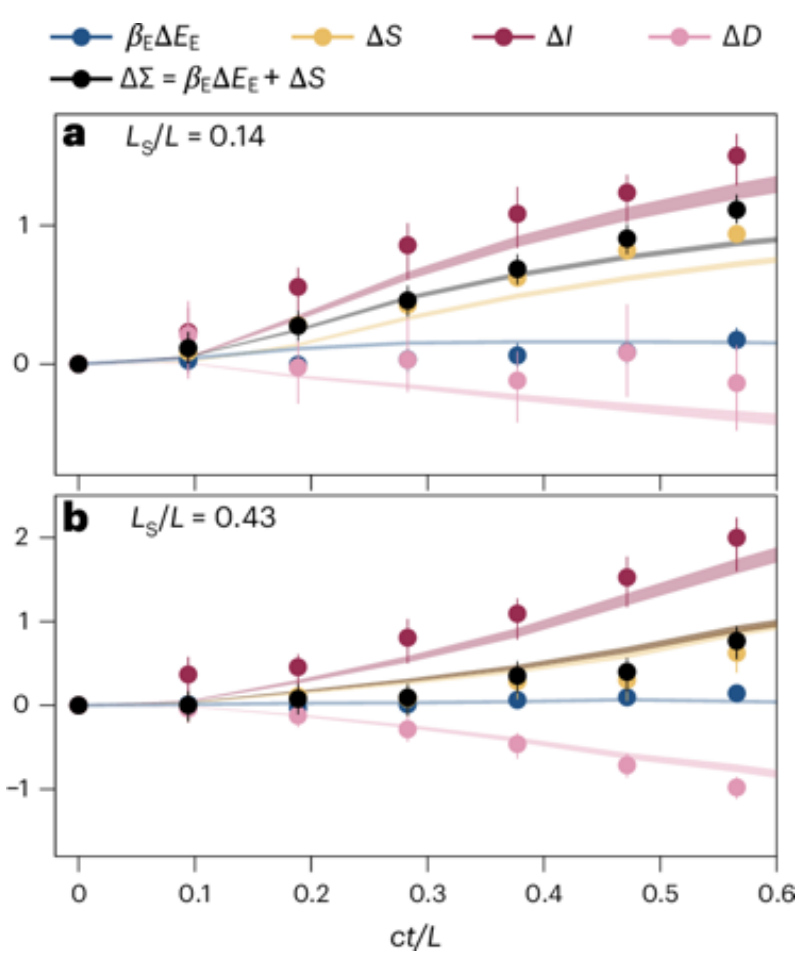

**Fig. 21** Quantities of Landauer's principle for the ultracold Bose gas in units of $k_{\mathrm{B}}$ as a function of time for subregion size ratios **(a)** $L_{\mathrm{S}}/L = 0.14$ and **(b)** $L_{\mathrm{S}}/L = 0.43$. The legend on the top applies to both panels. For each quantity, the experimental averages are represented by circles with error bars marking the 68% confidence intervals (equivalent to the standard error of the mean) obtained from bootstrapping with 999 samples. The shaded areas show quantum field theory simulation results using Neumann boundary conditions also with a 68% confidence interval from uncertainty in the estimated temperature and tunnelling rate and finite imaging resolution.[60]

The experimental results, shown in Fig. 21, reveal a striking departure from the classical case. The irreversible entropy production is not dominated by the simple dissipation of energy ($\beta_{\mathrm{E}}\Delta E_{\mathrm{E}}$), which remains small. Instead, it is dominated by the change in mutual information $\Delta I$.

This reveals that Landauer's principle operates at a deeper level than its original formulation suggested. In its classical form — as demonstrated in §7.1 — Landauer's principle equates the erasure of one bit stored in a physical register with a minimum entropy cost $k_{\mathrm{B}} \ln 2$ in an external thermal bath. The *bit* is discrete, the *bath* is macroscopic, and the *heat* generated is physical. Here, none of those features are present. There is no binary register, no macroscopic bath, and no measurable heat release. Yet the thermodynamic cost of irreversibility is unambiguously real and quantified. What has changed is not the principle but the identity of what it governs.

The information erased is not a binary register state but the subsystem $L_{\mathrm{S}}$ – unobserved environment $L_{\mathrm{E}}$ *mutual information* made inaccessible by the partition — the very quantity measured by the area-law experiments on the same platform. That result shows that, in equilibrium, quantum correlations are localised precisely at the partition boundary. The quench then acts as an erasure protocol: it launches entangled phonon pairs whose partners travel in opposite directions at the speed of sound. At the moment one partner of a pair crosses the partition boundary,

[65] T. Sagawa and M. Ueda, Minimal energy cost for thermodynamic information processing: measurement and information erasure. *Phys. Rev. Lett.* **102**, 250602 (2009)

the correlations it carried about its counterpart in $L_S$ become inaccessible to the chosen local reduced description of any local observer, even though the global quantum state of the entire field continues to evolve unitarily and no information is destroyed globally. The *bath* is not a macroscopic thermal reservoir — it is the unobserved portion of the same quantum wire. Tracing over $L_E$ converts entanglement into entropy, and it is this conversion that constitutes the erasure act in the Landauer sense.

The generalised Landauer decomposition (Eq. 19) makes this quantitative. In Fig. 21, the total irreversible entropy production $\Delta\Sigma$ is dominated by $\Delta I$, with a minimum thermodynamic cost of $k_B \ln 2$ per bit of mutual information change, not by the athermal term $\Delta D$ or by the energy dissipated into the environment. The Landauer cost of the quench is paid primarily in mutual information destroyed, not in heat generated. This is a direct experimental demonstration that, in the quantum many-body regime, $\Delta I$ is not merely a diagnostic of the process; in this experiment it is the dominant contribution to the generalized irreversible entropy production. Thus this extends the single-bit classical Landauer bound to a continuous quantum field.

Finally, there is a direct connection back to the trajectory-level fluctuation theorem (Eq. 3). The entropy production $\Delta S_{tot}[\Gamma] = k_B \ln P[\Gamma]/P[\tilde{\Gamma}]$ that biases the arrow of time is here dominated by $\Delta I$. It is therefore the irreversible spreading of entanglement — correlations leaking across the partition boundary — that makes forward trajectories exponentially more probable than reverse ones. The reverse process, in which phonon pairs reconverge and the original correlations rebuild, is not forbidden by energy conservation (the system is isolated and the global state evolves unitarily and its total von Neumann entropy is conserved), but it is exponentially suppressed by this informational bias. Disentangling quasiparticles is thermodynamically irreversible in precisely the same sense as resetting a bit register; both are processes in which information accessible to a local observer is permanently lost, and both carry a minimum entropic cost proportional to the information destroyed. The only difference is scale — from one bit to a quantum field.

## 8 There is plenty of room at the margins

Having explored the fundamental thermodynamic limits at the nanoscale, we now take a step back from specific experiments and turn to practical prospects. How do current technologies compare to these ultimate bounds?

Four days after Christmas in 1959, Richard Feynman gave a lecture at the Annual American Physical Society annual meeting at Caltech, entitled *There's Plenty of Room at the Bottom: An Invitation to Enter a New Field of Physics*.[66] Feynman articulated a vision for science and technology on very small scales, including data storage and electron microscopes with lenses that need not be stationary or axially symmetric. In the ensuing disciplines of nanoscience and nanotechnology, there is still plenty of room at the margins.

There was not much concern about global climate change in the nineteenth century. The energy issue then was about to get maximum power out of a steam engine. That led Sadi Carnot and his successors to consider how efficient an engine could be. The resulting thermodynamics turned out to have implications for science that extended far beyond steam engineering.

In the current century we are deeply concerned about energy efficiency for other reasons, because carbon emissions are having an impact on our planet that looks all set to escalate, causing changes that already affect human lives and may prove to be hard or in some cases impossible to reverse. What can nanoscale thermodynamics contribute?

Nanoscale electronic devices are powered by electricity. Global electricity demand in 2024 exceeded 30,000 TWh/year, with an annual increase of 4%.[67] Within that total, the electricity share of the information and communications technology (ICT) sector is hard to determine exactly, not least because it is increasing quite rapidly. In 2020 the ICT sector used about 4% of the global electricity in the use stage and represented about 1.4% of the global greenhouse gas emissions.[68] An estimate for ICT in 2024 was nearly 1,100 TWh/year.[69] Of that, between a quarter and a third is attributable to data centres, which are the fastest growing component of ICT. Whatever the changing numbers, gains in efficiency are more than offset by growth in use. For comparison, aviation contributes about 2.5% of the world's carbon emissions.[70]. Aero engines have progressed ever closer to the limit of thermodynamic efficiency permitted by the highest temperatures that the turbine blades can withstand. In 1990, the energy used per passenger-kilometre was 2.9 MJ, by 2019 this had more than halved to 1.3 MJ. ICT, using over 3.5% of the world's electricity, is far from the thermodynamic limit of efficiency. There is plenty of room at the margins.

---

[66] R. P. Feynman, There's plenty of room at the bottom. *Engineering and Science* 23, 22-36 (1960) https://calteches.library.caltech.edu/47/2/1960Bottom.pdf

[67] https://ember-energy.org/app/uploads/2025/04/Report-Global-Electricity-Review-2025.pdf

## 8.1 Clocks

How far are physical clocks from the thermodynamic limits established in §3? The answer depends on the class of clock. *Dissipative clockworks* — those whose ticks are intrinsically stochastic events — are directly constrained by the bounds (9) and (12), and the comparison is between the entropy they actually produce and the entropy they must produce to reach a stated accuracy. *Reference clocks* — atomic and optical lattice clocks, which slave a local oscillator to a natural quantum transition — are not constrained by (9) or (12), but rather by observation and reset.

Both classes nonetheless share the same structural decomposition. Operating clocks can be segmented into up to four nested shells, each a distinct layer at which entropy is actually produced: the clockwork itself, its quantum-to-classical readout, the supporting frequency-reference apparatus, and the biological natural oscillators that occupy a complementary regime. What differs between the two classes is *where the bounds apply*. In dissipative clockworks the bounds bite at Shell I — and, as we shall see, that is also where real devices are closest to saturating them. In reference clocks Shell I is essentially decoupled from the entropy budget, and the thermodynamic cost of timekeeping is paid almost entirely in Shells II–III.

The shells are not arbitrary: Shell I is where the maintenance cost resides, Shells II–III are where the readout cost and its supporting infrastructure reside, and the class distinction of §3.4 is precisely whether these costs are inextricably connected (dissipative) or separable (reference). Atomic clocks are a technology in which the separation is clean enough to be quantified shell by shell. Biological natural oscillators provide a humbling benchmark of achievable thermodynamic efficiency.

The clock coherence $N_{cc} \equiv (t_{tick}/\Delta t_{tick})^2$ introduced in §3 is an intrinsic property of the clockwork: it carries no observation time and the thermodynamic precision bound (9) constrains accuracy in terms of the entropy per tick for a dissipative clockwork. Practical clocks are usually characterised by the *Allan deviation* $\sigma_y(\tau)$, the standard time-domain measure of fractional frequency stability (square root of the Allan variance) at averaging time τ. For a given $N_{cc}$ under white frequency noise the Allan deviation is

$$\sigma_y(\tau) = \frac{1}{\sqrt{n \cdot N_{cc}}}, \qquad n \equiv \tau/t_{\text{tick}}, \tag{20}$$

The Allan deviation $\sigma_y(\tau)$ thus depends on both $N_{cc}$ and the length of the averaging chain $n \equiv \tau/t_{\text{tick}}$. For a fixed $N_{cc}$,, the Allan deviation varies as $\sqrt{t_{\text{tick}}/\tau}$. So for a *dissipative* clock a given $\sigma_y(\tau)$ depends jointly on the clockwork's intrinsic coherence and the length of the averaging chain. Combining this relation with the per-tick bound (9) and accumulating over n ticks gives a clean operational form,

$$\Delta S_{\text{total}}(\tau) \geq \frac{k_B}{2\pi^2\, \sigma_y^2(\tau)} \tag{21}$$

for the minimum entropy that must be produced over time $\tau$ to reach Allan deviation $\sigma_y(\tau)$. The dependence on the tick rate drops out: a given stability target costs a fixed minimum entropy regardless of how the clockwork is partitioned into ticks. Coherent quantum clockworks (12) and *reference* clocks based on intrinsic oscillators, like an atomic clock, allow dramatically smaller entropy budgets.

---

[68] J. Malmodin, N. Lövehagen, P. Bergmark, D. Lundén. ICT sector electricity consumption and greenhouse gas emissions – 2020 outcome. *Telecommunications Policy* **48**, 102701 (2024) https://www.sciencedirect.com/science/article/pii/S0308596123002124

[69] https://www.ericsson.com/en/reports-and-papers/mobility-report/dataforecasts/ict-carbon-footprint-decreasing?utm_source=chatgpt.com

[70] https://ourworldindata.org/global-aviation-emissions

The innermost *first shell* is the irreversible process whose periodic cycles constitute the ticks themselves, and is the layer which the bounds (9) and (12) directly address. The noise-driven membrane experiment (§3.1) confirmed the linear scaling of accuracy with entropy production within a numerical factor of the bound,[21] and the transmon clock (§3.2) verified the corresponding quantum kinetic-uncertainty bound.[23] Equation (12) shows that the linear bound (9) is itself not the ultimate limit: in coherent quantum architectures with dissipation confined to a single channel, accuracy can grow exponentially in the entropy budget per tick. Within Shell I, then, actual clocks are not far from the bound that constrains them, and that bound is itself more favourable than once thought.

The *second shell* is the quantum-to-classical readout, addressed by the double-quantum-dot experiment of §3.3. A clockwork, that ticks unobservably, produces no usable timing information; registering a tick as a macroscopically distinguishable event is itself an irreversible process, with its own entropy budget. The empirical separation reported there — roughly nine orders of magnitude between the entropy of the readout circuit and that of the clockwork — establishes that the dominant per-tick irreversibility lies in the quantum-to-classical transition rather than in the quantum mechanism itself. This shell was largely absent from the precision-bound literature that produced the bounds (9) and (12), and is where the gap to the limit first becomes large.

The *third shell* is the supporting frequency-reference apparatus, which is what laboratory frequency standards actually consist of once the clockwork and its readout are placed in context. State-of-the-art optical lattice clocks based on $^{87}$Sr or $^{171}$Yb routinely achieve Allan deviation $\sigma_y(\tau) \lesssim 10^{-18}$ at integration times of order $10^3$ s.[71] The full apparatus comprises laser cooling and trapping stages, an ultra-low-expansion reference cavity, vacuum systems, and active blackbody-radiation shielding, with continuous power dissipation of order $10^3$ W; over a 1 s integration window at room temperature this corresponds to $>10^{23}$ $k_B$ of entropy produced.

Applying the bound (21) at $\sigma_y = 10^{-18}$ would require at least $k_B/(2\pi^2 \times 10^{-36}) \approx 5\times10^{34}$ $k_B$ per second of integration — eleven orders of magnitude *above* the apparatus dissipation. The apparent inconsistency is the diagnosis, not a paradox: the linear bound (9) does not apply to *reference* clocks. Nevertheless, the apparatus still dissipates about ten orders of magnitude above the thermodynamic requirement but locating that gap unambiguously in the supporting infrastructure rather than in the clockwork itself. As with the very different quantum dot clock in §3.3, *the dominant entropic burden of modern frequency metrology is not in the physics of timekeeping but in the engineering that supports it*.

The *fourth shell* is much less profligate. The cyanobacterial KaiABC circadian oscillator and related biochemical clocks operate on ATP-hydrolysis budgets of order $k_BT$ per cycle per molecule, with periods on the scale of $10^5$ s and modest accuracy $N \sim 10^2 - 10^3$. Such clocks underperform their thermodynamic uncertainty bound (9) — $\Delta S_{tick} \geq k_B N_{cc}/(2\pi^2)$ — by several orders of magnitude,[72] attributable to the small number of internal molecular states combined with the high free-energy driving needed to sustain oscillation in living cells. Yet in absolute units of dissipation per tick, they are dramatically *thriftier* than any engineered timekeeper. A typical quartz wristwatch (32 kHz, ~1 μW) or the temperature-compensated crystal oscillator in a smartphone (~25 MHz, ~1 mW) each dissipate of order $10^{10}$ $k_B$ per tick; a laboratory optical lattice clock dissipates ~$10^{23}$ $k_B$ per averaging unit. The KaiABC oscillator and similar molecular timekeepers operate within a few $k_B$ per tick — ten orders of magnitude below consumer quartz, more than twenty below a laboratory frequency standard. Biological clocks are inefficient relative to their fundamental limit but absolutely parsimonious — a useful reminder that the figure of merit (efficiency relative to the bound versus absolute entropic cost) depends on the question being asked.

Where, then, is the biggest gap? Shell I gaps are small and may be closeable via Eq. (12). Shell II contributes nine orders of magnitude in the only system in which it has been directly resolved. Shell III contributes a further ten to twenty orders of magnitude in laboratory standards. Shell IV reverses the figure of merit. The headline lesson for nanoscale thermodynamics is that the bound (9) constrains the layer of the clock that is, in practice, the easiest to make efficient. The physics of *recording* time — making microscopic ticks legible to a macroscopic observer — is where the second law most heavily taxes the project of timekeeping. There is plenty of room at the margins, and most of it is in the readout and the apparatus rather than in the oscillator.

[71] T. Bothwell, et al., "Resolving the gravitational redshift across a millimetre-scale atomic sample," *Nature* **602**, 420–424 (2022). DOI: 10.1038/s41586-021-04349-7.

[72] R. Marsland III, W. Cui, and J. M. Horowitz, The thermodynamic uncertainty relation in biochemical oscillations, *J. R. Soc. Interface* **16**, 20190098 (2019). DOI: 10.1098/rsif.2019.0098.

## 8.2 CMOS

Complementary metal-oxide-semiconductor (CMOS) technology is ubiquitous in integrated circuits. How close are CMOS chips to the thermodynamic limit? Estimates vary, but it may be helpful to consider a worked example. Current CPUs such as ARM chips are fairly well optimised for floating point operations (FLOPs), and they use something in the range 8 – 15 pJ/FLOP.[73] Optimised GPUs do better, at about 3 – 8 pJ/FLOP. Depending on the particular arithmetic operation, each FLOP may involve about a thousand Boolean operations (BOs) such as NAND, XOR, etc. Thus current CMOS circuits are running at about 10 fJ/BO. Isolated transistors can have switching energies measured in attojoules,[74] but the total energy cost of a Boolean operation is higher.

A Boolean operation which involves two inputs and one output effectively involves the loss of one bit of information. This is therefore equivalent to a Landauer erasure with a minimum entropy cost of $k_B \ln 2$ and a minimum energy cost of $k_B T \ln 2$. At room temperature this amounts to 3 zJ ($3 \times 10^{-21}$ J). Current CMOS is dissipating about 3 million times that amount per Boolean operation,[75] which gives great thermodynamic scope for energy savings. If the gas turbine engines of an airliner were that far from their thermodynamic limit, it would get a few metres along the runway before running out of fuel. Even allowing for the increasing cost of Landauer erasure with realistic switching speeds and voltages, there is plenty of room at the margins.

Why has increasing the efficiency of silicon chips proved so elusive, where there is more than enough motivation to reduce their energy consumption? Field effect transistors operate by raising and lowering a barrier to the flow of electrons or holes between the source and the drain. At low enough temperatures this would require only tiny changes in the gate voltage to switch the transistor on or off. But electrons are subject to Fermi-Dirac statistics, and there is an exponential tail in their energy distribution. If the barrier voltage changes by $V$, then the current changes in the ratio $\exp(-qV/k_B T)$. The only relevant quantities apart from $V$ are the Boltzmann constant, the temperature, and the electron charge $q$. At room temperature, $V$ must change by 60 mV to give a factor 10 change in current. An ideal field effect transistor would thus have a so-called subthreshold swing of 60 mV/decade, and the best devices are close to this. No choice of materials or improvements to the design can beat this limit (though they help to approach it). The choice of on/off ratio is an engineering decision which determines the minimum gate voltage range required, and in practice it is not common to operate MOSFET circuits below 500 mV. There can be many sources of energy dissipation, and however much the others may be reduced the energy cost associated with charging and discharging the inevitable capacitance in any circuit sets a practical limit of about 10 aJ for switching a transistor in a computer.

The human brain does much better, running at about 20 W continuously. This is a relatively high rate of metabolic energy consumption; it is about 20% of the body's metabolic requirement for only 2% of the body's mass. But insofar as it is possible to make a comparison, it is orders of magnitude more energy efficient than a silicon computer for most of the tasks undertaken by the brain. It has been estimated that a human brain performs perhaps from $10^{15}$ to $10^{18}$ information processing events per second, and an average cost therefore in the range $2 \times 10^{-17}$ J – $2 \times 10^{-14}$ J,[76] The same analysis observes that the data centres of the USA consume ten times more energy than the inhabitants consume in food, and yet one individual human being has more information processing power, suitably measured, than all the data centres in the world. As with clocks, so with neural systems: biology can achieve thermodynamic efficiency that puts human technology to shame.

---

[73] Suárez, D., Almeida, F. & Blanco, V. Comprehensive analysis of energy efficiency and performance of ARM and RISC-V SoCs. *J Supercomput* **80**, 12771–12789 (2024). https://doi.org/10.1007/s11227-024-05946-9

[74] S. Datta, W. Chakraborty, and M. Radosavljevic, Towards attojoule switching in logic transistors. *Science*. 378, 733-740 (2022)

[75] Semiconductor Industry Association. *International Technology Roadmap for Semiconductors 2.0*, https://www.itrs2.org/itrs-reports.html (2015)

[76] Seth Lloyd and Michele Reilly. Natural Intelligence: the information processing power of life. https://arxiv.org/pdf/2506.16478

If the energy efficiency of MOSFETs is limited by the Fermi-Dirac exponential tail, could transistors be developed that work on a different principle? Yes, by using quantum confinement and quantum interference. Resonant tunnelling transistors are used for niche applications, mostly at microwave and terahertz frequencies. Quantum interference effects have been demonstrated in single-molecule transistors.[77] The manufacturability of silicon chips presents a very high bar to any disruptive technology. But there is plenty of thermodynamic room at the margins for innovative hardware for digital computing.

## 8.3 Quantum computing

There are at least five, possibly more, candidate platforms for building quantum computers. At the time of writing it seems still to be too early to say which, if any, will eventually dominate. This is not because any of them faces insuperable challenges, but rather the opposite—we are currently enjoying a steady stream of significant advances and breakthroughs in each candidate implementation.[78] Inevitably each manufacturer is adamant that their approach will win out. As quantum computers become increasingly scalable, they will enable predictions to be tested of the number of logical qubits that can be used in quantum information processing before reaching a fundamental limit, whether because of the finite information capacity of the Universe,[79] or because of a rational discretisation of Hilbert space.[80] Although these two approaches originate on utterly different grounds and are each speculative in their own way, they suggest similar computing limits of about 400 logical qubits. Before too long this may become amenable to experimental investigation.

Scalable quantum computers require error correction. They are then not a closed system any more due to the measurements required to extract the error syndrome. A further challenge in assessing the thermodynamic efficiency of quantum computers lies in the ill-definition of work and heat in a closed quantum system. One account lists a choice of four, with more to come.[81] Perhaps in part because of this there are conflicting accounts of whether an erasure in quantum computation generates more or less entropy than its binary equivalent. Since the evolution of the Schrödinger equation is time-reversible, after error correction only the final readout might necessarily generate entropy, but the mutual information in entanglement also needs to be taken into account. There are several factors contributing to the further entropic cost of practical quantum computing.

(i) Some methods of quantum computing with photons involve creating entanglement by a process of repeat until success. This will carry an entropic cost.

(ii) Some methods of quantum computing involve consuming entanglement by a series of measurements with feed-forward. This will carry an entropic cost.

(iii) All scalable methods of quantum computing require error correction, which in turn involves a series of measurements. This will carry an entropic cost of resetting memories after making the measurements (Landauer again), which is likely to be very high.

(iv) Most implementations of quantum computing require a controlled environment which may require a vacuum or cryogenic engineering or both, together with ancillary control and instrumentation. The energy cost of these may vastly exceed any energy necessarily involved in the actual quantum process.

To all this must be added the cost of ancillary classical computing. Error correction requires not only the measurement but also large amounts of classical computing to analyse the error syndrome and extract the correction feedback. To decode the surface code in real-time for a large-scale system, industry estimates suggest we need roughly 1 petaflop of classical compute power dedicated *solely* to error correction (10 kW power if using the numbers from CMOS given above). This highlights an irony: The "Zero Dissipation" Quantum Computer requires 10 kW dissipation of classical heat just to interpret its errors. And this 10 kW must be dissipated very close to the fridge (to minimize latency), fighting the cooling system. There is an understandable drive to put as much as possible of the CMOS circuitry within the cryostat, thereby minimising the otherwise excessive number of lines to the cold stage, but the resulting reduced heat load then needs to be pumped away at a cost determined by the ratio of ambient temperature to cryostat temperature.

---

[77] Z. Chen *et al*. Quantum interference enhances the performance of single-molecule transistors. *Nature Nanotechnology* **19**, 986–992 (2024).

[78] D. Castelvecchi, The stunning advances in quantum computing. *Nature* **650**, 24-26 (2026). https://www.nature.com/articles/d41586-026-00312-6.pdf

[79] P. C. W. Davies, The implications of a cosmological information bound for complexity, quantum information and the nature of physical law. *Fluctuation and Noise Letters* **7**, C37-C350 (2007)

[80] T. Palmer, Rational quantum mechanics: Testing quantum theory with quantum computers, *Proc. Natl. Acad. Sci. U.S.A.* 123 (12) e2523350123, https://doi.org/10.1073/pnas.2523350123 (2026)

[81] N. Yunger Halpern, *Quantum Steampunk* (Johns Hopkins University Press 2022)

As advances are made these requirements may be ameliorated. Some functions of the current classical control systems may eventually be performed by quantum machines.[82] Because quantum information processing is fundamentally different from classical processing, methods may become available whereby quantum computers can perform calculations much more efficiently than classical computers. As that comes about, there will be plenty of room at the margins.

## 8.4 Learning machines

Leaving aside the data centres required to support them, the most significant growth in information and communication technology (ICT) is to be found in learning machines. Before ChatGPT, no consumer technology had previously acquired over 100 million users within two months of its release. It is estimated that each LLM query uses about 0.34 Wh of electricity, or over 1,000 J for comparison with the quantities above for CMOS.[83] For a single query this is a small energy cost—not nearly enough to charge your phone—but with the vast number of queries a day handled by LLMs world-wide it all adds up, on top of the energy cost of training in the first place. A recent estimate is 2.5 billion queries per day for ChatGPT,[84] which already amounts to over 300 GWh per year for one provider alone. How much room is there at the margins?

In most electronic instrumentation and transmission noise is a nuisance. In 1948 Claude Shannon published a paper which he entitled "A Mathematical Theory of Communication". The following year he redesignated it "The Mathematical Theory of Communication" in his book of that title. He showed how the presence of noise provides a fundamental limit to the capacity of an information channel. The less noise the better.

Not so in a stochastic learning machine. If, *per impossibile*, you could entirely eliminate noise, the machine would learn at zero rate. Rather as a Carnot engine can generate no entropy only if it works infinitely slowly, thus producing no power, so a learning machine can generate no entropy only if it learns infinitely slowly, thus producing no knowledge (if knowledge can be attributed to a machine). Randomness is necessary for learning to occur. Just like humans, if the machine never makes a mistake, it never learns. This has been analysed with mathematical rigour, at least in theory and in simulations.[85] We may confidently expect that the predictions will soon be tested experimentally. The noise required for learning represents a thermodynamic cost distinct from simple erasure, but as with erasure we may expect the cost to increase with the speed of the operations and the voltage of practical devices. For a chosen rate of learning and required accuracy, there will be an optimal level of noise. This can change dynamically as learning proceeds, because as the learned model converges on the training model, so the level of noise can be reduced.

Progress is being made in developing quantum learning algorithms, though they will only be useful when large enough quantum computers are built.[86] It may then be that quantum learning machines will offer a thermodynamic advantage over classical learning machines. Here too there is plenty of room at the margins.

---

[82] José Antonio Marín Guzmán *et al.* Key issues review: useful autonomous quantum machines. *Rep. Prog. Phys.* **87** 122001 (2024)

[83] F. Oviedo *et al*. Energy use of AI inference: efficiency pathways and test-time compute (2025) https://arxiv.org/pdf/2509.20241

[84] ChatGPT Statistics 2025: Surprising Global Insights (22 November 2025) https://xtendedview.com/chatgpt-statistics/?utm_source=chatgpt.com

[85] G. J. Milburn and S. Basiri-Esfahani, The physics of learning machines. *Contemp. Phys.* **63**, 34-60 (2022)

[86] G. Acampora *et al.* Quantum computing and artificial intelligence: status and perspectives. https://arxiv.org/pdf/2505.23860

## 9 Summary and outlook: from ignorance to engineered irreversibility

Thermodynamics was born as the science of systems with an enormous number of microscopic constituents (Avogadro scale) by systematically ignoring microscopic degrees of freedom and retaining only a handful of emergent macroscopic variables such as volume, pressure and temperature. At the nanoscale this acquires a fresh poignancy: because experiments can access and control some microscopic variables, one must state explicitly which degrees of freedom are tracked and which are coarse-grained away. What is tracked defines the effective "system". Untracked degrees of freedom—including correlations with the system—constitute the "environment". Where the boundary lies is constrained by the available resources for measurement, control and modelling.

In our overview we illustrate these by a striking variety of experiments: nanomechanical and superconducting clocks in which accuracy is paid for in entropy per tick; single-electron devices where the entropy change of a charge transition can be measured directly; Szilárd engines and Maxwell demons that use information as a resource; levitated particles that test fluctuation theorems at the level of single trajectories; and, most recently, Landauer erasure in a continuous quantum field, where its consequence of disentangling quasiparticles is revealed as a fundamentally irreversible process. In all these cases, the Second Law survives intact, but its content is refined: irreversibility is not simply "heat over temperature", it is the loss of accessible information and correlations.

At the same time, nanoscale thermodynamics exposes how far our technologies sit from their ultimate limits. Clocks, whether mechanical, nanomechanical, or quantum, consume far more entropy in their read-out chains than is required by the fundamental cost per tick. Classical CMOS logic dissipates roughly six orders of magnitude more energy per Boolean operation than the Landauer bound, because transistors are constrained by the Fermi–Dirac tail rather than by thermodynamic necessity. Quantum computers face their own entropic overheads in state preparation, measurement, and error correction, while the energy cost of cryogenic and control infrastructure dwarfs the microscopic work done by the qubits themselves. Learning machines, classical and quantum, add a further layer in which randomness—once an enemy of reliable communication—becomes a necessary ingredient for training, with a corresponding thermodynamic signature. Taken together, these examples show that there is still plenty of room at the margins; the physical limits set by nanoscale thermodynamics lie far below today's practical implementations.

Looking ahead, we expect nanoscale thermodynamics to move from being a diagnostic tool to a design principle. At the fundamental level, experiments that measure quantities like generalized Landauer costs, mutual information, relative entropy and the relation between entanglement entropy and information in many-body systems will deepen our understanding of equilibration, thermalisation, decoherence and emergent quantities like the arrow of time, or new physical descriptions. At the applied level, the same concepts can guide the co-design of devices and architectures—clocks, memories, logic gates, quantum processors and learning machines.

Our central message when going from Avogadro scale to small and quantum is this: As we learn to engineer ignorance—what we do not measure, do not control, and deliberately trace out—we also learn to engineer irreversibility. We thus gain new emerging degrees of freedom describing and allowing us to understand and control complex many body systems which are at the heart of many of our new nano and quantum technologies. Nanoscale thermodynamics does not overturn classical thermodynamics, *μὴ γένοιτο* (God forbid![87]), but it transforms thermodynamics from a theory of steam engines into a framework for understanding, controlling and ultimately optimising the irreversible processes in complex many-body systems at the heart of future technologies.

## 10 Acknowledgements and thanks

This article was commissioned at the initiative of Sir Anthony Leggett, following a talk which one of us gave alongside him at a conference in 2023. We are grateful to him, and we lament that we were not able to share the final article with him and thank him personally while he was alive. We wish to express our thanks to friends and colleagues who read an early draft and offered helpful input and observations: Natalia Ares, Paul Davies, Jon Dexter, Jens Eisert, Marcus Huber, Gershon Kuritzky, Edward Laird, Gerard Milburn, Jan Mol, Tim Palmer, Juan

[87] Paul's letter to the Romans *passim*

Parrondo, Juka Perkola, and Nicole Yunger Halpern; and also the anonymous reviewers. We benefitted from many discussions on clocks with Paul Erker. We thank Diana Briggs for sourcing permissions for figures. JS acknowledges support from the EU through ERC-AdG 'Emergence in Quantum Physics' (EmQ). Finally, we thank Stephen Blundell, both for his expertise in thermal physics and for his unfailing support as Editor-in-Chief.